\documentstyle[preprint,pra,aps,psfig]{revtex}
\begin{document}
\tighten

\draft

\title{Coherent versus Incoherent Dynamics during 
       Bose-Einstein Condensation in Atomic Gases}

\author{H.T.C. Stoof}
\address{Institute for Theoretical Physics, University of Utrecht, \\   
         Princetonplein 5, 3584 CC  Utrecht, The Netherlands}

\maketitle

\begin{abstract}
We review and extend the theory of the dynamics of Bose-Einstein condensation in 
weakly interacting atomic gases. We present in a unified way both the 
semiclassical theory as well as the full quantum theory. This is achieved by 
deriving a Fokker-Planck equation that incorporates both the coherent and 
incoherent effects of the interactions in a dilute Bose gas. In first instance 
we focus our attention on the nonequilibrium dynamics of a homogeneous Bose gas 
with a positive interatomic scattering length. After that we discuss how our 
results can be generalized to the inhomogeneous situation that exists in the 
present experiments with magnetically trapped alkali gases, and how we can deal 
with a negative interatomic scattering length in that case as well. We also show 
how to arrive at a discription of the collective modes of the gas that obeys the 
Kohn theorem at all temperatures. The theory is based on the many-body T-matrix 
approximation throughout, since this approximation has the correct physical 
behavior near the critical temperature and also treats the coherent and 
incoherent processes taking place in the gas on an equal footing. 
\end{abstract}


\section{INTRODUCTION}
\label{int}
In the second half of the 1920's the founding fathers of quantum mechanics were 
mainly concerned with the understanding of the physical world on an atomic 
scale. In particular the study of the spectra of atomic hydrogen and helium 
led, besides the formulation of the famous Schr\"odinger equation \cite{schr} 
and the interpretation of the wavefunction as a probability amplitude 
\cite{born}, to the exclusion principle \cite{pauli1} or more generally the 
(anti)symmetrization of the wavefunction in the case of particles with 
(half)integer spin \cite{pauli2}. This important postulate of modern quantum 
theory divides all particles into two distinct families, which are now of course 
known as fermions and bosons, respectively. 

Already before the discovery of the (anti)symmetrization requirement, it was 
shown on statistical grounds only \cite{bec1,bec2,fermi} that under extreme 
conditions ideal gases of these indistinguishable particles have also 
remarkable features on a macroscopic scale. Furthermore, in the case of an 
interacting gas even more stunning phenomena may occur, however weak the 
interaction. A well-known example in this context is superconductivity. In a 
superconducting metal an attractive interaction between two electrons with 
opposite momenta causes an instability of the Fermi-surface and the formation of 
Cooper pairs \cite{cooper}. The latter are allowed to move freely through the 
lattice resulting in a superconducting current and a vanishing resistance. 
According to the sucessful BCS-theory \cite{BCS} describing this, the attractive 
interaction is the result of a phonon exchange process. It is, however, believed 
that due to the weakness of this interaction only critical temperatures up to 
about 30 K are obtainable. Therefore, the underlying mechanism for the 
high-temperature superconductors cannot be explained in this way and is today 
still a very active field of investigation.

Although the BCS-theory shows that superconductivity can in a certain sense be 
regarded as a result of a Bose-Einstein condensation \cite{bec2,bec3,bec4} of 
the Cooper pairs, the most striking example of this condensation process was 
until recently always associated with the superfluid phase of liquid $^4$He 
\cite{london}. Phenomenologically, the characteristics of superfluidity can be 
explained if the dispersion relation of the elementary excitations differs from 
the particle-like dispersion $\epsilon({\bf k}) = \hbar^2{\bf k}^2/2m$ and is 
linear for small momenta \cite{landau}. To see that this feature indeed leads 
to superfluidity, we consider the situation of liquid $^4$He in a very long 
cylindrical pipe moving with velocity {\bf v} along its symmetry axis 
\cite{baym}. Describing the strongly interacting Bose liquid in the pipe as an 
assembly of noninteracting quasiparticles in a state of thermal equilibrium, we 
find from the usual methods of statistical mechanics that in the laboratory 
frame the number of quasiparticles with momentum $\hbar {\bf k}$ and energy 
$\hbar\omega({\bf k})$ is given by
\begin{equation}
N({\bf k}) = 
 \frac{1}{e^{( \hbar\omega({\bf k}) - \hbar {\bf k} \cdot {\bf v} )/k_BT} -1}~.
\end{equation}

The velocity {\bf v} is still arbitrary at this point, but its magnitude 
has an upper bound because the occupation numbers $N({\bf k})$ 
must be positive. Therefore, we require for all ${\bf k} \neq {\bf 0}$ that
$\hbar\omega({\bf k}) > \hbar {\bf k} \cdot {\bf v}$. In this case the total 
momentum ${\bf P} = \sum_{\bf k} \hbar {\bf k} N({\bf k})$ 
carried along with the walls of the pipe, is at low temperatures $T$ clearly 
much smaller than the momentum $Nm{\bf v}$ that we obtain when the whole liquid 
containing $N$ atoms is moving rigidly with the walls. Hence, if the dispersion 
relation is linear, i.e., $\hbar\omega({\bf k}) = \hbar c k$, we conclude by a 
Galilean transformation that the fluid can have a stationary (frictionless) 
flow if the velocity is small enough and obeys $v < c$ \cite{helium}. Herewith, 
both the existence of superfluidity as well as a critical velocity above which 
the phenomenon cannot take place is explained. It is important to note, 
however, that the above argument is semiclassical and predicts that a superflow 
is absolutely stable if the condition $v < c$ is satisfied. This is incorrect 
in principle, because the superflow can actually decay by means of quantum and 
thermal fluctuations and is therefore only metastable \cite{langer}. As we will 
see below, the effect of quantum and thermal fluctuations on the semiclassical 
picture is a recurrent theme throughout the following.  

At zero temperature, the required linear behavior of the dispersion relation 
was first derived on the basis of a microscopic theory by Bogoliubov \cite{bog} 
and was subsequently put on a firm basis by Beliaev \cite{bel}, Hugenholtz and 
Pines \cite{HP}, and Gavoret and Nozi\`eres \cite{GN}. Simultaneously, it was 
also  shown to be valid at nonzero temperatures by Lee and Yang \cite{lee} 
using the pseudopotential method \cite{huang1}, and more rigorously by 
Hohenberg and Martin \cite{HM} and Popov \cite{popov} using field-theoretical 
methods. At this point it is important to mention also the work of 
Nepomnyashchi\u{\i} and Nepomnyashchi\u{\i}, who explicitly showed that the 
linear dispersion of the elementary excitations also survives if the infrared 
divergencies in the perturbation theory are taken into account exactly 
\cite{NN}. 

From these famous papers the linear dispersion at long wavelengths in a Bose 
system with short-range interactions is therefore well established. However, to 
actually calculate its velocity $c$ from first principles is not feasible in 
general and in particular not for the strongly interacting $^4$He liquid. To 
achieve that we have to consider a dilute Bose gas, for which we can rely on 
the weakness of the interactions or more precisely on the smallness of the gas 
parameter $(na^3)^{1/2}$, where $n$ is the density and $a$ the positive $s$-wave 
scattering length of the particles in the gas. Unfortunately however, until 
recently Bose-Einstein condensation in a gas was never realized experimentally 
and the theory of the weakly interacting Bose gas \cite{bog,bel,lee,popov} 
could therefore not be applied to any real system. This situation changed 
dramatically in the summer of 1995 when within a few months three experimental 
groups reported the observation of Bose-Einstein condensation in a vapor of  
spin-polarized $^{87}$Rb \cite{JILA}, $^7$Li \cite{Rice}, and $^{23}$Na 
\cite{MIT} atoms.

Historically, the first step towards the achievement of Bose-Einstein 
condensation in a dilute gas was made in 1980 by Silvera and Walraven 
\cite{ike}, when they demonstrated the stabilization of spin-polarized atomic 
hydrogen. After this pioneering experiment, the combined efforts of several 
experimental and theoretical groups around the world led to a detailed 
understanding of the atomic and statistical properties of this new quantum 
system \cite{H}. Moreover, just before the reports on the above mentioned 
experiment with the rubidium vapor, doubly spin-polarized atomic hydrogen gases 
still held the record in achieving the required high densities and low 
temperatures and were generally considered to be the most promising candidates 
for the observation of Bose-Einstein condensation \cite{allan1}. Although, the 
latter expectation has clearly been proven wrong, it is fair to say that the 
sucessful experiments with the alkali vapors would not have been possible 
without the knowledge acquired by the atomic hydrogen community. In particular 
the development of magnetic traps \cite{H1,H2} and evaporative cooling 
techniques \cite{H3,H4} has turned out to be of crucial importance for the 
actual achievement of Bose-Einstein condensation \cite{H5}. 

At present, the most important reason for the interest in degenerate Bose gases 
is that we can study the nonequilibrium dynamics of these trapped gases in such 
detail that experimental results can be directly confronted to microscopic 
many-body theories. In particular, the dynamics of condensate formation and the 
collective modes of an inhomogeneous Bose condensed gas are two topics that have 
received considerable attention in the last three years and on which the present 
paper is also focussed. More generally, however, the aim of this paper is to 
develop with the techniques of quantum field theory a unified nonequilibrium 
theory for trapped atomic gases that is capable of treating the dynamics of the 
gas, not only in the normal and superfluid phases, but also during the phase 
transition itself. In agreement with the above goal of the paper, we refrain 
from reviewing other approaches that can and actually have been applied to the 
topics of interest here. Nevertheless, we have included a rather large amount of 
references. In this manner we aim to give, to the best of our knowledge, proper 
reference to important contributions to this field of physics. Furthermore, we 
also want to provide the reader with additional (and sometimes more advanced) 
literature that may be useful for a deeper understanding of various technical 
details, even though this is not necessary for a first reading, since the paper 
is intended to be self contained and essentially only assumes a working 
knowledge of the methods of second quantization, the path-integral formulation 
of quantum mechanics and the Hartree-Fock theory of interacting many-particle 
systems.     

The theoretical challenge that is posed by trapped atomic Bose gases can be 
stated as follows. On the one hand the interactions between the atoms influence 
the one-particle wavefunctions and energies of the gas. On the other hand the 
same interactions also change the occupation numbers of these states. As a 
result of the strong coupling between these, respectively, coherent and 
incoherent processes, the nonequilibrium dynamics of the gas is in principle 
quite complicated and requires a quantum kinetic theory that simultaneously 
incorporates both effects of the interactions. Such a quantum kinetic theory has 
recently been developed by Gardiner and Zoller on the basis of a perturbative 
treatment of the master equation for the many-body density matrix \cite{peter2}. 
Below we ultimately also arrive at a quantum kinetic theory, but this is 
achieved by using field-theoretical tools to derive a single Fokker-Planck 
equation that nevertheless describes both the coherent and incoherent dynamics 
of the gas. In principle we could, of course, also first derive a master 
equation and subsequently the same Fokker-Planck equation by making use of 
well-known procedures from quantum optics. However, we believe that the approach 
presented here is more direct and, most importantly, much more convenient if we 
systematically want to go beyond the pseudopotential method that is not 
sufficiently accurate for our purposes and that is used by Gardiner and Zoller. 

The paper is organized as follows. In Sec.~\ref{CL} we start with a discussion 
of an ideal Bose gas in a trap. In Sec.~\ref{BQD} we consider the isolated case, 
but in Sec.~\ref{BQDR} we couple the gas in the trap also to a thermal 
reservoir. The reason for treating the ideal Bose gas is that in this way we are 
able to introduce most easily all the mathematical tools that are required for 
the discussion of an interacting Bose gas. The necessary extensions for an 
interacting gas are then developed in Secs.~\ref{semicl}, \ref{quant} and 
\ref{inhom}. In these sections we mainly focus on the problem of understanding 
the dynamics of condensate formation, because the theory needs to be the most 
accurate and therefore the most general for this problem. In Secs.~\ref{semicl} 
and \ref{quant} we first treat a homogeneous Bose gas and in Sec.~\ref{inhom} we 
then turn to the inhomogeneous case. Discussing the homogeneous Bose gas first 
is a useful intermediate step, because as a result of the translational symmetry 
the coherent dynamics of the gas involves only the one-particle energies and not 
their wavefunctions. Nevertheless, the dynamics is nontrivial. To get a better 
understanding of the timescales involved in Bose-Einstein condensation we 
therefore first consider in Sec.~\ref{semicl} the semiclassical dynamics of the 
gas, in which fluctuations of the order parameter are neglected. Roughly 
speaking, the semiclassical theory is the nonequilibrium generalization of the 
Bogoliubov theory. Next, we discuss in Sec.~\ref{quant} the full quantum theory 
that incorporates also fluctuations of the order parameter.

Sec.~\ref{inhom} is devoted to the quantum kinetic theory for inhomogeneous Bose 
gases. After our experience with the homogeneous Bose gas we can easily derive 
the desired Fokker-Planck equation for an interacting Bose gas trapped in an 
external potential. In contrast to homogeneous gases there are now, however, two 
cases to consider. If the density of the gas is sufficiently small, the average 
interaction between the atoms is always smaller than the energy splitting 
between the one-particle states in the trap and the coherent dynamics of the gas 
can be neglected. This most simple situation corresponds to the weak-coupling 
limit and is presented in Sec.~\ref{wc}. Experimentally, the weak-coupling limit 
is essentially only realized for atomic $^7$Li due to the effectively attractive 
interatomic interactions of this element. As an application of the weak-coupling 
theory, we therefore discuss in Sec.~\ref{neg} the dynamics of Bose-Einstein 
condensation in a gas with effectively attractive interations in some detail. It 
is only after that, that we turn our attention to the most complicated 
strong-coupling limit and discuss three other applications of our general 
nonequilibrium approach. Since it follows most naturally from the theory 
presented in Sec.~\ref{neg}, we first briefly consider in Sec.~\ref{diff} an 
interesting (zero temperature) property of a condensate, namely the so-called 
`diffusion' of its phase. In Sec.~\ref{sc} we then return to the dynamics of 
Bose-Einstein condensation and in Sec.~\ref{cm} we finally present also a theory 
for the collective modes of a trapped Bose gas. We end the paper in 
Sec.~\ref{conc} with some conclusions and an outlook.

\section{CALDEIRA-LEGGETT LIKE TOY MODEL}
\label{CL}
Before we start with the full problem of Bose-Einstein condensation in an 
interacting gas, we first consider in this section the case of an ideal Bose 
gas. In particular, we consider an isolated, noninteracting Bose gas in an 
external trapping potential $V^{ex}({\bf x})$ with one-particle states 
$\chi_{\alpha}({\bf x})$ and one-particle energies $\epsilon_{\alpha}$. In this 
manner we try to give an introduction to the Schwinger-Keldysh formalism for 
nonequilibrium processes \cite{schwinger,keldysh} that we are going to use 
throughout this paper. A review of the Schwinger-Keldysh formalism has already 
been given by Danielewicz \cite{D}, but he uses the operator language which 
turns out to be less convenient for our purposes. Instead, we need a 
formulation in terms of functional integrals \cite{C}. In addition, we also 
couple this system to a reservoir in a way that is familiar from the 
Caldeira-Leggett model for a particle experiencing friction \cite{tony1}. This 
gives us the opportunity to explain how dissipation, and as a result 
relaxation, is incorporated in the Schwinger-Keldysh formalism. This is 
especially important in Sec.~\ref{quant}, when we discuss the effect of thermal 
fluctuations on the semiclassical theory of Bose-Einstein condensation. 

\subsection{Isolated Ideal Bose Gas}
\label{BQD}
In textbooks an ideal Bose gas is generally discussed in terms of the average 
occupation numbers of the one-particle states $\chi_{\alpha}({\bf x})$ 
\cite{huang2}. Given the density matrix $\hat{\rho}(t_0)$ of the gas at an 
initial time $t_0$, these occupation numbers obey
\begin{equation}
N_{\alpha}(t) = {\rm Tr} \left[ \hat{\rho}(t_0) 
\hat{\psi}^{\dagger}_{\alpha}(t)
                                                \hat{\psi}_{\alpha}(t)
                         \right]~, 
\end{equation}
with $\hat{\psi}^{\dagger}_{\alpha}(t)$ and $\hat{\psi}_{\alpha}(t)$ the usual 
(Heisenberg picture) creation and annihilation operators of second quantization 
\cite{fetter1}, respectively. Because the hamiltonian of the gas 
\begin{equation}
\hat{H} = \sum_{\alpha} \epsilon_{\alpha} \hat{\psi}^{\dagger}_{\alpha}(t)
                                          \hat{\psi}_{\alpha}(t)
\end{equation}
commutes with the number operators 
$\hat{N}_{\alpha}(t) = \hat{\psi}^{\dagger}_{\alpha}(t) \hat{\psi}_{\alpha}(t)$, 
the nonequilibrium dynamics of the system is trivial and the average
occupation numbers are at all times equal to their value at the initial time 
$t_0$. If we are also interested in fluctuations, it is convenient to introduce 
the eigenstates of the number operators, i.e., 
\begin{equation}
|\{N_{\alpha}\};t\rangle = \prod_{\alpha} 
    \frac{ \left( \hat{\psi}^{\dagger}_{\alpha}(t) \right)^{N_{\alpha}} }
         { \sqrt{ N_{\alpha}! } } |0\rangle~,
\end{equation}
and to consider the full probability distribution
\begin{equation}
\label{PN}
P(\{N_{\alpha}\};t) = {\rm Tr} \left[ \hat{\rho}(t_0) 
                             |\{N_{\alpha}\};t\rangle \langle\{N_{\alpha}\};t| 
                           \right]~,
\end{equation}
which is again independent of time for an ideal Bose gas. The average 
occupation numbers are then determined by
\begin{equation}
N_{\alpha}(t) = \sum_{\{N_{\alpha}\}} N_{\alpha} P(\{N_{\alpha}\};t)
\end{equation}
and the fluctuations can be obtained from similar expressions. 

As mentioned in the introduction, we are in the following also interested in 
the phase dynamics of the gas. To study that it is necessary to introduce 
so-called coherent states, which are eigenstates of the annihilation operators 
$\hat{\psi}_{\alpha}(t)$ and are much used both in the field of quantum optics 
\cite{mandel} as well as condensed matter physics \cite{NO}. For our purposes 
it is even more convenient to consider eigenstates of the field operator
\begin{equation}
\hat{\psi}({\bf x},t) = 
         \sum_{\alpha} \hat{\psi}_{\alpha}(t) \chi_{\alpha}({\bf x})~.
\end{equation}
An eigenstate of $\hat{\psi}({\bf x},t)$ with eigenvalue 
$\phi({\bf x}) = \sum_{\alpha} \phi_{\alpha} \chi_{\alpha}({\bf x})$ is given 
by
\begin{equation}
|\phi;t\rangle = 
      \exp \left\{ \int d{\bf x}~ 
              \phi({\bf x}) \hat{\psi}^{\dagger}({\bf x},t)
           \right\} |0\rangle
    = \exp \left\{ \sum_{\alpha} 
              \phi_{\alpha} \hat{\psi}^{\dagger}_{\alpha}(t)
           \right\} |0\rangle
\end{equation}
and is clearly also an eigenstate of $\hat{\psi}_{\alpha}(t)$ with eigenvalue 
$\phi_{\alpha}$. The latter can also be proved immediately by noting that the 
equal-time commutation relation 
$[\hat{\psi}_{\alpha}(t),\hat{\psi}^{\dagger}_{\alpha}(t)] = \hat{1}$ implies 
that $\hat{\psi}_{\alpha}(t)$ acts as $d/d\hat{\psi}^{\dagger}_{\alpha}(t)$ on 
these states. Moreover, from their definition it can easily be shown that these 
eigenstates obey the inner product 
\begin{equation}
\langle\phi;t|\phi';t\rangle =
      \exp \left\{ \int d{\bf x}~ 
              \phi^*({\bf x}) \phi'({\bf x})
           \right\}
    = \exp \left\{ \sum_{\alpha} \phi^*_{\alpha} \phi'_{\alpha} \right\}
\end{equation}
and the completeness relation
\begin{equation}
\label{compl}
\int d[\phi^*]d[\phi]~ \frac{|\phi;t\rangle \langle\phi;t|}
                            {\langle\phi;t|\phi;t\rangle} = \hat{1}~,
\end{equation}
if the integration measure of the functional integral is taken to be 
\begin{equation}
\int d[\phi^*]d[\phi] = \int \prod_{\alpha} 
                             \frac{d\phi^*_{\alpha} d\phi_{\alpha}}{2\pi i}~.
\end{equation}                               

In analogy with the occupation number representation in Eq.~(\ref{PN}), we can 
now obtain a completely different description of the Bose gas by making use of 
the above coherent states and considering the probability distribution
\begin{equation}
\label{prob1}
P[\phi^*,\phi;t] = 
   {\rm Tr} \left[ \hat{\rho}(t_0) 
              \frac{|\phi;t\rangle \langle\phi;t|}
                   {\langle\phi;t|\phi;t\rangle}
            \right]~.
\end{equation}
Although we expect that this probability distribution is again independent of 
time, let us nevertheless proceed to derive its equation of motion in a way 
that can be generalized in Secs.~\ref{semicl} and \ref{quant} when we consider 
an interacting Bose gas. First, we need to expand the density matrix 
$\hat{\rho}(t_0)$ in the above coherent states. For an isolated Bose gas it is 
appropriate to take an initial density matrix that commutes with the total 
number operator $\hat{N} = \sum_{\alpha} \hat{N}_{\alpha}(t)$ and we then find 
the expression
\begin{equation}
\hat{\rho}(t_0) =  \int d[\phi^*_0]d[\phi_0]~ 
                   \rho[|\phi_0|^2;t_0] 
                   \frac{|\phi_0;t_0\rangle \langle\phi_0;t_0|}
                        {\langle\phi_0;t_0|\phi_0;t_0\rangle}~,
\end{equation}
in which the expansion coefficients $\rho[|\phi_0|^2;t_0]$ only depend on the 
amplitude of the field $\phi_0({\bf x})$ but not on its phase. This is 
equivalent to saying that the initial state of the gas does not have a 
spontaneously broken $U(1)$ symmetry. Since we are ultimately interested in the 
dynamics of Bose-Einstein condensation it is of course crucial not to consider 
an initial state in which this symmetry is already broken and we therefore 
never use such initial states in the following.

Next, we substitute this expansion into Eq.~(\ref{prob1}) to obtain 
\begin{equation}
P[\phi^*,\phi;t] = \int d[\phi^*_0]d[\phi_0]~ 
                      \rho[|\phi_0|^2;t_0]
                      \frac{|\langle\phi;t|\phi_0;t_0\rangle|^2}
                           {\langle\phi;t|\phi;t\rangle
                              \langle\phi_0;t_0|\phi_0;t_0\rangle}~.
\end{equation}
This is a particularly useful result, because the time dependence is now 
completely determined by the matrix element $\langle\phi;t|\phi_0;t_0\rangle$ 
for which the functional integral representation is well known \cite{NO}. It is 
given by a `path' integral over all complex field evolutions 
$\psi({\bf x},t_+) = \sum_{\alpha} \psi_{\alpha}(t_+) \chi_{\alpha}({\bf x})$ 
from $t_0$ to $t$. More precisely we have
\begin{equation}
\langle\phi;t|\phi_0;t_0\rangle = 
  \int_{\psi({\bf x},t_0)=\phi_0({\bf x})}^{\psi^*({\bf x},t)=\phi^*({\bf x})}
     d[\psi^*]d[\psi]~ \exp \left\{ \frac{i}{\hbar} S_+[\psi^*,\psi] \right\}~,
\end{equation}                           
with the forward action $S_+[\psi^*,\psi]$ given by
\begin{equation}
S_+[\psi^*,\psi] = \sum_{\alpha} 
   \left\{ -i\hbar \psi^*_{\alpha}(t) \psi_{\alpha}(t) 
   + \int_{t_0}^{t} dt_+~ \psi^*_{\alpha}(t_+) 
        \left( i\hbar \frac{\partial}{\partial t_+} - \epsilon_{\alpha} \right)
                                                \psi_{\alpha}(t_+) \right\}~.
\end{equation}
In the same manner the matrix element 
$\langle\phi;t|\phi_0;t_0\rangle^* = \langle\phi_0;t_0|\phi;t\rangle$ can be 
written as a `path' integral over all field configurations 
$\psi({\bf x},t_-) = \sum_{\alpha} \psi_{\alpha}(t_-) \chi_{\alpha}({\bf x})$  
evolving backward in time from $t$ to $t_0$, i.e.,
\begin{equation}
\langle\phi;t|\phi_0;t_0\rangle^* = 
  \int_{\psi({\bf x},t)=\phi({\bf x})}^{\psi^*({\bf x},t_0)=\phi^*_0({\bf x})}
     d[\psi^*]d[\psi]~ \exp \left\{ \frac{i}{\hbar} S_-[\psi^*,\psi] \right\}~,
\end{equation}   
with a backward action
\begin{eqnarray}
S_-[\psi^*,\psi] &=& \sum_{\alpha} 
   \left\{ -i\hbar \psi^*_{\alpha}(t_0) \psi_{\alpha}(t_0) 
   + \int_{t}^{t_0} dt_-~ \psi^*_{\alpha}(t_-) 
        \left( i\hbar \frac{\partial}{\partial t_-} - \epsilon_{\alpha} \right)
                                                \psi_{\alpha}(t_-) \right\}
                                                                 \nonumber \\
 &=& \sum_{\alpha} 
   \left\{ -i\hbar \psi^*_{\alpha}(t) \psi_{\alpha}(t) 
   + \int_{t}^{t_0} dt_-~ \psi_{\alpha}(t_-) 
        \left( -i\hbar \frac{\partial}{\partial t_-} - \epsilon_{\alpha} 
\right)
                                                \psi^*_{\alpha}(t_-) \right\}~.
\end{eqnarray}

Putting all these results together, we see that the probability distribution 
$P[\phi^*,\phi;t]$ can in fact be represented by a functional integral over all 
fields $\psi({\bf x},t)$ that evolve backwards from $t$ to $t_0$ and then 
forward in time from $t_0$ to $t$. Absorbing the factor $\rho[|\phi_0|^2;t_0]$ 
into the measure of the functional integral, we thus arrive at the desired 
result that
\begin{equation}
\label{prob2}
P[\phi^*,\phi;t] = 
   \int_{\psi({\bf x},t)=\phi({\bf x})}^{\psi^*({\bf x},t)=\phi^*({\bf x})} 
            d[\psi^*]d[\psi]~ 
                     \exp \left\{ \frac{i}{\hbar} S[\psi^*,\psi] \right\}~,
\end{equation}
where the total (backward-forward) action in first instance obeys
\begin{eqnarray}
\label{action}
S[\psi^*,\psi] = S_-[\psi^*,\psi] + S_+[\psi^*,\psi] 
  = -i\hbar \sum_{\alpha} \left( \psi^*_{\alpha}(t) \psi_{\alpha}(t) -
                                                 |\phi_{\alpha}|^2 \right)
                                                 \hspace*{1.5in} \nonumber \\  
   + \sum_{\alpha} \int_{{\cal C}^t} dt'~ \left\{ \frac{1}{2} \left(
      \psi^*_{\alpha}(t') i\hbar \frac{\partial}{\partial t'} \psi_{\alpha}(t')
     - \psi_{\alpha}(t') i\hbar \frac{\partial}{\partial t'} 
\psi^*_{\alpha}(t')
                                          \right)
       - \epsilon_{\alpha} \psi^*_{\alpha}(t') \psi_{\alpha}(t')  \right\}~,
\end{eqnarray}
and the integration along the Schwinger-Keldysh contour ${{\cal C}^t}$ is 
defined by $\int_{{\cal C}^t} dt' =\int_{t}^{t_0} dt_- + \int_{t_0}^{t} dt_+ $ 
\cite{schwinger,keldysh,D}. Note also that in Eq.~(\ref{prob2}) we have 
explicitly specified the boundary conditions on the functional integral. It is 
interesting to mention that these boundary conditions are essentially dictated 
by the topological terms in the action $S[\psi^*,\psi]$, which is a general 
feature of the path-integral formulation of quantum mechanics due to the fact 
that the quantum theory should have the correct (semi)classical limit 
\cite{bernard}. Making use of the periodicity of the field $\psi({\bf x},t)$ on 
the Schwinger-Keldysh contour, the variational principle 
$\delta S[\psi^*,\psi]/\delta \psi^*_{\alpha}(t_{\pm}) = 0$ indeed leads not 
only to the Euler-Lagrange equation 
\begin{equation}
i\hbar \frac{\partial}{\partial t_{\pm}}\psi_{\alpha}(t_{\pm}) =
                             \epsilon_{\alpha} \psi_{\alpha}(t_{\pm})~,
\end{equation}
which agrees with the Heisenberg equation of motion 
$i\hbar \partial\hat{\psi}_{\alpha}(t_{\pm})/\partial t_{\pm} = 
                                             [\hat{\psi}_{\alpha}(t_{\pm}),H]$
for the annihilation operators, but also to the appropriate boundary condition 
$\delta \psi^*_{\alpha}(t) = 0$. In the same way we find the complex conjugate 
results if we require that 
$\delta S[\psi^*,\psi]/\delta \psi_{\alpha}(t_{\pm}) = 0$. Substituting the 
boundary conditions in Eq.~(\ref{action}) and performing a partial integration, 
we then finally obtain for the action
\begin{equation}
S[\psi^*,\psi] = 
  \sum_{\alpha} \int_{{\cal C}^t} dt'~
         \psi^*_{\alpha}(t') \left( i\hbar \frac{\partial}{\partial t'} 
                                - \epsilon_{\alpha} \right) \psi_{\alpha}(t')~,
\end{equation}
which we for completeness sake also rewrite as
\begin{equation}  
S[\psi^*,\psi] = \int_{{\cal C}^t} dt' \int d{\bf x}~
         \psi^*({\bf x},t') \left( i\hbar \frac{\partial}{\partial t'} 
            + \frac{\hbar^2 \nabla^2}{2m} - V^{ex}({\bf x}) \right)   
            \psi({\bf x},t')~.
\end{equation}
                                
We are now in a position to derive the equation of motion, i.e., the 
Fokker-Planck equation, for the probability distibution $P[\phi^*,\phi;t]$. 
This is most easily achieved by performing the variable transformation 
$\psi({\bf x},t_{\pm}) = \phi({\bf x},t') \pm \xi({\bf x},t')/2$ in 
Eq.~(\ref{prob2}). In this manner the fields $\psi({\bf x},t_-)$ and 
$\psi({\bf x},t_+)$ that live on the backward and forward branch of the 
Schwinger-Keldysh contour, respectively, are `projected' onto the real time 
axis. Moreover, at the same time we effect a separation between the 
(semi)classical dynamics described by $\phi({\bf x},t')$ and the quantum 
fluctuations determined by $\xi({\bf x},t')$. After the transformation we have
\begin{equation}
P[\phi^*,\phi;t] = 
   \int_{\phi({\bf x},t)=\phi({\bf x})}^{\phi^*({\bf x},t)=\phi^*({\bf x})}    
     d[\phi^*]d[\phi] \int d[\xi^*]d[\xi]~ 
         \exp \left\{ \frac{i}{\hbar} S[\phi^*,\phi;\xi^*,\xi] \right\}~,
\end{equation}
with 
\begin{equation}
S[\phi^*,\phi;\xi^*,\xi] =
  \sum_{\alpha} \int_{t_0}^t dt'~ \left\{
      \phi^*_{\alpha}(t') \left( i\hbar \frac{\partial}{\partial t'} 
                                - \epsilon_{\alpha} \right) \xi_{\alpha}(t')
     + \xi^*_{\alpha}(t') \left( i\hbar \frac{\partial}{\partial t'} 
                                - \epsilon_{\alpha} \right) \phi_{\alpha}(t')
                         \right\}~.
\end{equation}
Because this action is linear in $\xi_{\alpha}(t')$ and $\xi^*_{\alpha}(t')$, 
the integration over the quantum fluctuations leads just to a constraint and we 
find that
\begin{equation}
P[\phi^*,\phi;t] = 
   \int_{\phi({\bf x},t)=\phi({\bf x})}^{\phi^*({\bf x},t)=\phi^*({\bf x})} 
   d[\phi^*]d[\phi]~ \prod_{\alpha} \delta \left[
     \left( -i \frac{\partial}{\partial t'} 
             - \frac{\epsilon_{\alpha}}{\hbar} \right) \phi^*_{\alpha}(t') 
\cdot
      \left( i \frac{\partial}{\partial t'} 
             - \frac{\epsilon_{\alpha}}{\hbar} \right) \phi_{\alpha}(t')         
                                \right] 
\end{equation}
or equivalently that
\begin{equation}
\label{prob3}
P[\phi^*,\phi;t] = 
      \int d[\phi^*_0]d[\phi_0]~ P[|\phi_0|^2;t_0]
        \prod_{\alpha} \delta \left( |\phi_{\alpha} - \phi^{cl}_{\alpha}(t)|^2
                              \right)~,
\end{equation}
where we used that $P[\phi^*,\phi;t_0]$ is only a function of the amplitude 
$|\phi|^2$ and also introduced the quantity $\phi^{cl}_{\alpha}(t)$ obeying the 
semiclassical equation of motion
\begin{equation}
i\hbar \frac{\partial}{\partial t} \phi^{cl}_{\alpha}(t) =
                             \epsilon_{\alpha} \phi^{cl}_{\alpha}(t)
\end{equation}
and the initial condition $\phi^{cl}_{\alpha}(t_0) = \phi_{0,\alpha}$.

The latter equation is thus solved by 
$\phi^{cl}_{\alpha}(t) = \phi_{0,\alpha} e^{-i\epsilon_{\alpha} (t-t_0)/\hbar}$ 
and we conclude from a simple change of variables in Eq.~(\ref{prob3}) that for 
an ideal Bose gas $P[\phi^*,\phi;t] = P[|\phi|^2;t_0]$, as expected. We also 
see 
from Eq.~(\ref{prob3}) that the desired equation of motion for 
$P[\phi^*,\phi;t]$ reads
\begin{equation}
\label{FP1}
i\hbar \frac{\partial}{\partial t} P[\phi^*,\phi;t] =
  - \left( \sum_{\alpha} \frac{\partial}{\partial \phi_{\alpha}} 
      \epsilon_{\alpha} \phi_{\alpha} \right) P[\phi^*,\phi;t] 
  + \left( \sum_{\alpha} \frac{\partial}{\partial \phi^*_{\alpha}} 
      \epsilon_{\alpha} \phi^*_{\alpha} \right) P[\phi^*,\phi;t]~.
\end{equation}
This is indeed the correct Fokker-Planck equation for an ideal Bose gas 
\cite{carruthers}. To see that we first consider the average 
$\langle \phi_{\alpha} \rangle(t) = 
                       \int d[\phi^*]d[\phi]~ \phi_{\alpha} P[\phi^*,\phi;t]$.
Multiplying Eq.~(\ref{FP1}) with $\phi_{\alpha}$ and integrating over 
$\phi({\bf x})$, we easily find after a partial integration that
\begin{equation}
i\hbar \frac{\partial}{\partial t} \langle \phi_{\alpha} \rangle(t) =
                          \epsilon_{\alpha} \langle \phi_{\alpha} \rangle(t)~,
\end{equation}
which precisely corresponds to the equation of motion of 
$\langle \hat{\psi}_{\alpha}(t) \rangle = 
                             {\rm Tr}[\hat{\rho}(t_0) \hat{\psi}_{\alpha}(t)]$
in the operator formalism. Similarly, we find that 
\begin{equation}
i\hbar \frac{\partial}{\partial t} \langle \phi^*_{\alpha} \rangle(t) =
                      - \epsilon_{\alpha} \langle \phi^*_{\alpha} \rangle(t)~,
\end{equation}
in agreement with the result for 
$\langle \hat{\psi}^{\dagger}_{\alpha}(t) \rangle = 
            {\rm Tr}[\hat{\rho}(t_0) \hat{\psi}^{\dagger}_{\alpha}(t)]$. 
            
Next we consider the average of $|\phi_{\alpha}|^2$, for which we immediately 
obtain
\begin{equation}
\label{phisquare}
i\hbar \frac{\partial}{\partial t} \langle |\phi_{\alpha}|^2 \rangle(t) = 0.
\end{equation}
We expect this result to be related to the fact that in the operator formalism 
the occupation numbers $N_{\alpha}(t)$ are independent of time. Although this 
turns out to be true, to give the precise relation between 
$\langle |\phi_{\alpha}|^2 \rangle(t)$ and $N_{\alpha}(t)$ is 
complicated by the fact that at equal times the operators 
$\hat{\psi}_{\alpha}(t)$ and $\hat{\psi}^{\dagger}_{\alpha}(t)$ do not commute. 
However, the path-integral formulation of quantum mechanics is only capable of 
calculating time-ordered operator products \cite{hagen2}. In our case this 
implies that $\langle |\phi_{\alpha}|^2 \rangle(t)$ is the value at $t' = t$ of
\begin{eqnarray}
{\rm Tr} \left[ \hat{\rho}(t_0)
         T_{{\cal C}^t} \left( \hat{\psi}_{\alpha}(t) 
                           \hat{\psi}^{\dagger}_{\alpha}(t') \right) \right] =
    \Theta(t,t') {\rm Tr} \left[ \hat{\rho}(t_0)
        \hat{\psi}_{\alpha}(t) \hat{\psi}^{\dagger}_{\alpha}(t') \right]
  + \Theta(t',t) {\rm Tr} \left[ \hat{\rho}(t_0)
        \hat{\psi}^{\dagger}_{\alpha}(t') \hat{\psi}_{\alpha}(t) \right]~,
                                                           \nonumber
\end{eqnarray}
with $T_{{\cal C}^t}$ the time-ordening operator on the Schwinger-Keldysh 
contour and $\Theta(t,t')$ the corresponding Heaviside function. Since the 
Heaviside function is equal to $1/2$ at equal times \cite{half}, we conclude 
that 
\begin{equation}
\langle |\phi_{\alpha}|^2 \rangle(t) = N_{\alpha}(t) + \frac{1}{2}
\end{equation}
and that Eq.~(\ref{phisquare}) is thus fully consistent with the operator 
formalism. An intuitive understanding of the relation between 
$\langle |\phi_{\alpha}|^2 \rangle(t)$ and the occupation numbers can be 
obtained by noting that, since all our manipulations up to now have been exact, 
we expect $\langle |\phi_{\alpha}|^2 \rangle(t)$ to contain both classical and 
quantum fluctuations. These correspond precisely to the contributions 
$N_{\alpha}(t)$ and $1/2$, respectively.

Finally we need to discuss the stationary solutions of the Fokker-Planck 
equation. It is not difficult to show that any functional that only depends on 
the amplitudes $|\phi_{\alpha}|^2$ is a solution. As it stands the 
Fokker-Planck equation, therefore, does not lead to a unique equilibrium 
distribution. This is not surprising, because for an isolated, ideal Bose gas 
there is no mechanism for redistributing the particles over the various energy 
levels and thus for relaxation towards equilibrium. However, the situation 
changes when we allow the bosons in the trap to tunnel back and forth to a 
reservoir at a temperature $T$. The corrections to the Fokker-Planck equation 
that are required to describe the physics in this case are considered next. 
However, to determine these corrections in the most convenient way, we have to 
slightly generalize the above theory because with the probability distibution 
$P[\phi^*,\phi;t]$ we are only able to study spatial, but not temporal 
correlations in the Bose gas.

To study also those we follow a well-known procedure in quantum field theory 
\cite{amit,zinn} and construct a generating functional $Z[J,J^*]$ for all 
(time-ordered) correlation functions. It is obtained by performing two steps. 
First, we introduce the probability distribution $P_J[\phi^*,\phi;t]$ for a 
Bose gas in the presence of the external currents $J({\bf x},t)$ and 
$J^*({\bf x},t)$ by adding to the hamiltonian the terms
\begin{eqnarray}
- \hbar \int d{\bf x}~ \left( \hat{\psi}({\bf x},t) J^*({\bf x},t) +
                              J({\bf x},t) \hat{\psi}^{\dagger}({\bf x},t)
                       \right) =
- \hbar \sum_{\alpha} \left( \hat{\psi}_{\alpha}(t) J^*_{\alpha}(t) +
                              J_{\alpha}(t) \hat{\psi}^{\dagger}_{\alpha}(t)
                       \right)~.                                    \nonumber
\end{eqnarray}
As a result we have
\begin{eqnarray}
P_J[\phi^*,\phi;t] = 
   \int_{\psi({\bf x},t)=\phi({\bf x})}^{\psi^*({\bf x},t)=\phi^*({\bf x})} 
            d[\psi^*]d[\psi]~ \exp \left\{ \frac{i}{\hbar} S[\psi^*,\psi]
                               \right\} \hspace*{2.0in}  \nonumber \\
   \times \exp \left\{ i \int_{{\cal C}^t} dt' \int d{\bf x}~ \left(
            \psi({\bf x},t') J^*({\bf x},t') + J({\bf x},t') \psi^*({\bf x},t')
                                                    \right) \right\}~.
\end{eqnarray}
Second, we integrate this expression over $\phi({\bf x})$ to obtain the desired 
generating functional. Hence
\begin{eqnarray}
\label{z}
Z[J,J^*] = \int d[\phi^*]d[\phi]~ P_J[\phi^*,\phi;t]
         = \int d[\psi^*]d[\psi]~ \exp \left\{ \frac{i}{\hbar} S[\psi^*,\psi]
                               \right\} \hspace*{1.0in}  \nonumber \\
   \times \exp \left\{ i \int_{{\cal C}^{\infty}} dt \int d{\bf x}~ \left(
            \psi({\bf x},t) J^*({\bf x},t) + J({\bf x},t) \psi^*({\bf x},t)
                                                    \right) \right\}~.
\end{eqnarray}

It is important to realize that $Z[J,J^*]$ is indeed independent of the time 
$t$ because of the fact that $P_J[\phi^*,\phi;t]$ is a probability distribution 
(cf. Eqs.~(\ref{compl}) and (\ref{prob1})) and thus properly normalized. We are 
therefore allowed to deform the contour ${\cal C}^t$ to any closed contour that 
runs through $t_0$. Since we are in principle interested in all times $t \ge 
t_0$, the most convenient choice is the countour that runs backward from 
infinity to $t_0$ and then forwards from $t_0$ to infinity. This contour is 
denoted by ${\cal C}^{\infty}$ and also called the Schwinger-Keldysh contour in 
the following because there is in practice never confusion with the more 
restricted contour ${\cal C}^t$ that is required when we consider a probability 
distribution. With this choice it is also clear that $Z[J,J^*]$ is a generating 
functional. Indeed, Eq.~(\ref{z}) shows explicitly that all time-ordered 
correlation functions can be obtained by functional differentiation with 
respect to the currents $J({\bf x},t)$ and $J^*({\bf x},t)$. We have, for 
instance, that
\begin{equation}
{\rm Tr}[\hat{\rho}(t_0) \hat{\psi}({\bf x},t)] =
   \left. \frac{1}{i} \frac{\delta}{\delta J^*({\bf x},t)} Z[J,J^*] 
\right|_{J,J^*=0}
\end{equation}
and similarly that
\begin{equation}
\label{corr}
{\rm Tr} \left[ \hat{\rho}(t_0)
         T_{{\cal C}^{\infty}} \left( \hat{\psi}({\bf x},t) 
                           \hat{\psi}^{\dagger}({\bf x}',t') \right) \right] =
   \left. \frac{1}{i^2}
          \frac{\delta^2}{\delta J^*({\bf x},t) \delta J({\bf x}',t')} 
                                                    Z[J,J^*] 
\right|_{J,J^*=0}~.
\end{equation} 
Note that the times $t$ and $t'$ always have to be larger or equal to $t_0$ for 
these identities to be valid.
            
\subsection{Ideal Bose Gas coupled to a Reservoir}
\label{BQDR}
For the reservoir we take an ideal gas of $N$ bosons in a box with volume $V$. 
The states in this box are labeled by the momentum $\hbar{\bf k}$ and equal to 
$\chi_{\bf k}({\bf x}) = e^{i {\bf k} \cdot {\bf x} }/\sqrt{V}$. They have an 
energy $\epsilon({\bf k}) = \hbar^2 {\bf k}^2/2m + \Delta V^{ex}$, where 
$\Delta V^{ex}$ accounts for a possible bias between the potential energies of 
a particle in the center of the trap and a particle in the reservoir. The 
reservoir is also taken to be sufficiently large that it can be treated in the 
thermodynamic limit and is in an equilibrium with temperature $T$ and chemical 
potential $\mu$ for times $t < t_0$. At $t_0$ it is brought into contact with 
the trap by means of a tunnel hamiltonian
\begin{equation}
\hat{H}^{int} = \frac{1}{\sqrt{V}}
  \sum_{\alpha} \sum_{\bf k}~
  \left( t_{\alpha}({\bf k}) 
           \hat{\psi}_{\alpha}(t) \hat{\psi}^{\dagger}_{\bf k}(t)
       + t^*_{\alpha}({\bf k}) 
           \hat{\psi}_{\bf k}(t) \hat{\psi}^{\dagger}_{\alpha}(t) \right)~,
\end{equation}
with complex tunneling matrix elements $t_{\alpha}({\bf k})$ that for 
simplicity are assumed to be almost constant for momenta $\hbar k$ smaller that 
a cutoff $\hbar k_c$ but to vanish rapidly for momenta larger than this 
ultraviolet cutoff. Moreover, we consider here only the low-temperature regime 
in which the thermal de Broglie wavelength $\Lambda = (2\pi\hbar^2/mk_BT)^{1/2}$ 
of the particles obeys $\Lambda \gg 1/k_c$, since this is the most appropriate 
limit for realistic atomic gases.  

To study the evolution of the combined system for times $t \ge t_0$ we thus 
have to deal with the action
\begin{eqnarray}
S[\psi^*,\psi;\psi_R^*,\psi_R] = 
 - \frac{1}{\sqrt{V}} \sum_{\alpha} \sum_{\bf k} \int_{{\cal C}^{\infty}} dt~ 
      \left( t_{\alpha}({\bf k}) \psi_{\alpha}(t) \psi^*_{\bf k}(t)
         + t^*_{\alpha}({\bf k}) \psi_{\bf k}(t) \psi^*_{\alpha}(t) \right)
                                          \hspace*{0.75in} \nonumber \\
 + \sum_{\alpha} \int_{{\cal C}^{\infty}} dt~
         \psi^*_{\alpha}(t) \left( i\hbar \frac{\partial}{\partial t} 
                         - \epsilon_{\alpha} + \mu \right) \psi_{\alpha}(t)                                     
 + \sum_{\bf k} \int_{{\cal C}^{\infty}} dt~ 
         \psi^*_{\bf k}(t) \left( i\hbar \frac{\partial}{\partial t} 
                         - \epsilon({\bf k}) + \mu \right) \psi_{\bf k}(t)~,
\end{eqnarray}
if we measure all energies relative to the chemical potential and also 
introduce the complex field 
$\psi_R({\bf x},t) = \sum_{\bf k} \psi_{\bf k}(t) \chi_{\bf k}({\bf x})$ for 
the degrees of freedom of the reservoir. However, we are only interested in the 
evolution of the Bose gas in the trap and therefore only in the time-ordered 
correlation functions of this part of the system. The corresponding generating 
functional
\begin{eqnarray}
Z[J,J^*] = \int d[\psi^*]d[\psi] \int d[\psi_R^*]d[\psi_R]~ 
   \exp \left\{ \frac{i}{\hbar} S[\psi^*,\psi;\psi_R^*,\psi_R]
                               \right\} \hspace*{1.5in}  \nonumber \\
   \times \exp \left\{ i \int_{{\cal C}^{\infty}} dt \int d{\bf x}~ \left(
            \psi({\bf x},t) J^*({\bf x},t) + J({\bf x},t) \psi^*({\bf x},t)
                                                    \right) \right\}~.
\end{eqnarray} 
is of the same form as the functional integral in Eq.~(\ref{z}), but now with 
an effective action that is defined by
\begin{equation}
\exp \left\{ \frac{i}{\hbar} S^{eff}[\psi^*,\psi] \right\} \equiv
  \int d[\psi_R^*]d[\psi_R]~
          \exp \left\{ \frac{i}{\hbar} S[\psi^*,\psi;\psi_R^*,\psi_R] 
\right\}~.
\end{equation}
Hence, our next task is to integrate out the field $\psi_R({\bf x},t)$, which 
can be done exactly because it only requires the integration of a gaussian.

To familiarize ourselves with the Schwinger-Keldysh formalism, we perform the 
gaussian integration here in some detail. In principle, this can of course be 
done explicitly by making use of the fact that for the initial density matrix 
$\rho_R(t_0)$ of the reservoir we have that
\begin{equation}
\rho_R[|\phi_R|^2;t_0] = \prod_{\bf k} 
   \frac{1}{N({\bf k})} e^{ - |\phi_{\bf k}|^2/N({\bf k}) }~,
\end{equation}
with $N({\bf k}) = 1/(e^{\beta(\epsilon({\bf k})-\mu)}-1)$ the appropriate 
Bose distribution function and $\beta = 1/k_BT$. However, in practice it is 
much more easy to use a different procedure. It is based on the observation that 
if we introduce the $\delta$ function on the Schwinger-Keldysh contour defined 
by$\int_{{\cal C}^{\infty}} dt'~ \delta(t,t') = 1$ and the Green's function 
$G({\bf k};t,t')$ obeying
\begin{equation}
\label{green}
\left( i\hbar \frac{\partial}{\partial t} 
                         - \epsilon({\bf k}) + \mu \right) G({\bf k};t,t') =
                                           \hbar \delta(t,t')~,
\end{equation}
the action $S[\psi^*,\psi;\psi_R^*,\psi_R]$ can be written as a complete 
square,or more precisely as the sum of two squares $S_1[\psi^*,\psi]$ and 
$S_2[\psi^*,\psi;\psi_R^*,\psi_R]$ that are given by 
\begin{eqnarray}
S_1[\psi^*,\psi] = \sum_{\alpha} \int_{{\cal C}^{\infty}} dt~    
         \psi^*_{\alpha}(t) 
            \left( i\hbar \frac{\partial}{\partial t} 
                   - \epsilon_{\alpha} + \mu \right) \psi_{\alpha}(t)
                                               \hspace*{2.4in} \nonumber \\
 - \frac{1}{\hbar V} \sum_{\alpha,\alpha'} \sum_{\bf k}
   \int_{{\cal C}^{\infty}} dt \int_{{\cal C}^{\infty}} dt'~
      \psi^*_{\alpha}(t) t^*_{\alpha}({\bf k}) G({\bf k};t,t') 
                               t_{\alpha'}({\bf k}) \psi_{\alpha'}(t')  
\end{eqnarray}
and 
\begin{eqnarray}
S_2[\psi^*,\psi;\psi_R^*,\psi_R] = \sum_{\bf k} \int_{{\cal C}^{\infty}} dt~
    \left( \psi_{\bf k}^*(t) 
         - \frac{1}{\hbar\sqrt{V}} \sum_{\alpha} \int_{{\cal C}^{\infty}} dt'~
                 t^*_{\alpha}({\bf k}) \psi^*_{\alpha}(t') G({\bf k};t',t)
    \right)                    \hspace*{0.5in} \nonumber \\
  \times \left( i\hbar \frac{\partial}{\partial t} 
                         - \epsilon({\bf k}) + \mu \right)
         \left( \psi_{\bf k}(t) 
         - \frac{1}{\hbar\sqrt{V}} \sum_{\alpha} \int_{{\cal C}^{\infty}} dt'~
                 G({\bf k};t,t') \psi_{\alpha}(t') t_{\alpha}({\bf k}) 
         \right)~,      
\end{eqnarray}
respectively. Since the first term is independent of the field 
$\psi_R({\bf x},t)$, we only need to evaluate the functional integral 
$\int d[\psi_R^*]d[\psi_R]~
                \exp \left( i S_2[\psi^*,\psi;\psi_R^*,\psi_R]/\hbar \right)$. 
Performing a shift in the integration variables, we however see that this 
functional integral is equal to 
\begin{equation}
\int d[\psi_R^*]d[\psi_R]~
   \exp \left\{ \frac{i}{\hbar} 
          \sum_{\bf k} \int_{{\cal C}^{\infty}} dt~ 
             \psi^*_{\bf k}(t) \left( i\hbar \frac{\partial}{\partial t} 
                         - \epsilon({\bf k}) + \mu \right) \psi_{\bf k}(t)
        \right\}
  = {\rm Tr}[\rho_R(t_0)] = 1~.
\end{equation}
As a result the effective action $S^{eff}[\psi^*,\psi]$ is just equal to 
$S_1[\psi^*,\psi]$, which can be slightly rewritten to read
\begin{eqnarray}
\label{Seff}
S^{eff}[\psi^*,\psi]                        \hspace*{5.2in} \nonumber \\
  = \sum_{\alpha,\alpha'} 
    \int_{{\cal C}^{\infty}} dt \int_{{\cal C}^{\infty}} dt'~
         \psi^*_{\alpha}(t) \left\{ 
            \left( i\hbar \frac{\partial}{\partial t} 
                   - \epsilon_{\alpha} + \mu \right) \delta_{\alpha,\alpha'}
                                                     \delta(t,t')
                   - \hbar \Sigma_{\alpha,\alpha'}(t,t') 
                            \right\} \psi_{\alpha'}(t')~,
\end{eqnarray}
with the so-called selfenergy $\Sigma_{\alpha,\alpha'}(t,t')$ obeying
\begin{equation}
\hbar \Sigma_{\alpha,\alpha'}(t,t') =
  \frac{1}{\hbar V} \sum_{\bf k} 
                 t^*_{\alpha}({\bf k}) G({\bf k};t,t') t_{\alpha'}({\bf k})~. 
\end{equation}  

This is our first example of an effective action describing the dynamics of a 
Bose gas. Before we can study its consequences we clearly first need to 
determine the Green's function $G({\bf k};t,t')$ in terms of which the 
selfenergy is expressed. Although we know that this Green's function fulfills 
Eq.~(\ref{green}), we cannot directly solve this equation because we do not 
know the appropriate boundary condition at $t = t'$. To calculate 
$G({\bf k};t,t')$ 
we therefore have to proceed differently. It is however clear from 
Eq.~(\ref{green}) that $G({\bf k};t,t')$ is a property of the reservoir and we 
thus expect that it can somehow be related to a time-ordered correlation 
function of this reservior. To see that explicitly we consider again the 
generating functional of these correlation functions, i.e., 
\begin{eqnarray}
Z_R[J,J^*] = \int d[\psi_R^*]d[\psi_R]~ 
   \exp \left\{ \frac{i}{\hbar} 
          \sum_{\bf k} \int_{{\cal C}^{\infty}} dt~
             \psi^*_{\bf k}(t) \left( i\hbar \frac{\partial}{\partial t} 
                         - \epsilon({\bf k}) + \mu \right) \psi_{\bf k}(t)
                               \right\}           \hspace*{0.6in} \nonumber \\
  \times \exp \left\{ i \sum_{\bf k} \int_{{\cal C}^{\infty}} dt~
            \left(
             \psi_{\bf k}(t) J^*_{\bf k}(t) + J_{\bf k}(t) \psi^*_{\bf k}(t)
            \right)
              \right\}~.
\end{eqnarray} 
It is again a gaussian integral and can thus be evaluated in the same manner as 
before. The result is now
\begin{equation}
Z_R[J,J^*] = \exp \left\{ -i \sum_{\bf k} \int_{{\cal C}^{\infty}} dt 
                                         \int_{{\cal C}^{\infty}} dt'~ 
               J^*_{\bf k}(t) G({\bf k};t,t') J_{\bf k}(t') \right\}~,
\end{equation}
which shows by means of Eq.~(\ref{corr}) that
\begin{equation}
i G({\bf k};t,t') = 
  {\rm Tr} \left[ \hat{\rho}_R(t_0)
         T_{{\cal C}^{\infty}} \left( \hat{\psi}_{\bf k}(t) 
                           \hat{\psi}^{\dagger}_{\bf k}(t') \right) \right]~. 
\end{equation}

Note first of all that this is indeed consistent with Eq.~(\ref{green}), 
because the right-hand side obeys
\begin{eqnarray}
i\hbar \frac{\partial}{\partial t} && \hspace*{-0.2in}
   {\rm Tr} \left[ \hat{\rho}_R(t_0) 
     T_{{\cal C}^{\infty}} \left( \hat{\psi}_{\bf k}(t) 
                           \hat{\psi}^{\dagger}_{\bf k}(t') \right) \right]
                                                          \nonumber \\ 
  &=& i\hbar \delta(t,t') {\rm Tr} \left[ \hat{\rho}_R(t_0) 
                [\hat{\psi}_{\bf k}(t),\hat{\psi}^{\dagger}_{\bf k}(t)] \right]
    + {\rm Tr} \left[ \hat{\rho}_R(t_0)
     T_{{\cal C}^{\infty}} \left( 
                      i\hbar \frac{\partial}{\partial t} \hat{\psi}_{\bf k}(t) 
                           \hat{\psi}^{\dagger}_{\bf k}(t') \right) \right]
                                                          \nonumber \\
  &=& i\hbar \delta(t,t') + (\epsilon({\bf k}) - \mu)
                          {\rm Tr} \left[ \hat{\rho}_R(t_0)
                            T_{{\cal C}^{\infty}} \left( \hat{\psi}_{\bf k}(t) 
                            \hat{\psi}^{\dagger}_{\bf k}(t') \right) \right]
\end{eqnarray}
in the operator formalism. Moreover, from this identification we see that the 
desired solution fulfilling the appropriate boundary conditions is apparently
\begin{equation}
\label{greens}
G({\bf k};t,t') = -i e^{-i(\epsilon({\bf k}) - \mu)(t-t')/\hbar}
 \left\{ \Theta(t,t') \left( 1+ N({\bf k}) \right) +
                                         \Theta(t',t) N({\bf k}) \right\}~.
\end{equation}
The specific dependence on the backward and forward branches of the 
Schwinger-Keldysh contour is thus solely determined by the Heaviside function 
$\Theta(t,t')$. As a result it is convenient to decompose the Green's function 
into its analytic pieces $G^>({\bf k};t-t')$ and $G^<({\bf k};t-t')$ 
\cite{kadanoff} by means of
\begin{equation}
G({\bf k};t,t') = \Theta(t,t') G^>({\bf k};t-t') 
                                         + \Theta(t',t) G^<({\bf k};t-t')~. 
\end{equation} 
Due to the fact that we are always dealing with time-ordered correlation 
functions, such a decomposition turns out to be a generic feature of all the 
functions on the Schwinger-Keldysh contour that we will encounter in the 
following \cite{langreth}. For a general function $F(t,t')$, it is however also 
possible to have $\delta$-function singularities. If that happens the correct 
decomposition is
\begin{equation}
F(t,t') = F^{\delta}(t) \delta(t,t') + \Theta(t,t') F^>(t,t') 
                                         + \Theta(t',t) F^<(t,t')~. 
\end{equation}
This more general decomposition is not required here, but will be needed in 
Secs.~\ref{semicl} and \ref{quant} when we determine the effective interaction 
between two atoms in a gas.

Having obtained the Green's function of the reservoir, we can now return to our 
discussion of the effective action $S^{eff}[\psi^*,\psi]$ for the Bose gas in 
the trap. Although we have chosen to derive in Eq.~(\ref{Seff}) the effective 
action for the generating functional $Z[J,J^*]$, it is straightforward to show 
that the effective action for the probability distribution $P[\phi^*,\phi;t]$ is 
obtained by only replacing the contour ${\cal C}^{\infty}$ by ${\cal C}^{t}$. 
In the following we therefore no longer always specify all the boundary 
conditions on the time integration, if the precise details of this integration 
are not important and the discussion applies equally well to both cases. Keeping 
this in mind, we now again perform the transformation 
$\psi_{\alpha}(t_{\pm}) = \phi_{\alpha}(t) \pm \xi_{\alpha}(t)/2$ to explicitly 
separate the (semi)classical dynamics from the effect of fluctuations. It leads 
in first instance to \cite{remark}
\begin{eqnarray}
\label{Sxi}
S^{eff}[\phi^*,\phi;\xi^*,\xi] &=&
  \sum_{\alpha} \int_{t_0} dt~ \left\{
      \phi^*_{\alpha}(t) \left( i\hbar \frac{\partial}{\partial t} 
                           - \epsilon_{\alpha} + \mu \right) \xi_{\alpha}(t)
     + \xi^*_{\alpha}(t) \left( i\hbar \frac{\partial}{\partial t} 
                           - \epsilon_{\alpha} + \mu \right) \phi_{\alpha}(t)
                         \right\}                        \nonumber \\
&-& \sum_{\alpha,\alpha'} 
    \int_{t_0} dt \int_{t_0} dt'~
          \left( \phi^*_{\alpha}(t) 
                    \hbar \Sigma^{(-)}_{\alpha,\alpha'}(t-t') \xi_{\alpha'}(t')
                + \xi^*_{\alpha}(t)
                    \hbar \Sigma^{(+)}_{\alpha,\alpha'}(t-t') 
\phi_{\alpha'}(t')
          \right)                                       \nonumber \\
&-& \frac{1}{2} \sum_{\alpha,\alpha'} 
    \int_{t_0} dt \int_{t_0} dt'~ 
                  \xi^*_{\alpha}(t)
                    \hbar \Sigma^{K}_{\alpha,\alpha'}(t-t') \xi_{\alpha'}(t')~,
\end{eqnarray}
where we introduced the retarded and advanced components of the selfenergy 
\begin{equation}
\Sigma^{(\pm)}_{\alpha,\alpha'}(t-t') = \pm \Theta(\pm(t-t')) 
  \left( \Sigma^{>}_{\alpha,\alpha'}(t-t') 
                               - \Sigma^{<}_{\alpha,\alpha'}(t-t') \right)
\end{equation} 
that affect the terms in the action that are linear in $\xi^*_{\alpha}(t)$ and 
$\xi_{\alpha}(t)$, respectively, and also the Keldysh component
\begin{equation}
\Sigma^{K}_{\alpha,\alpha'}(t-t') = 
  \left( \Sigma^{>}_{\alpha,\alpha'}(t-t') 
                               + \Sigma^{<}_{\alpha,\alpha'}(t-t') \right)
\end{equation}
that is associated with the part quadratic in the fluctuations. 

The physical content of these various components of the selfenergy is 
understood most clearly if we now apply the beautiful functional-integral 
procedure that is due to Hubbard and Stratonovich 
\cite{stratonovich,hubbard,NO}. The basic idea is to write the factor 
\begin{eqnarray}
\exp \left\{ - \frac{i}{2} \sum_{\alpha,\alpha'} 
               \int_{t_0} dt \int_{t_0} dt'~ 
                 \xi^*_{\alpha}(t)
                    \Sigma^{K}_{\alpha,\alpha'}(t-t') \xi_{\alpha'}(t')
     \right\}                                               \nonumber
\end{eqnarray}
in the integrant of the functional integral 
$\int d[\phi^*]d[\phi] \int d[\xi^*]d[\xi]~ 
                   \exp\left( iS^{eff}[\phi^*,\phi;\xi^*,\xi]/\hbar \right)$
as a gaussian integral over a complex field $\eta({\bf x},t)$. It is 
equivalent, but in practice more convenient, to just multiply the integrant by a 
factor $1$ that is written as a gaussian integral 
$\int d[\eta^*]d[\eta]~ \exp\left( iS^{eff}[\eta^*,\eta]/\hbar \right)$ 
with
\begin{eqnarray}
S^{eff}[\eta^*,\eta] =
  \frac{1}{2} \sum_{\alpha,\alpha'} \int_{t_0} dt \int_{t_0} dt'~ 
     \left(2 \eta^*_{\alpha}(t) - \sum_{\alpha''} \int_{t_0} dt''~
              \xi^*_{\alpha''}(t'') \hbar \Sigma^K_{\alpha'',\alpha}(t''-t)
     \right)                                    \hspace*{0.8in} \nonumber \\
  \times (\hbar\Sigma^K)^{-1}_{\alpha,\alpha'}(t-t') 
     \left(2 \eta_{\alpha'}(t') - \sum_{\alpha''} \int_{t_0} dt''~
              \hbar \Sigma^K_{\alpha',\alpha''}(t'-t'') \xi_{\alpha''}(t'') 
     \right) 
\end{eqnarray}
a complete square. Adding this to $S^{eff}[\phi^*,\phi;\xi^*,\xi]$ the total 
effective action becomes
\begin{eqnarray}
\label{snoise}
& & \hspace*{-0.4in} S^{eff}[\phi^*,\phi;\xi^*,\xi;\eta^*,\eta]  \nonumber \\
&=& \sum_{\alpha,\alpha'} \int_{t_0} dt \int_{t_0} dt'~
      \phi^*_{\alpha}(t) 
       \left\{ \left( i\hbar \frac{\partial}{\partial t} 
                      - \epsilon_{\alpha} + \mu - \eta^*_{\alpha}(t) \right)
                                \delta_{\alpha,\alpha'} \delta(t-t')
               - \hbar \Sigma^{(-)}_{\alpha,\alpha'}(t-t')
       \right\} \xi_{\alpha'}(t')                          \nonumber \\
&+& \sum_{\alpha,\alpha'} \int_{t_0} dt \int_{t_0} dt'~
      \xi^*_{\alpha}(t) 
       \left\{ \left( i\hbar \frac{\partial}{\partial t} 
                      - \epsilon_{\alpha} + \mu - \eta_{\alpha}(t) \right)
                                \delta_{\alpha,\alpha'} \delta(t-t')
               - \hbar \Sigma^{(+)}_{\alpha,\alpha'}(t-t')
       \right\} \phi_{\alpha'}(t')                          \nonumber \\
&+& 2 \sum_{\alpha,\alpha'} \int_{t_0} dt \int_{t_0} dt'~ 
        \eta^*_{\alpha}(t) (\hbar\Sigma^K)^{-1}_{\alpha,\alpha'}(t-t') 
        \eta_{\alpha'}(t')
\end{eqnarray}
and is thus linear in $\xi_{\alpha}(t)$ and $\xi^*_{\alpha}(t)$. Integrating 
over these fluctuations we conclude from this action that the field 
$\phi({\bf x},t)$ is constraint to obey the Langevin equations \cite{nico}
\begin{equation}
\label{lan1}
i\hbar \frac{\partial}{\partial t} \phi_{\alpha}(t)
  = (\epsilon_{\alpha} - \mu)\phi_{\alpha}(t) 
    +  \sum_{\alpha'} \int_{t_0}^{\infty} dt'~
           \hbar \Sigma^{(+)}_{\alpha,\alpha'}(t-t') \phi_{\alpha'}(t') 
    + \eta_{\alpha}(t)
\end{equation}
and
\begin{equation}
\label{lan2}
-i\hbar \frac{\partial}{\partial t} \phi^*_{\alpha}(t)
  = (\epsilon_{\alpha} - \mu)\phi^*_{\alpha}(t) 
    +  \sum_{\alpha'} \int_{t_0}^{\infty} dt'~
           \phi^*_{\alpha'}(t') \hbar \Sigma^{(-)}_{\alpha',\alpha}(t'-t)  
    + \eta^*_{\alpha}(t)
\end{equation}
with gaussian noise terms $\eta_{\alpha}(t)$ and $\eta^*_{\alpha}(t)$ that from 
the last term in the right-hand side of Eq.~(\ref{snoise}) are seen to have the 
time correlations
\begin{equation}
\langle \eta^*_{\alpha}(t) \eta_{\alpha'}(t') \rangle =
                      \frac{i\hbar^2}{2} \Sigma^K_{\alpha,\alpha'}(t-t')
    = \frac{1}{2} \int \frac{d{\bf k}}{(2\pi)^3}~ \left( 1+2N({\bf k}) \right)
                 t^*_{\alpha}({\bf k}) 
                    e^{-i(\epsilon({\bf k}) - \mu)(t-t')/\hbar}
                 t_{\alpha'}({\bf k})
\end{equation}
in the thermodynamic limit. Moreover, in that limit the retarded selfenergy is 
equal to
\begin{equation}
\hbar \Sigma^{(+)}_{\alpha,\alpha'}(t-t') =
  - \Theta(t-t') \frac{i}{\hbar} \int \frac{d{\bf k}}{(2\pi)^3}~  
                 t^*_{\alpha}({\bf k}) 
                    e^{-i(\epsilon({\bf k}) - \mu)(t-t')/\hbar}
                 t_{\alpha'}({\bf k})
\end{equation}
and the advanced selfenergy can be found from the relation
$\Sigma^{(-)}_{\alpha',\alpha}(t'-t) 
                     = \left( \Sigma^{(+)}_{\alpha,\alpha'}(t-t') \right)^*$.
Note that the Heaviside functions in these retarded and advanced selfenergies 
are precisely such that Eqs.~(\ref{lan1}) and (\ref{lan2}) are causal, i.e., 
the evolution of $\phi({\bf x},t)$ and $\phi^*({\bf x},t)$ depends only on their 
value at previous times. This is clearly a sensible result and explains why 
these components of the full selfenergy enter into the terms in the action that 
are linear in the fluctuations.

To understand why the Keldysh component enters into the terms that are 
quadratic in the fluctuations is more complicated and is related to the famous 
fluctuation-dissipation theorem \cite{kadanoff,nico,forster}. Physically, the 
fluctuation-dissipation theorem ensures in our case that, due to the 
interactions with the reservoir, the gas in the trap now relaxes towards 
thermal equilibrium in the limit $t \rightarrow \infty$. Let us therefore study 
this limit in somewhat more detail. However, before we do so we would like to 
mention for completeness that if we apply the above Hubbard-Stratonovich 
transformation to the functional integral for the probability distribution 
$P[\phi^*,\phi;t]$,we immediately find that analogous to Eq.~(\ref{prob3})     
\begin{equation}
P[\phi^*,\phi;t] =
      \int d[\eta^*]d[\eta]~ P[\eta^*,\eta]
      \int d[\phi^*_0]d[\phi_0]~ P[|\phi_0|^2;t_0]
        \prod_{\alpha} \delta \left( |\phi_{\alpha} - \phi^{cl}_{\alpha}(t)|^2
                              \right)~,
\end{equation}
where $\phi^{cl}_{\alpha}(t)$ now obeys the Langevin equation in 
Eq.~(\ref{lan1}) for a particular realization of the noise $\eta_{\alpha}(t)$ 
and with the initial condition $\phi^{cl}_{\alpha}(t_0)=\phi_{0,\alpha}$. 
Furthermore, the probability distribution for each realization of the noise is 
given by
\begin{equation}
P[\eta^*,\eta] = \exp \left\{ \frac{2i}{\hbar}
   \sum_{\alpha,\alpha'} \int_{t_0}^t dt' \int_{t_0}^t dt''~ 
        \eta^*_{\alpha}(t') (\hbar\Sigma^K)^{-1}_{\alpha,\alpha'}(t'-t'') 
        \eta_{\alpha'}(t'') \right\}~.
\end{equation}
This result shows explicitly that, as expected, the Fokker-Planck equation 
associated with the Langevin equations in Eqs.~(\ref{lan1}) and (\ref{lan2}) is 
in fact the desired Fokker-Planck equation for the probability distribution 
$P[\phi^*,\phi;t]$. Hence, this Fokker-Planck equation can be obtained either 
from these Langevin equations or directly from the effective action in 
Eq.~(\ref{Sxi}), without performing a Hubbard-Stratonovich transformation. We 
now first follow the former route, but lateron also explain the more direct 
procedure because that is interesting in its own right and also shows the 
consistency of the theory. 

After this slight digression, we now return to the long-time behaviour of the 
gas. In the long-time limit two important simplifications occur. Looking at the 
expresions for the selfenergies, we see that their width as a function of the 
time difference $t-t'$ is at most of order ${\cal O}(\hbar/k_BT)$ in the 
low-temperature regime of interest where the thermal de Broglie wavelength of 
the particles obeys $\Lambda \gg 1/k_c$. If therefore 
$t \gg t_0 + {\cal O}(\hbar/k_BT)$ we are allowed to take the limit 
$t_0 \rightarrow -\infty$. Taking this limit means physically that we neglect 
the initial transients that are due to the precise way in which the contact 
between the trap and the reservoir is made, and focus on the `universal' 
dynamics that is independent of these details. In addition, at long times the 
dynamics of the gas is expected to be sufficiently slow that we can neglect the 
memory effects altogether. Finally, we consider here first the case of a 
reservoir that is so weakly coupled to the gas in the trap that we can treat 
the coupling with second order perturbation theory. As a result we can also 
neglect the nondiagonal elements of the selfenergies. In total we are then 
allowed to put (but see below) 
\begin{equation}
\Sigma^{(\pm),K}_{\alpha,\alpha'}(t-t') \simeq 
    \Sigma^{(\pm),K}_{\alpha,\alpha}(\epsilon_{\alpha}-\mu)
                \delta_{\alpha,\alpha'} \delta(t-t')
     \equiv \Sigma^{(\pm),K}_{\alpha} \delta_{\alpha,\alpha'} \delta(t-t')~,
\end{equation}
where the Fourier transform of the selfenergies is defined by
\begin{equation}                                                               
\Sigma^{(\pm),K}_{\alpha,\alpha'}(t-t') 
   = \int d\epsilon~ \Sigma^{(\pm),K}_{\alpha,\alpha'}(\epsilon) 
                                 \frac{e^{-i\epsilon 
(t-t')/\hbar}}{2\pi\hbar}~.
\end{equation}
Note that this implies in general that
$\Sigma^{(-)}_{\alpha',\alpha}(\epsilon) 
                 = \left( \Sigma^{(+)}_{\alpha,\alpha'}(\epsilon) \right)^*$, 
and in particular therefore that 
$\Sigma^{(-)}_{\alpha} = \left( \Sigma^{(+)}_{\alpha} \right)^*$. 

With these simplifications our Langevin equations become 
\begin{equation}
i\hbar \frac{\partial}{\partial t} \phi_{\alpha}(t)
  = (\epsilon_{\alpha} + \hbar\Sigma^{(+)}_{\alpha} - \mu)\phi_{\alpha}(t) 
    + \eta_{\alpha}(t)
\end{equation}
and just the complex conjugate equation for $\phi^*_{\alpha}(t)$. The retarded 
selfenergy in this equation is given by
\begin{equation}
\hbar\Sigma^{(+)}_{\alpha} 
  = \int \frac{d{\bf k}}{(2\pi)^3}~  
                 t^*_{\alpha}({\bf k}) 
                    \frac{1}{\epsilon_{\alpha}^+ - \epsilon({\bf k})} 
                 t_{\alpha}({\bf k})~,
\end{equation}
with $\epsilon^+_{\alpha}$ the usual notation for the limiting procedure
$\epsilon_{\alpha} + i0$. It clearly has real and imaginary parts that we 
denote by $S_{\alpha}$ and $- R_{\alpha}$, respectively. Denoting the Cauchy 
principle value part of an integral by ${\cal P}$, they obey
\begin{equation}
\label{shift}
S_{\alpha}
  = \int \frac{d{\bf k}}{(2\pi)^3}~
                 t^*_{\alpha}({\bf k}) 
                    \frac{{\cal P}}{\epsilon_{\alpha} - \epsilon({\bf k})} 
                 t_{\alpha}({\bf k})
\end{equation}
and
\begin{equation}
\label{decay}
R_{\alpha}
  = \pi \int \frac{d{\bf k}}{(2\pi)^3}~
        \delta( \epsilon_{\alpha} - \epsilon({\bf k}) )         
                                          |t_{\alpha}({\bf k})|^2 ~.
\end{equation} 
The interpretation of these results is quite obvious if we consider the average 
of the Langevin equation, i.e.,
\begin{equation}
i\hbar \frac{\partial}{\partial t} \langle \phi_{\alpha} \rangle(t)
  = (\epsilon_{\alpha} + S_{\alpha} - iR_{\alpha} - \mu)
                                   \langle \phi_{\alpha} \rangle(t)~,
\end{equation}
which is solved by
\begin{equation}
\langle \phi_{\alpha} \rangle(t) = \langle \phi_{\alpha} \rangle(0)
    e^{-i(\epsilon_{\alpha} + S_{\alpha} - \mu)t/\hbar} 
e^{-R_{\alpha}t/\hbar}~.
\end{equation}
Hence the real part of the retarded selfenergy $S_{\alpha}$ represents the 
shift in the energy of state $\chi_{\alpha}({\bf x})$ due to the coupling with 
the reservoir. Indeed, Eq.~(\ref{shift}) agrees precisely with the value of this 
shift in second order perturbation theory. Moreover, the fact that
$|\langle \phi_{\alpha} \rangle(t)|^2 = 
         |\langle \phi_{\alpha} \rangle(0)|^2 e^{-2R_{\alpha}t/\hbar}$
shows that the average rate of decay $\Gamma_{\alpha}$ of the state 
$\chi_{\alpha}({\bf x})$ is equal to $2R_{\alpha}/\hbar$, which, together with 
Eq.~(\ref{decay}), exactly reproduces Fermi's Golden Rule.

From the equation of motion for the average $\langle \phi_{\alpha} \rangle(t)$ 
we can also conclude that the right-hand side of the Fokker-Planck equation 
must contain the `streaming' terms
\begin{eqnarray}
- \left( \sum_{\alpha} \frac{\partial}{\partial \phi_{\alpha}} 
     (\epsilon_{\alpha} + \hbar\Sigma^{(+)}_{\alpha} - \mu)
                                      \phi_{\alpha} \right) P[\phi^*,\phi;t] 
  + \left( \sum_{\alpha} \frac{\partial}{\partial \phi^*_{\alpha}} 
     (\epsilon_{\alpha} + \hbar\Sigma^{(-)}_{\alpha} - \mu)
                                    \phi^*_{\alpha} \right) P[\phi^*,\phi;t]~.
                                                            \nonumber
\end{eqnarray}
However, there must now also be a `diffusion' term due to the fact that the 
average $\langle |\phi_{\alpha}|^2 \rangle(t)$ is no longer independent of time 
since the interactions with the reservoir can lead to changes in the occupation 
numbers $N_{\alpha}(t)$. To obtain the `diffusion' term we thus need to 
determine $i\hbar \partial \langle |\phi_{\alpha}|^2 \rangle(t)/\partial t$. To 
do so we first formally solve the Langevin equation by
\begin{equation}
\phi_{\alpha}(t) = 
       e^{-i(\epsilon_{\alpha} + \hbar\Sigma^{(+)}_{\alpha} - \mu)t/\hbar}
   \left\{ \phi_{\alpha}(0)
     - \frac{i}{\hbar} \int_0^t dt'~\eta(t') 
          e^{i(\epsilon_{\alpha} + \hbar\Sigma^{(+)}_{\alpha} - \mu)t'/\hbar} 
                                                                 \right\}~.
\end{equation}
Multiplying this with the complex conjugate expression and taking the average, 
we obtain first of all
\begin{equation}
\langle |\phi_{\alpha}|^2 \rangle(t) = e^{-2R_{\alpha}t/\hbar}
   \left\{ \langle |\phi_{\alpha}|^2 \rangle(0)
           + \frac{i}{2} \Sigma^K_{\alpha} 
                    \int_0^t dt'~ e^{2R_{\alpha}t'/\hbar} \right\}~,  
\end{equation}
which shows that 
\begin{equation}
\label{fluc}
i\hbar \frac{\partial}{\partial t} \langle |\phi_{\alpha}|^2 \rangle(t)
  = -2iR_{\alpha} \langle |\phi_{\alpha}|^2 \rangle(t)
    - \frac{1}{2} \hbar \Sigma^K_{\alpha}~.
\end{equation}
Moreover, the Keldysh component of the selfenergy is given by
\begin{equation}
\hbar \Sigma^K_{\alpha}
= -2\pi i \int \frac{d{\bf k}}{(2\pi)^3}~
                 (1 + 2N({\bf k}))
                    \delta( \epsilon_{\alpha} - \epsilon({\bf k}) )
                                                |t_{\alpha}({\bf k})|^2 
\end{equation} 
and therefore obeys
\begin{equation}
\hbar \Sigma^K_{\alpha} = -2i(1 + 2N_{\alpha}) R_{\alpha}
\end{equation}
with $N_{\alpha} = (e^{\beta(\epsilon_{\alpha} -\mu)} -1 )^{-1}$ the Bose 
distribution function. This is in fact the fluctuation-dissipation theorem, 
because it relates the strength of the fluctuations determined by $\hbar 
\Sigma^K_{\alpha}$, to the amount of dissipation that is given by $R_{\alpha}$.
As mentioned previously, the fluctuation-dissipation theorem ensures that the 
gas relaxes to thermal equilibrium. This we can already see from 
Eq.~(\ref{fluc}), because substitution of the fluctuation-dissipation theorem 
leads to 
\begin{equation}
\label{fluct}
i\hbar \frac{\partial}{\partial t} \langle |\phi_{\alpha}|^2 \rangle(t)
  = -2iR_{\alpha} \langle |\phi_{\alpha}|^2 \rangle(t)
    + iR_{\alpha}(1 + 2N_{\alpha})~,
\end{equation}
which tells us that in equilibrium 
$\langle |\phi_{\alpha}|^2 \rangle = N_{\alpha} + 1/2$. In fact, we have argued 
in Sec.~\ref{BQD} that in general 
$\langle |\phi_{\alpha}|^2 \rangle(t) = N_{\alpha}(t) + 1/2$. Substituting this
identity in Eq.~(\ref{fluct}) and realizing that 
$\Gamma_{\alpha} N_{\alpha} = \Gamma_{\alpha} N({\bf k})$ due to the 
energy-conserving $\delta$ function in $\Gamma_{\alpha}$, we indeed obtain the 
correct rate equation for the average occupation numbers
\begin{equation}
\label{qdBE}
\frac{\partial}{\partial t} N_{\alpha}(t) 
   = - \Gamma_{\alpha} N_{\alpha}(t)  + \Gamma_{\alpha} N_{\alpha}
   = - \Gamma_{\alpha} N_{\alpha}(t)(1+N({\bf k})) 
                    + \Gamma_{\alpha} (1+N_{\alpha}(t)) N({\bf k})~,
\end{equation}
that might justly be called the quantum Boltzmann equation for the gas, because 
the right-hand side contains precisely the rates for scattering into and out of 
the reservoir.  

Furthermore, Eq.~(\ref{fluc}) shows that the `diffusion' term in the 
Fokker-Planck equation is
\begin{eqnarray}
- \left( \frac{1}{2} \sum_{\alpha} 
    \frac{\partial^2}{\partial \phi^*_{\alpha} \partial \phi_{\alpha}} 
     \hbar \Sigma^K_{\alpha} \right) P[\phi^*,\phi;t]~,  \nonumber
\end{eqnarray}
because the first term in the right-hand side of Eq.~(\ref{fluc}) is also 
present without the noise and is therefore already accounted for in the 
`streaming' terms. In total we have thus arrived at
\begin{eqnarray}
\label{FP2}
&& \hspace*{-0.3in} i\hbar \frac{\partial}{\partial t} P[\phi^*,\phi;t] =
                                                         \nonumber \\
&-& \left( \sum_{\alpha} \frac{\partial}{\partial \phi_{\alpha}} 
       (\epsilon_{\alpha} + \hbar\Sigma^{(+)}_{\alpha} - \mu)
                                      \phi_{\alpha} \right) P[\phi^*,\phi;t] 
  + \left( \sum_{\alpha} \frac{\partial}{\partial \phi^*_{\alpha}} 
     (\epsilon_{\alpha} + \hbar\Sigma^{(-)}_{\alpha} - \mu)
                                    \phi^*_{\alpha} \right) P[\phi^*,\phi;t]
                                                         \nonumber \\
&-& \left( \frac{1}{2} \sum_{\alpha} 
    \frac{\partial^2}{\partial \phi^*_{\alpha} \partial \phi_{\alpha}} 
      \hbar \Sigma^K_{\alpha} \right) P[\phi^*,\phi;t]~.                                                                                                       
\end{eqnarray}
Using again the fluctuation-dissipation theorem, it is not difficult to show 
that the stationary solution of this Fokker-Planck equation is 
\begin{equation}
\label{stat}
P[\phi^*,\phi;\infty] = \prod_{\alpha} \frac{1}{N_{\alpha} + 1/2}
     \exp \left\{ - \frac{1}{N_{\alpha} +1/2} |\phi_{\alpha}|^2
          \right\}~.
\end{equation}
Although this appears to be quite reasonable, it is important to note that this 
result is not fully consistent because the `streaming' terms in Eq.~(\ref{FP2}) 
show that the energies of the states $\chi_{\alpha}({\bf x})$ are shifted and 
equal to $\epsilon_{\alpha} + S_{\alpha}$. We therefore expect that the 
equilibrium occupation numbers are equal to the Bose distribution evaluated at 
those energies and not at the unshifted values $\epsilon_{\alpha}$. The reason 
for this inconsistency is, of course, that we have taken 
$\Sigma^{(\pm),K}_{\alpha}$ equal to 
$\Sigma^{(\pm),K}_{\alpha,\alpha}(\epsilon_{\alpha}-\mu)$, which was only 
justified on the basis of second-order perturbation theory. We now, however, 
see that to obtain a fully consistent theory we must take instead
$\Sigma^{(\pm),K}_{\alpha} \equiv  
        \Sigma^{(\pm),K}_{\alpha,\alpha}(\epsilon_{\alpha}+S_{\alpha}-\mu)$~.
This implies that to obtain the retarded selfenergy, we first have to solve
for the real part  
\begin{equation}
S_{\alpha}
  = \int \frac{d{\bf k}}{(2\pi)^3}~
         t^*_{\alpha}({\bf k}) 
            \frac{{\cal P}}{\epsilon_{\alpha} + S_{\alpha} - \epsilon({\bf k})} 
                t_{\alpha}({\bf k})
\end{equation}
and then substitute this into 
\begin{equation}
R_{\alpha}
  = \pi \int \frac{d{\bf k}}{(2\pi)^3}~
        \delta( \epsilon_{\alpha} + S_{\alpha} - \epsilon({\bf k}) )         
                                                   |t_{\alpha}({\bf k})|^2 
\end{equation}      
to obtain the imaginary part. The fluctuation-dissipation theorem then again 
reads $\hbar \Sigma^K_{\alpha} = -2i(1 + 2N_{\alpha}) R_{\alpha}$ but now with 
the equilibrium distribution
$N_{\alpha} = (e^{\beta(\epsilon_{\alpha} + S_{\alpha} -\mu)} -1 )^{-1}$. As a 
consequence the quantum Boltzmann equation and the stationary state of the 
Fokker-Planck equation will now also contain these correct equilibrium 
occupation numbers \cite{PQW}.

This almost completes our discussion of this Caldeira-Leggett like toy model, 
that nevertheless already shows many of the features that we will encounter in 
Sec.~\ref{semicl} and \ref{quant} when we study a realistic interacting Bose 
gas. However, we promised to derive the Fokker-Planck equation also derictly 
from the effective action $S^{eff}[\phi^*,\phi;\xi^*,\xi]$, without making use 
of a Hubberd-Stratonovich transformation and the Langevin equations that are an 
immediate result of that transformation. Making the same approximations on the 
selfenergies as before, the effective action for the probability distribution 
$P[\phi^*,\phi;t]$ reads
\begin{eqnarray}
S^{eff}[\phi^*,\phi;\xi^*,\xi] = 
  \sum_{\alpha} \int_{t_0}^t dt'~ 
      \phi^*_{\alpha}(t') \left( i\hbar \frac{\partial}{\partial t'} 
                       - \epsilon_{\alpha} - \hbar \Sigma^{(-)}_{\alpha} + \mu
                         \right) \xi_{\alpha}(t')
                                                 \hspace*{1.3in} \nonumber \\
 + \sum_{\alpha} \int_{t_0}^t dt'~ 
      \xi^*_{\alpha}(t') \left( i\hbar \frac{\partial}{\partial t'} 
                       - \epsilon_{\alpha} - \hbar \Sigma^{(+)}_{\alpha} + \mu 
                         \right) \phi_{\alpha}(t')                    
 - \frac{1}{2} \sum_{\alpha} \int_{t_0}^t dt'~ 
         \xi^*_{\alpha}(t') \hbar \Sigma^{K}_{\alpha} \xi_{\alpha}(t')~, 
\end{eqnarray}
and is quadratic in the fluctuation field $\xi({\bf x},t)$. We can thus again 
perform the integration over this field exactly. The result is
\begin{equation}
S^{eff}[\phi^*,\phi] =
  \sum_{\alpha} \int_{t_0}^t dt'~
    \frac{2}{\hbar \Sigma^{K}_{\alpha}}
       \left| \left( i\hbar \frac{\partial}{\partial t'} 
                       - \epsilon_{\alpha} - \hbar \Sigma^{(+)}_{\alpha} + \mu 
                         \right) \phi_{\alpha}(t') \right|^2
  \equiv \int_{t_0}^t dt'~ L(t')~.
\end{equation}
Since the probability distribution $P[\phi^*,\phi;t]$ is equal to the 
functional integral 
\begin{equation}
P[\phi^*,\phi;t] = 
   \int_{\phi({\bf x},t)=\phi({\bf x})}^{\phi^*({\bf x},t)=\phi^*({\bf x})}    
     d[\phi^*]d[\phi]~ 
         \exp \left\{ \frac{i}{\hbar} S^{eff}[\phi^*,\phi] \right\}~,
\end{equation}         
we know from the general connection between classical mechanics and the 
functional formulation of quantum mechanics, that $P[\phi^*,\phi;t]$ must obey 
the Schr\"odinger equation that results from quantizing the classical theory 
with the lagrangian $L(t)$.       
         
Fortunately, the quantization of this theory is straightforward. The momentum 
conjugate to $\phi_{\alpha}(t)$ is 
\begin{equation}
\pi_{\alpha}(t) = \frac{\partial L(t)}
                       {\partial (\partial \phi_{\alpha}(t)/\partial t)}
    = \frac{2i}{\Sigma^{K}_{\alpha}} 
         \left( -i\hbar \frac{\partial}{\partial t} 
                       - \epsilon_{\alpha} - \hbar \Sigma^{(-)}_{\alpha} + \mu 
                         \right) \phi^*_{\alpha}(t)~,
\end{equation}
whereas the momentum conjugate to $\phi^*_{\alpha}(t)$, i.e., 
$\pi^*_{\alpha}(t)$, is given by the complex conjugate expression. The 
corresponding hamiltonian is therefore
\begin{eqnarray}
\label{FPham}
H &=& \sum_{\alpha} \left\{ \pi_{\alpha}(t) 
                          \frac{\partial \phi_{\alpha}(t)}{\partial t}
                        + \pi^*_{\alpha}(t) 
                          \frac{\partial \phi^*_{\alpha}(t)}{\partial t}
                  \right\} - L(t)                             \nonumber \\
  &=& \sum_{\alpha} \left\{
        - \frac{i}{\hbar} \pi_{\alpha}(t)
             \left( \epsilon_{\alpha} + \hbar \Sigma^{(+)}_{\alpha} - \mu 
                         \right) \phi_{\alpha}(t) 
        + \frac{i}{\hbar} \pi^*_{\alpha}(t)
             \left( \epsilon_{\alpha} + \hbar \Sigma^{(-)}_{\alpha} - \mu 
                         \right) \phi^*_{\alpha}(t) \right\} \nonumber \\
  &+& \sum_{\alpha} \frac{\Sigma^{K}_{\alpha}}{2\hbar} |\pi_{\alpha}(t)|^2~.   
\end{eqnarray}
Applying now the usual quantum mechanical recipe of demanding nonvanishing 
commutation relations between the coordinates and their conjugate momenta, we 
can put in this case 
$\pi_{\alpha} = (\hbar/i) \partial/\partial \phi_{\alpha}$ and similarly
$\pi^*_{\alpha} = (\hbar/i) \partial/\partial \phi^*_{\alpha}$. The 
Schr\"odinger equation
\begin{equation}
i \hbar \frac{\partial}{\partial t} P[\phi^*,\phi;t] = H P[\phi^*,\phi;t]
\end{equation}
then indeed exactly reproduces the Fokker-Planck equation in Eq.~(\ref{FP2}). 
It should be noted at this point that the hamiltonian in Eq.~(\ref{FPham}) has 
in principle ordening problems, due to the fact that the coordinates and the 
conjugate momenta no longer commute after quantization. These ordening problems 
are, however, resolved by realizing that in the path-integral formulation of 
quantum mechanics we always deal with so-called normally ordered hamiltonians 
in which the momenta are positioned to the left of the coordinates. For that 
reason we have to use the same ordening in Eq.~(\ref{FPham}).
         
Having arrived at the same Fokker-Planck equation, we have now demonstrated the 
use of essentially all the theoretical tools that are required for a discussion 
of a weakly interacting Bose gas. Before we start with this discussion, 
however, we want to make a final remark about the effect of the nondiagonal 
elements of the selfenergies. Physically, including these nondiagonal elements 
accounts for the change in the wavefunctions $\chi_{\alpha}({\bf x})$ due to the 
interaction with the reservoir. It can clearly be neglected if the coupling with 
the reservoir is sufficiently weak, or more precisely if 
$|\hbar\Sigma_{\alpha',\alpha}^{(+)}(\epsilon_{\alpha}+S_{\alpha}-\mu)|$
is much smaller than the energy splitting 
$|\epsilon_{\alpha'}-\epsilon_{\alpha}|$ between the states in the trap. For a 
quantum dot in solid-state phyisics \cite{kastner} this is often the case and 
for that reason we might call our toy model a bosonic quantum dot. However, 
in magnetically trapped atomic gases the average interaction energy between the 
atoms is already slightly below the Bose-Einstein condensation critical 
temperature larger than the typical energy splitting and the nondiagonal 
elements of the selfenergies are very important. Such a strong-coupling 
situation can, of course, also be treated in our Caldeira-Leggett like model. 
The main difference is that we need to expand our various fields not in the 
eigenstates of the trapping potential $\chi_{\alpha}({\bf x})$ but in the 
eigenstates $\chi'_{\alpha}({\bf x})$ of the nonlocal Schr\"odinger equation
\begin{equation}
\label{nonloc}
\epsilon'_{\alpha} \chi'_{\alpha}({\bf x}) = 
  \left( - \frac{\hbar^2 \nabla^2}{2m} + V^{ex}({\bf x}) \right)
                                               \chi'_{\alpha}({\bf x})
  + \int d{\bf x}'~ 
    {\rm Re}\left[\hbar\Sigma^{(+)}({\bf x},{\bf x}';\epsilon'_{\alpha}-\mu)  
                                      \right] \chi'_{\alpha}({\bf x}')
\end{equation}
where $\epsilon'_{\alpha}$ are the new eigenvalues and
$\hbar\Sigma^{(+)}({\bf x},{\bf x}';\epsilon) 
  = \sum_{\alpha,\alpha'} \chi_{\alpha}({\bf x})
      \hbar\Sigma_{\alpha,\alpha'}^{(+)}(\epsilon) \chi^*_{\alpha'}({\bf x}')$.
In this new basis the nondiagonal elements of the selfenergies can now be 
neglected and we find essentially the same results as before. We only need to 
replace $\epsilon_{\alpha}+S_{\alpha}$ by $\epsilon'_{\alpha}$ and 
$t_{\alpha}({\bf k})$ by 
$\sum_{\alpha'} \left( \int d{\bf x}~ \chi'_{\alpha}({\bf x}')  
                      \chi^*_{\alpha'}({\bf x}) \right) t_{\alpha'}({\bf k})$.
Neglecting the nondiagonal elements in this basis only requires that the real 
part of the retarded selfenergy is much larger that its imaginary part, which 
is always fulfilled in our case because $k_c \Lambda \gg 1$.

Summarizing, the coherent and incoherent dynamics of the gas in the trap is for 
our toy model quite generally solved by
\begin{equation}
\langle \phi_{\alpha} \rangle(t) = 
  \langle \phi_{\alpha} \rangle(0) e^{-i(\epsilon'_{\alpha}-\mu)t/\hbar} 
                                                   e^{-\Gamma_{\alpha}t/2}
\end{equation}
and
\begin{equation}
N_{\alpha}(t) = N_{\alpha}(0) e^{-\Gamma_{\alpha}t}
              + N_{\alpha} \left( 1 - e^{-\Gamma_{\alpha}t} \right)~,
\end{equation}
respectively. In the limit $t \rightarrow \infty$, the average of the 
annihilation operators $\langle \phi_{\alpha} \rangle(t)$ thus always vanishes 
but the average occupation numbers $N_{\alpha}(t)$ relax to the equilibrium 
distribution $N_{\alpha} = (e^{\beta(\epsilon'_{\alpha}-\mu)}-1)^{-1}$. 
Although this appears to be an immediately obvious result, its importance stems 
from the fact that it is also true if we tune the potential energy bias 
$\Delta V^{ex}$ 
such that at low temperatures the groundstate $\chi'_g({\bf x})$ acquires a 
macroscopic occupation, i.e., $N_g \gg 1$. The gas therefore never shows a 
spontaneous breaking of the $U(1)$ symmetry, in agreement with the notion that 
we are essentially dealing with an ideal Bose gas in the grand-canonical 
ensemble \cite{henk2}. The reason of the absense of the spontaneous symmetry 
breaking can also be understood from our stationary solution of the 
Fokker-Planck equation in Eq.~(\ref{stat}), which shows that the probability 
distribution for $|\phi_g|$ is proportional to the Boltzmann factor 
$e^{-\beta(\epsilon'_g-\mu) |\phi_g|^2}$ in the degenerate regime of interest. 
Hence, the corresponding Landau free energy  
$F(|\phi_g|) = (\epsilon'_g-\mu) |\phi_g|^2$ never shows an instability towards 
the formation of a nonzero average of $|\phi_g|$ due to the fact that 
$\epsilon'_g-\mu$ can never become less then zero. Once we introduce 
interactions between the atoms in the gas, this picture fully changes.

\section{SEMICLASSICAL THEORY OF BOSE-EINSTEIN CONDENSATION}
\label{semicl}
In the present and the folowing section we turn our attention to an interacting 
but homogeneous Bose gas. However, in Sec.~\ref{inhom} we generalize our 
results and include also the effects of an external trapping potential. There 
are several reason for considering the homogeneous case first. One important 
reason is that for a homogeneous gas we are allowed to take the thermodynamic 
limit in which Bose-Einstein condensation becomes a true second-order phase 
transition. We are then in fact studying the dynamics of a spontaneous symmetry 
breaking under the most ideal circumstances, which may nevertheless have 
important implications for cosmology \cite{kibble,guth,linde,zurek} and also 
lead to a better understanding of the recent experiments performed with liquid 
$^4$He \cite{hendry}. Furthermore, the homogeneous case is in a sense also the 
most difficult one because thermalizing collisions cannot lead to a macroscopic 
occupation of the one-particle ground state \cite{LY,sergei}. 
Using the language of quantum optics, we also cannot compare Bose-Einstein 
condensation of a homogeneous gas with a single-mode laser and we are 
essentially dealing with the (in principle infinitely many) multi-mode 
situation. Since realistic experiments with atomic rubidium and sodium gases 
\cite{JILA,MIT,rest} are usually also in a multi-mode regime, it even turns out 
that an understanding of the homogeneous gas is important for magnetically 
trapped gases as well. This is less so for experiments with atomic lithium 
\cite{Rice}, because then the inhomogeneity plays a much more fundamental role. 

In the semiclassical theory of Bose-Einstein condensation in a homogeneous, 
interacting gas, the main emphasis is on the study of the coherent dynamics of 
the gas \cite{henk1}. The reason for this is that if Bose-Einstein condensation 
is possible at all, it has to be caused by coherent processes since incoherent, 
i.e., kinetic, processes cannot lead to Bose-Einstein condensation in the 
thermodynamic limit. The reason why Bose-Einstein condensation 
cannot be achieved in this way is easily understood. From the appropriate 
quantum Boltzmann equation (cf.\ Eq.~(\ref{qdBE})) we obtain that the production 
rate for the number of particles in the zero-momentum state is given by
\begin{equation}
\frac{d N({\bf 0};t)}{dt} = \frac{C}{\tau_{el}} (1 + N({\bf 0};t))~,
\end{equation}
where $\tau_{el}$ is the average time between two elastic collisions and $C$ is 
a constant of ${\cal O}(1)$. For initial conditions close to equilibrium but 
with $N({\bf 0};t_0)=0$, we can in first instance neglect the evolution of the 
noncondensed part of the gas, i.e., the time dependence of $\tau_{el}$, and we 
find that
\begin{equation}
N({\bf 0};t) = \left( e^{C(t-t_0)/\tau_{el}} -1 \right)~.
\end{equation}
Hence, a macroscopic occupation of the one-particle ground state is only 
achieved at times $t > t_0 + {\cal O}( \tau_{el} \ln(N) )$, which indeed 
diverge if $N \rightarrow \infty$. For the initial conditions of interest this 
condition can also be written as 
$t > t_0 + {\cal O}( \tau_{el} \ln(L/\Lambda_c) )$, with 
$L$ the (linear) system size and $\Lambda_c = (2\pi\hbar^2/mk_BT_c)^{1/2}$ the 
thermal de Broglie wavelength of the atoms at the critical temperature $T_c$. 

Physically, we thus envisage the dynamics of the gas in 
response to evaporative cooling in the following way. In the nondegenerate 
regime the gas just thermalizes by elastic collisions in such a manner that the 
occupation numbers $N({\bf k};t)$ are in a good approximation at all times 
equal to a truncated equilibrium distribution \cite{jom,wolfgang1}. Since the 
temperature of the gas is decreasing, this distribution becomes more sharply 
peaked at low momenta and we enter at some point the degenerate regime where 
$n\Lambda^3 = {\cal O}(1)$ but no condensate has been formed yet because we 
know that this is impossible by means of elastic collisions. At this stage the 
most important question to be answered is: Is it nevertheless possible that for 
a noncondensed but degenerate distribution function $N({\bf k};t)$, the energy 
$\epsilon'({\bf 0};t)$ of the zero-momentum state becomes less than the 
instantaneous chemical potential $\mu(t)$ and the gas becomes unstable towards 
Bose-Einstein condensation? If the answer to this question is no, Bose-Einstein 
condensation is impossible and the study of the dynamics of the gas 
is restricted to the normal phase and relatively well understood 
\cite{dave,boris,ST}. However, if the answer is yes, the dynamics of the gas 
after the point of instability is clearly more complicated and deserves further 
investigation. 

In view of the central importance of the above question for the topic of this 
paper, we now first of all show in Sec.~\ref{unstable} that the gas indeed 
develops the required instability for Bose-Einstein condensation. We then turn 
to the dynamics of the unstable gas. In Sec.~\ref{sdyn} the dynamics of the 
phase transition is first studied in the semiclassical approximation, in which 
fluctuations in the order parameter are neglected. Due to this neglect of 
fluctuations, the time at which the instability occurs in the gas is equivalent 
to the time at which Bose-Einstein condensation takes place. In 
Sec.~\ref{quant}, however, we include the fluctuations into our theory and 
discuss in detail how this affects the semiclassical picture.    

\subsection{Instability of the Bose gas}
\label{unstable}
From our Caldeira-Leggett toy model we remember that the (renormalized) 
energies $\epsilon'_{\alpha}$ and the corresponding eigenstates 
$\chi'_{\alpha}({\bf x})$ can be determined from a nonlocal Schr\"odinger 
equation once we know the retarded selfenergy 
$\hbar \Sigma^{(+)}({\bf x},{\bf x}';t-t')$. The same is 
true for a homogeneous atomic Bose gas, only with this exception that the 
nonzero selfenergy is now due to the interatomic interactions and not due to 
the presence of a reservoir. In a sense an interacting gas plays also the role 
of its own reservoir. Moreover, the homogeneity of the gas leads to an important 
simplification because translational invariance requires that the retarded 
selfenergy is only a function of the relative distance 
${\bf x}-{\bf x}'$. Therefore, the Schr\"odinger equation in Eq.~(\ref{nonloc}) 
is solved by $\chi'_{\bf k}({\bf x}) = \chi_{\bf k}({\bf x})$ and
\begin{equation}
\label{energy}
\epsilon'({\bf k}) = \epsilon({\bf k}) 
             + {\rm Re}[\hbar\Sigma^{(+)}({\bf k};\epsilon'({\bf k})-\mu)]~,
\end{equation}
where from now on $\epsilon({\bf k}) = \hbar^2{\bf k}^2/2m$ again and
\begin{equation}
\Sigma^{(+)}({\bf x}-{\bf x}';t-t') =
   \int \frac{d{\bf k}}{(2\pi)^3} \int \frac{d\epsilon}{2\pi\hbar}~
     \Sigma^{(+)}({\bf k};\epsilon) 
          e^{i({\bf k} \cdot ({\bf x}-{\bf x}')-\epsilon(t-t')/\hbar)}~.
\end{equation}
Having arrived at this result, it is important to briefly remind ourselves what 
it implies physically. Mathematically, we have assumed that the gas has at the 
initial time $t_0$ the occupation numbers $N({\bf k})$ that correspond to a 
truncated equilibrium distribution with chemical potential $\mu$. Subsequently, 
we have taken the limit of $t_0 \rightarrow -\infty$. The energy 
$\epsilon'({\bf k})$ is therefore the energy of the one-particle state 
$\chi_{\bf k}({\bf x})$ after all the transients have died out. However, as is 
shown explicitly in Ref.~\cite{henk1}, the memory time in the selfenergy is for 
the same reason as in our Caldeira-Leggett toy model only of 
${\cal O}(\hbar/k_BT)$ and hence much smaller than all the other relevant time 
scales in the problem. As a consequence we are for all practical purposes 
allowed to consider $\epsilon'({\bf k})$ as the energy of the state 
$\chi_{\bf k}({\bf x})$, given all the occupation numbers $N({\bf k})$. The 
time dependence of $\epsilon'({\bf k};t)$ can then be seen as solely due to the 
time dependence of the occupation numbers $N({\bf k};t)$.

Keeping the last remark in mind, we now proceed to solve Eq.~(\ref{energy}). To 
do that we of course need an expression for the retarded selfenergy of a 
weakly interacting Bose gas, which follows once we know the full selfenergy 
$\hbar\Sigma({\bf k};t,t')$ defined on the Schwinger-Keldysh contour
${\cal C}^{\infty}$. Unfortunately, even for a dilute system this quantity 
cannot be calculated exactly and some approximation is called for. The 
approximation that we will make here is the so-called many-body T-matrix 
approximation that has recently been reviewed in the literature 
\cite{henk4,allan2}. The main motivation for this approximation is that due to 
the smallness of the gasparameter $(na^3)^{1/2}$ it is very unlikely for three 
or more particles to be within the range of the interaction and we only need to 
take account of all possible two-body processes taking place in the gas. It 
goes beyond the much used Popov approximation \cite{popov,allan3} by including 
also quite accurately the effect of the surrounding medium on the effective 
interaction between the atoms in the gas \cite{michel1,michel2}, which turns 
out to be of crucial importance for our purposes. 

Given this effective interaction $V({\bf k},{\bf k}',{\bf K};t,t')$ for the 
scattering of two atoms that at time $t'$ have the momenta 
$\hbar({\bf K}/2 \pm {\bf k}')$ and at time $t$ the momenta 
$\hbar({\bf K}/2 \pm {\bf k})$, respectively, the exact selfenergy obeys the 
Hartree-Fock-like relation
\begin{equation}
\hbar\Sigma({\bf k};t,t') = 2i \int \frac{d{\bf k}'}{(2\pi)^3}~
  V({\bf k}-{\bf k}',{\bf k}-{\bf k}',{\bf k}+{\bf k}';t,t') G({\bf k}';t',t)~,
\end{equation} 
where the Green's function equals again Eq.~(\ref{greens}) but with 
$\epsilon({\bf k})$ replaced by $\epsilon'({\bf k})$ to make the theory 
selfconsistent. The retarded selfenergy therefore obeys
\begin{eqnarray}
\hbar\Sigma^{(+)}({\bf k};t-t') = 2i \int \frac{d{\bf k}'}{(2\pi)^3}~
 \left(
   V^{(+)}({\bf k}-{\bf k}',{\bf k}-{\bf k}',{\bf k}+{\bf k}';t-t') 
              G^<({\bf k}';t'-t)  \right. \hspace*{0.3in} \nonumber \\
 \left.                                                         
 + V^<({\bf k}-{\bf k}',{\bf k}-{\bf k}',{\bf k}+{\bf k}';t-t') 
                                                     G^{(-)}({\bf k}';t'-t)
 \right)~,
\end{eqnarray}
leading to 
\begin{eqnarray}
\label{splus}
\hbar\Sigma^{(+)}({\bf k};\epsilon'({\bf k})-\mu) = 
                                     2 \int \frac{d{\bf k}'}{(2\pi)^3}~
   V^{(+)}({\bf k}-{\bf k}',{\bf k}-{\bf k}',{\bf k}+{\bf k}';
           \epsilon'({\bf k}) + \epsilon'({\bf k}') - 2\mu) N({\bf k}')
                                                                 \nonumber \\
 + 2i \int \frac{d{\bf k}'}{(2\pi)^3} \int \frac{d\epsilon}{2\pi\hbar}~                                                          
   V^<({\bf k}-{\bf k}',{\bf k}-{\bf k}',{\bf k}+{\bf k}';\epsilon) 
           \frac{\hbar}{\epsilon^- - (\epsilon'({\bf k})+ 
                                                  \epsilon'({\bf k}')-2\mu)}~,
\end{eqnarray}
if we also introduce the notation $\epsilon^-$ for the limiting procedure 
$\epsilon - i0$. Note that the two terms in the right-hand side correspond 
roughly speaking to the contributions to the selfenergy from scattering 
processes out and into the state $\chi_{\bf k}({\bf x})$, respectively. This 
will be more clear in Sec.~\ref{quant} when we discuss the imaginairy part of 
this expression that describes the (incoherent) dynamics of the occupation 
numbers. To see the instability of the gas, however, we only need to consider 
the real part. 

Hence, our next task is to determine the effective interaction 
$V({\bf k},{\bf k}',{\bf K};t,t')$ by considering all possible two-body 
scattering processes. Because we are dealing with bosons, the effective 
interaction is a sum of a direct and an exchange term and can be written as
$V({\bf k},{\bf k}',{\bf K};t,t') = (T({\bf k},{\bf k}',{\bf K};t,t') 
                                    + T(-{\bf k},{\bf k}',{\bf K};t,t'))/2$
in terms of the many-body T-matrix that obeys the Lippmann-Schwinger equation
\cite{lippmann}
\begin{eqnarray}
\label{LS}
T({\bf k},{\bf k}',{\bf K};t,t') = V({\bf k}-{\bf k}') \delta(t,t')
                                                  \hspace*{3.6in} \nonumber \\
  + \frac{i}{\hbar} \int_{{\cal C}^{\infty}} dt'' 
                    \int\frac{d{\bf k}''}{(2\pi)^3}~
     V({\bf k}-{\bf k}'') G({\bf K}/2 + {\bf k}'';t,t'')                               
        G({\bf K}/2 - {\bf k}'';t,t'') T({\bf k}'',{\bf k}',{\bf K};t'',t')~,
\end{eqnarray}
with $V({\bf k}-{\bf k}')$ the Fourier transform of the interatomic interaction 
potential. By iterating this equation, we immediately see that the many-body 
T-matrix indeed sums all possible collisions between two particles. Moreover, 
the Green's functions $G({\bf K}/2 \pm {\bf k}'';t,t'')$ describe the 
propagation of an atom with momentum $\hbar({\bf K}/2 \pm {\bf k}'')$ from time 
$t''$ to time $t$ in the gas. Therefore, we also see that the many-body 
T-matrix incorporates the effect of the surrounding gaseous medium on the 
propagation of the atoms between two collisions. This in contrast to the 
two-body T-matrix, that is well-known from scattering theory \cite{walter} and 
obeys the same Lippman-Schwinger equation but with the Green's functions 
$G({\bf K}/2 \pm {\bf k}'';t,t'')$ replaced by 
\begin{eqnarray}
G_0({\bf K}/2 \pm {\bf k}'';t,t'') =
   -ie^{-i\epsilon({\bf K}/2 \pm {\bf k}'')(t-t'')/\hbar} \Theta(t,t'')
                                                                  \nonumber
\end{eqnarray}
describing the propagation of an atom in the absence of a surrounding gas, 
i.e., in a vacuum. Note also that the many-body T-matrix has a $\delta$-function 
singularity on the Schwinger-Keldysh contour. As a result the retarded and 
advanced components are now given by the more general expression
\begin{eqnarray}
T^{(\pm)}({\bf k},{\bf k}',{\bf K};t-t') =
  T^{\delta}({\bf k},{\bf k}',{\bf K};t) \delta(t-t') 
                                                 \hspace*{2.5in} \nonumber \\
  \pm \Theta(\pm(t-t')) 
    \left( T^{>}({\bf k},{\bf k}',{\bf K};t-t')
           - T^{<}({\bf k},{\bf k}',{\bf K};t-t') \right)
\end{eqnarray}
but the Keldysh component still obeys
\begin{equation}
T^{K}({\bf k},{\bf k}',{\bf K};t-t') =
    \left( T^{>}({\bf k},{\bf k}',{\bf K};t-t')
           + T^{<}({\bf k},{\bf k}',{\bf K};t-t') \right)~.
\end{equation}

Before we turn to the solution of the Lippmann-Schwinger equation in 
Eq.~(\ref{LS}), it is instructive to briefly mention what we obtain if we now 
apply the pseudopotential method \cite{huang1}. In this method it is assumed 
that the effective interaction equals 
\begin{equation}
V({\bf k},{\bf k}',{\bf K};t,t') = \frac{4\pi a\hbar^2}{m} \delta(t,t')~,
\end{equation}
which implies that only the retarded and advanced components of the effective 
interaction are nonzero and both equal to $4\pi a\hbar^2 \delta(t-t')/m$. 
Substituting this in Eq.~(\ref{splus}), we simply find that
\begin{equation}
\hbar\Sigma^{(+)}({\bf k};\epsilon'({\bf k})-\mu) = \frac{8\pi na\hbar^2}{m}
\end{equation}
and thus that $\epsilon'({\bf k};t) = \epsilon({\bf k}) + 8\pi n(t)a\hbar^2/m$. 
On the basis of the pseudopotential method we therefore conclude that the Bose 
gas never shows an instability towards Bose-Einstein condensation because a 
momentum independent selfenergy only leads to a shift in the chemical 
potential, which plays no role in the dynamics of the gas, even if we are 
evaporatively cooling the gas and the density $n(t)$ is a time-dependent 
quantity. For our purposes it is thus important to determine a more accurate 
expresion for the effective interaction. 

As a first step towards this goal, we consider the retarded and advanced parts 
of Eq.~(\ref{LS}). They are given by
\begin{eqnarray}
T^{(\pm)}({\bf k},{\bf k}',{\bf K};t-t') = V({\bf k}-{\bf k}') \delta(t-t')
                                                  \hspace*{3.0in} \nonumber \\
  + \frac{i}{\hbar} \int dt'' 
                    \int\frac{d{\bf k}''}{(2\pi)^3}~
     V({\bf k}-{\bf k}'') G^{(\pm)}({\bf k}'',{\bf K};t-t'')                               
                             T^{(\pm)}({\bf k}'',{\bf k}',{\bf K};t''-t')~,
\end{eqnarray} 
where $G^{(\pm)}({\bf k}'',{\bf K};t-t'')$ are the retarded and advanced 
components of the two-particle Green's function  
$G({\bf k}'',{\bf K};t,t'') \equiv 
              G({\bf K}/2 + {\bf k}'';t,t'') G({\bf K}/2 - {\bf k}'';t,t'')$ 
describing the propagation of two (independent) atoms with momenta 
$\hbar({\bf K}/2 + {\bf k}'')$ and $\hbar({\bf K}/2 - {\bf k}'')$ from time 
$t''$ to time $t$ in the gas. Explicitly, we have that
\begin{eqnarray}
G^{(\pm)}({\bf k}'',{\bf K};t-t'')  \hspace*{4.5in} \nonumber \\
  = \pm \Theta(\pm(t-t'')) 
     \left( 1 + N({\bf K}/2 + {\bf k}'') + N({\bf K}/2 - {\bf k}'') \right)
     e^{-i(\epsilon'({\bf k}'',{\bf K}) - 2\mu)(t-t'')/\hbar}~,
\end{eqnarray}
with $\epsilon'({\bf k}'',{\bf K}) 
      = \epsilon'({\bf K}/2 + {\bf k}'') + \epsilon'({\bf K}/2 - {\bf k}'')$ 
the appropriate two-particle energy. Performing a Fourier transformation we then 
obtain 
\begin{eqnarray}
T^{(\pm)}({\bf k},{\bf k}',{\bf K};\epsilon) = V({\bf k}-{\bf k}')
                                                  \hspace*{3.9in} \nonumber \\
  + \int \frac{d{\bf k}''}{(2\pi)^3}~ V({\bf k}-{\bf k}'') 
         \frac{1 + N({\bf K}/2 + {\bf k}'') + N({\bf K}/2 - {\bf k}'')}
              {\epsilon^{\pm} - (\epsilon'({\bf k}'',{\bf K}) - 2\mu)}                              
                             T^{(\pm)}({\bf k}'',{\bf k}',{\bf K};\epsilon)~,
\end{eqnarray} 
which is well-known in nuclear physics as the Bethe-Salpeter equation and was 
first used in the context of Bose-Einstein condensation by Brueckner and Sawada 
\cite{brueckner}. Compared with the usual Lippmann-Schwinger equation 
\begin{equation}
T^{(\pm)}({\bf k},{\bf k}';\epsilon) = V({\bf k}-{\bf k}')
    + \int \frac{d{\bf k}''}{(2\pi)^3}~ V({\bf k}-{\bf k}'') 
            \frac{1}
                 {\epsilon^{\pm} - 2\epsilon({\bf k}'')}                              
                                      T^{(\pm)}({\bf k}'',{\bf k}';\epsilon)
\end{equation} 
for the two-body T matrix, we see that the Bethe-Salpeter equation incorporates 
the effect that if the intermediate states are already occupied it is more 
likely for bosons to scatter into these states. Indeed the factor 
$1 + N({\bf K}/2 + {\bf k}'') + N({\bf K}/2 - {\bf k}'')$ arises as the net 
difference between two atoms scattering into and scattering out of the 
intermediate states $\chi_{{\bf K}/2 + {\bf k}''}({\bf x})$ and 
$\chi_{{\bf K}/2 - {\bf k}''}({\bf x})$, which are proportional to 
$(1 + N({\bf K}/2 + {\bf k}''))(1 + N({\bf K}/2 - {\bf k}''))$ and 
$N({\bf K}/2 + {\bf k}'') N({\bf K}/2 - {\bf k}'')$, respectively. In a real 
space picture the effect of the Bose enhancement factors can be seen as a 
result of the fact that the scattering wavefunction of the two colliding atoms 
has to be matched to the many-body wavefunction of the gas at large interatomic 
distances \cite{paul}. Seen in this manner, it will be clear that this effect 
is not small in the gasparameter $(na^3)^{1/2}$ and cannot be neglected 
beforehand if we aim to include all two-body processes in the gas.    

In general the solution of a Bethe-Salpeter equation is quite complicated. 
However, for the quantum gases of interest we can basically solve it 
analytically by making use of the fact in the quantum regime the temperatures 
are always such that the thermal de Broglie wavelength obeys $a/\Lambda \ll 1$. 
As a result the many-body T matrix at the relevant momenta and energies that 
are much smaller than $\hbar/a$ and $\hbar^2/2ma^2$, respectively, can in a very 
good approximation be expressed as \cite{henk4}
\begin{equation}
\label{tmb}
T^{(\pm)}({\bf k},{\bf k}',{\bf K};\epsilon) =
  \frac{T^{(\pm)}({\bf 0},{\bf 0};0)}
       {1-T^{(\pm)}({\bf 0},{\bf 0};0) \Xi({\bf K};\epsilon)}~,
\end{equation}
where $T^{(\pm)}({\bf 0},{\bf 0};0) = 4\pi a\hbar^2/m$ and we have also 
introduced the function
\begin{equation} 
\Xi({\bf K};\epsilon) =
   \int \frac{d{\bf k}''}{(2\pi)^3}~ 
       (N({\bf K}/2 + {\bf k}'') + N({\bf K}/2 - {\bf k}''))
         \frac{{\cal P}}
              {\epsilon - (\epsilon'({\bf k}'',{\bf K}) - 2\mu)}~.
\end{equation} 
Because $\Xi({\bf K};\epsilon)$ contains only the principle value part of the 
integral, the above approximation gives only the real part of the many-body T 
matrix. This is all we need for the semiclassical theory, but for the full 
quantum theory we need also the imaginary part. The latter is found from the 
optical theorem \cite{walter}, which in the many-body case reads
\begin{eqnarray}
\label{ot}
T^{(+)}({\bf k},{\bf k}',{\bf K};\epsilon) 
 - T^{(-)}({\bf k},{\bf k}',{\bf K};\epsilon)  
  = -2\pi i \int \frac{d{\bf k}''}{(2\pi)^3}~ 
     \delta(\epsilon - (\epsilon'({\bf k}'',{\bf K}) - 2\mu)) 
                                                  \hspace*{0.6in} \nonumber \\
   \times T^{(+)}({\bf k},{\bf k}'',{\bf K};\epsilon)                            
       (1 + N({\bf K}/2 + {\bf k}'') + N({\bf K}/2 - {\bf k}''))
                             T^{(-)}({\bf k}'',{\bf k}',{\bf K};\epsilon)~.
\end{eqnarray} 
Note that no approximation is made to obtain this result.

We are now in the position to evaluate the contribution of the 
$V^{(+)}G^<$-term to the real part of the retarded selfenergy. Combining 
Eqs.~(\ref{splus}) and (\ref{tmb}) we obtain in first instance
\begin{eqnarray}
2 \int \frac{d{\bf k}'}{(2\pi)^3}~
 \frac{T^{(+)}({\bf 0},{\bf 0};0)}
      {1-T^{(+)}({\bf 0},{\bf 0};0) 
        \Xi({\bf k}+{\bf k}';\epsilon'({\bf k}) + \epsilon'({\bf k}') - 2\mu)}
                                                     N({\bf k}')~.   \nonumber
\end{eqnarray} 
However, the dominant contribution to the integral comes from thermal momenta 
due to the presence of the distribution function $N({\bf k}')$. It is not 
difficult to show that for these momenta 
$T^{(+)}({\bf 0},{\bf 0};0) 
        \Xi({\bf k}+{\bf k}';\epsilon'({\bf k}) + \epsilon'({\bf k}') - 2\mu) =
                                                        {\cal O}(na\Lambda^2)$
and therefore, for temperatures near the critical temperature where 
$n\Lambda^3 = {\cal O}(1)$, always much smaller than one. Under these 
conditions this contribution to the retarded selfenergy is essentially equal to 
$8\pi na\hbar^2/m$, i.e., the same as the full result obtained by the 
pseudopotential method. At this point we thus conclude that if the gas becomes 
unstable, it is due to the second term in the right-hand side of 
Eq.~(\ref{splus}). This contribution we consider next.

We first have to determine the component 
$T^<({\bf k},{\bf k}',{\bf K};\epsilon)$ of the many-body T matrix. In general 
Eq.~(\ref{LS}) shows that
\begin{eqnarray}
T^{<,>}({\bf k},{\bf k}',{\bf K};t-t')   
 &=& \frac{i}{\hbar} \int dt'' 
                    \int\frac{d{\bf k}''}{(2\pi)^3}~
     V({\bf k}-{\bf k}'') G^{<,>}({\bf k}'',{\bf K};t-t'')                               
                             T^{(-)}({\bf k}'',{\bf k}',{\bf K};t''-t')  
                                                               \nonumber \\
 & & \hspace*{-1.0in} + \frac{i}{\hbar} \int dt'' 
                    \int\frac{d{\bf k}''}{(2\pi)^3}~
     V({\bf k}-{\bf k}'') G^{(+)}({\bf k}'',{\bf K};t-t'')                               
                           T^{<,>}({\bf k}'',{\bf k}',{\bf K};t''-t')~,
\end{eqnarray} 
which is immediately solved by
\begin{eqnarray}
T^{<,>}({\bf k},{\bf k}',{\bf K};t-t') 
 = \frac{i}{\hbar} \int dt'' \int dt''' 
                    \int\frac{d{\bf k}''}{(2\pi)^3}~
  T^{(+)}({\bf k},{\bf k}'',{\bf K};t-t'')     \hspace*{1.0in} \nonumber \\
    \times G^{<,>}({\bf k}'',{\bf K};t''-t''')                               
                             T^{(-)}({\bf k}'',{\bf k}',{\bf K};t'''-t')~,  
\end{eqnarray} 
if we use the Bethe-Salpeter equation for the retarded component of the 
many-body T matrix. A Fourier transformation leads, therefore, in particular to
\begin{eqnarray}
\label{tsmall}
T^<({\bf k},{\bf k}',{\bf K};\epsilon)  
  = -2\pi i \int \frac{d{\bf k}''}{(2\pi)^3}~ 
     \delta(\epsilon - (\epsilon'({\bf k}'',{\bf K}) - 2\mu)) 
                                                  \hspace*{2.0in} \nonumber \\
   \times T^{(+)}({\bf k},{\bf k}'',{\bf K};\epsilon)                            
       N({\bf K}/2 + {\bf k}'') N({\bf K}/2 - {\bf k}'')
                             T^{(-)}({\bf k}'',{\bf k}',{\bf K};\epsilon)
\end{eqnarray}  
but also to
\begin{eqnarray}
\label{tkel}
T^K({\bf k},{\bf k}',{\bf K};\epsilon)  
  &=& -2\pi i \int \frac{d{\bf k}''}{(2\pi)^3}~ 
     \delta(\epsilon - (\epsilon'({\bf k}'',{\bf K}) - 2\mu)) 
         T^{(+)}({\bf k},{\bf k}'',{\bf K};\epsilon) 
         T^{(-)}({\bf k}'',{\bf k}',{\bf K};\epsilon)          \nonumber \\
  & & \hspace*{-0.5in}
   \times (1 + N({\bf K}/2 + {\bf k}'') + N({\bf K}/2 - {\bf k}'')
            + 2N({\bf K}/2 + {\bf k}'') N({\bf K}/2 - {\bf k}'') )~,
\end{eqnarray} 
which we need lateron in Sec.~\ref{quant}.

The real part of the $V^<G^{(-)}$-term in the expression for the retarded 
selfenergy thus becomes
\begin{eqnarray}
-2 \int \frac{d{\bf k}'}{(2\pi)^3} \int \frac{d{\bf k}''}{(2\pi)^3}~
  \left| V^{(+)}({\bf k},{\bf k}',{\bf k}'') \right|^2 
         N({\bf k}') N({\bf k}'')
     \frac{{\cal P}}{\epsilon'({\bf k}) + \epsilon'({\bf k}'+{\bf k}''-{\bf k})
                   - \epsilon'({\bf k}') - \epsilon'({\bf k}'')}~, \nonumber
\end{eqnarray}
with $V^{(\pm)}({\bf k},{\bf k}',{\bf k}'')$ a convenient shorthand notation 
for the following matrix element of the effective interaction 
$V^{(\pm)}({\bf k} - ({\bf k}'+{\bf k}'')/2,({\bf k}'-{\bf k}'')/2, 
      {\bf k}'+{\bf k}''; \epsilon'({\bf k}') + \epsilon'({\bf k}'') - 2\mu)$.
For momenta $\hbar{\bf k}$ that are of the order of the thermal momenta 
$\hbar/\Lambda$, this contribution to the retarded selfenergy is easily 
estimated to be only of ${\cal O}(2nT^{(+)}({\bf 0},{\bf 0};0)(na\Lambda^2))$ 
and therefore negligible compared tot the contribution from the 
$V^{(+)}G^<$-term that we considered previously and was shown to be equal to 
$2nT^{(+)}({\bf 0},{\bf 0};0)$. For thermal momenta we thus conclude that
\begin{equation}
\label{eth}
\epsilon'({\bf k};t) = \epsilon({\bf k}) + \frac{8\pi n(t)a\hbar^2}{m}~.
\end{equation}
However, this conclusion is not valid for momenta $\hbar{\bf k}$ that are much 
smaller than the thermal momenta, because in that case the energy denominator 
in the integrant favors the small momenta where the occupation numbers are 
esspecially large in the degenerate regime. For such momenta, and in particular 
for the momenta obeying $\hbar k < \hbar \sqrt{8\pi na} \ll \hbar/\Lambda$ that 
are shown in Sec.~\ref{sdyn} to be the most important states for the dynamics 
of the gas after the occurrence of the instability, we find in a good 
approximation that 
\begin{eqnarray}
\label{e0}
\epsilon'({\bf k};t) = \epsilon({\bf k}) &+& \frac{8\pi n(t)a\hbar^2}{m}
                                                     \nonumber \\
 &-& 2 \int \frac{d{\bf k}'}{(2\pi)^3} \int \frac{d{\bf k}''}{(2\pi)^3}~
  \left| V^{(+)}({\bf 0},{\bf k}',{\bf k}'') \right|^2 
         N({\bf k}';t) N({\bf k}'';t)
          \frac{{\cal P}}{\hbar^2 {\bf k}' \cdot {\bf k}''/m}~.
\end{eqnarray} 
Note that in the right-hand side $V^{(+)}({\bf 0},{\bf k}',{\bf k}'')$ also 
implicily depends on $t$, because it depends on the distribution function 
$N({\bf k};t)$ as shown for example in detail in Eq.~(\ref{tmb}).       
      
In principle, Eqs.~(\ref{eth}) and (\ref{e0}) already clearly show the tendency 
of the gas to become unstable towards Bose-Einstein condensation, because the 
energy of the one-particle ground state is shifted less upwards as compared to 
the one-particle states with thermal energies. To actually show when the gas is 
unstable we, however, need to compare the energy of the zero-momentum state 
with the instantaneous chemical potential. Using Eq.~(\ref{eth}) and our picture 
that during evaporative cooling the occupation numbers of the gas are almost 
equal to a truncated equilibrium distribution, the chemical potential becomes
\begin{equation}
\mu(t) \simeq \frac{8\pi n(t)a\hbar^2}{m} + \mu_0(t)~,
\end{equation}
where the time dependence of the ideal gas chemical potential 
$\mu_0(t) \equiv \mu_0(n(t),T(t))$ depends on the precise path in the 
density-temperature plane that is followed during the cooling process. An 
instability therefore occurs once the quantity
\begin{eqnarray}
\epsilon'({\bf 0};t) - \mu(t) \simeq
- \mu_0(t)                                
- 2 \int \frac{d{\bf k}'}{(2\pi)^3} \int \frac{d{\bf k}''}{(2\pi)^3}~
  \left| V^{(+)}({\bf 0},{\bf k}',{\bf k}'') \right|^2 
         N({\bf k}';t) N({\bf k}'';t)
          \frac{{\cal P}}{\hbar^2 {\bf k}' \cdot {\bf k}''/m}  \nonumber
\end{eqnarray} 
becomes less than zero. It was the single most important conclusion of 
Ref.~\cite{henk1} that the gas indeed develops the required instability for 
Bose-Einstein condensation if $a > 0$ and the temperature is less than a 
critical temperature $T_c = T_0(1 + {\cal O}(a/\Lambda_0))$ that is slightly 
higher than the critical temperature $T_0$ of the ideal Bose gas. Although we 
can argue that this might be an artefact of the many-body T-matrix 
approximation, which certainly does not take proper account of all the critical 
fluctuations near the critical temperature, it should be noted that all the 
predictions of the many-body T-matrix approximation are fully confirmed by a 
renormalization group calculation that does take critical fluctuations into 
account and only neglects the effects of a mass renormalization \cite{michel2}. 
Moreover, the upward shift of the interacting critical temperature by an amount 
of ${\cal O}((a/\Lambda_0)T_0)={\cal O}((na^3)^{1/3}T_0)$ has recently also 
been found by path-integral Monte-Carlo simulations \cite{david}. We therefore 
believe that the results obtained here give an, even more than qualitatively, 
correct picture of the behavior of the gas near the critical temperature.

\subsection{Semiclassical Dynamics}
\label{sdyn}
Having positively answered the question if an interacting Bose gas can be 
experimentally quenched into the regime in which it is unstable for a 
spontaneous breaking of the $U(1)$ symmetry, our next task is to consider the 
dynamics of the gas after it has developed the instability. At the 
semiclassical level our results obtained in Sec.~\ref{unstable} show that the 
effective action for the long-wavelength dynamics of the gas, i.e., for the 
states with momenta $\hbar{\bf k} < \hbar\sqrt{8\pi na}$, is given by
\begin{eqnarray}
S^{cl}[\phi^*,\phi] = \int dt~ \left\{
   \sum_{\bf k} \phi^*_{\bf k}(t) 
       \left( i\hbar \frac{\partial}{\partial t} 
                    - \epsilon'({\bf k};t) + \mu(t) \right) \phi_{\bf k}(t)
                                \right.          \hspace*{1.7in} \nonumber \\
   \left. - \frac{1}{2V} \sum_{{\bf k},{\bf k}',{\bf K}} 
       T^{(+)}({\bf 0},{\bf 0},{\bf 0};0) 
         \phi^*_{{\bf K}/2+{\bf k}}(t) \phi^*_{{\bf K}/2-{\bf k}}(t)
         \phi_{{\bf K}/2-{\bf k}'}(t) \phi_{{\bf K}/2+{\bf k}'}(t)
   \right\}~.
\end{eqnarray}
We can understand that this is the correct semiclassical action in two 
different ways. First, it is found from our Schwinger-Keldysh formalism if we 
realize that at the semiclassical level the `path' followed by the gas on the 
backward branch of the Schwinger-Keldysh contour is identical to the `path' 
followed on the forward branch. Applying this observation to the full effective 
action $S^{eff}[\psi^*,\psi]$ immediately leads to the action 
$S^{cl}[\phi^*,\phi]$ describing the semiclassical `path' on either branch. 
Second, we can make use of the fact that $\phi({\bf x},t)$ can also be seen as 
the order parameter of the Bose gas. The corresponding time-dependent 
Landau-Ginzburg theory can easily be derived in the many-body T-matrix 
approximation, if we use the imaginary-time approach to the equilibrium theory 
of interacting Bose gases \cite{NO,zinn}. Peforming then the usual Wick rotation 
to real time, we again obtain $S^{cl}[\phi^*,\phi]$ for the effective action of 
the order parameter. 

In particular, we thus find that the dynamics of the zero-momentum part of the 
order parameter, i.e., the condensate, is in first instance determined by
\begin{equation}
S^{cl}_0[\phi^*_{\bf 0},\phi_{\bf 0}] = \int dt~ \left\{
   \phi^*_{\bf 0}(t) 
       \left( i\hbar \frac{\partial}{\partial t} 
                    - \epsilon'({\bf 0};t) + \mu(t) \right) \phi_{\bf 0}(t)
   - \frac{T^{(+)}({\bf 0},{\bf 0},{\bf 0};0)}{2V} |\phi_{\bf 0}(t)|^4 
   \right\}~.
\end{equation}
Introducing the density $n_0(t)$ of the condensate and its phase $\theta_0$ by 
means of the relation 
$\phi_{\bf 0}(t) \equiv \sqrt{n_0(t)V} e^{i\theta_0}$, this simply leads to
\begin{equation}
S^{cl}_0[n_0,\mu] =  V \int dt~ 
  \left( - \epsilon'({\bf 0};t) + \mu(t)
         - \frac{T^{(+)}({\bf 0},{\bf 0},{\bf 0};0)}{2} n_0(t) \right) n_0(t) 
~.
\end{equation}
Notice that in the process of deriving the last equation, we have omitted the 
topological term $\int dt \left( i\hbar \partial n_0(t)/\partial t \right)$ 
that does not affect the equations of motion and is therefore irrelevant at the 
semiclassical level. It is in principle only important when we also want to 
consider the quantum fluctuations of the condensate. Note also that we have 
made use of the definition of the chemical potential, or equivalently of the 
Josephson relation \cite{joseph}, to take the condensate phase independent of 
time. If we do not measure our energies relative to the chemical potential, we 
must use $\phi_{\bf 0}(t) \equiv \sqrt{n_0(t)V} e^{i\theta_0(t)}$ and  
$\hbar\partial\theta_0/\partial t = - \mu(t)$ instead, which then leads to 
exactly the same action. Clearly, the above action is minimized by 
$n_0(t) = 0$ if $\epsilon'({\bf 0};t) - \mu(t) > 0$. However, in agreement with 
our previous stability analysis, we have a nontrivial minimum at
\begin{equation}
\label{evol}
n_0(t) = 
  - \frac{\epsilon'({\bf 0};t) - \mu(t)}{T^{(+)}({\bf 0},{\bf 0},{\bf 0};0)}~,
\end{equation}
if $\epsilon'({\bf 0};t) - \mu(t) < 0$. This result gives the desired evolution 
of the condensate density after the gas has become unstable in terms of the 
time-dependent chemical potential $\mu(t)$. Our next task is therefore to 
determine an equation of motion for the chemical potential.

To achieve this we now also have to consider the interactions between the 
condensed and the noncondensed parts of the gas, which have been neglected 
thusfar. Substituting 
$\phi_{\bf k}(t) = \phi_{\bf 0}(t) \delta_{{\bf k},{\bf 0}} 
                           + \phi'_{\bf k}(t)(1 - \delta_{{\bf k},{\bf 0}})$
into the semiclassical action  and integrating over the fluctuations 
$\phi'_{\bf k}(t)$ describing the noncondensed part of the gas, 
we find that the exact semiclassical action for the condensate obeys
\begin{equation}
S^{cl}[n_0,\mu] = S_0^{cl}[n_0,\mu] -i\hbar {\rm ln}( Z^{cl}[n_0,\mu] )~,
\end{equation}
where $Z^{cl}[n_0,\mu]$ represents the functional integral over the 
fluctuations 
for given evolutions of the condensate density and the chemical potential. 
Writing $S^{cl}[\phi_{\bf 0}^*+\phi'^*,\phi_{\bf 0}+\phi']$ as 
$S_0^{cl}[\phi^*_{\bf 0},\phi_{\bf 0}] + S_1^{cl}[\phi'^*,\phi';n_0,\mu]$ we 
thus have explicitly that
\begin{equation}
Z^{cl}[n_0,\mu] = \int d[\phi'^*]d[\phi']~
        \exp \left\{ \frac{i}{\hbar} S_1^{cl}[\phi'^*,\phi';n_0,\mu] \right\}~.
\end{equation} 
With this action the total density of the gas is calculated in the 
thermodynamic limit as
\begin{equation}
n(t) = \frac{1}{V} \frac{\delta S^{cl}[n_0,\mu]}{\delta \mu(t)}
     = n_0(t) + \int \frac{d{\bf k}}{(2\pi)^3}~ N({\bf k};t)~,
\end{equation}
where the occupation numbers are found from                                                     
\begin{equation}
N({\bf k};t) = \frac{ 
   \int d[\phi'^*]d[\phi']~ \phi'^*_{\bf k}(t) \phi'_{\bf k}(t)
             \exp \left\{ iS_1^{cl}[\phi'^*,\phi';n_0,\mu]/\hbar \right\}
                     }{Z^{cl}[n_0,\mu]}
             \equiv \langle \phi'^*_{\bf k}(t) \phi'_{\bf k}(t) \rangle^{cl}~.
\end{equation}
The latter two equations, together with Eq.~(\ref{evol}), in principle give 
both the condensate density $n_0(t)$ and the chemical potential $\mu(t)$ as a 
function of the total density $n(t)$ and formally thus completely solve the 
semiclassical dynamics of the gas. However, the above derivation of the 
semiclassical theory appears to be not fully consistent. On the one hand we 
have in the calculation of the total density varied the full semiclassical 
action $S^{cl}[n_0,\mu]$ with respect to the chemical potential, but on the 
other hand we have varied only $S_0^{cl}[n_0,\mu]$ with respect to $n_0(t)$ to 
obtain the equation for the condensate. The following question therefore arises: 
Why does the term $-i\hbar {\rm ln}( Z^{cl}[n_0,\mu] )$ not contribute to the 
dynamics of the condensate? Fortunately, the reason for this is well-known 
\cite{weichmann}. Expanding $-i\hbar {\rm ln}( Z^{cl}[n_0,\mu] )$ in powers of 
$n_0(t)$ we can easily convince ourselves that the fluctuations only lead to 
corrections to the linear and quadratic terms in $S_0^{cl}[n_0,\mu]$ that in 
fact have already been accounted for by using the renormalized energy 
$\epsilon'({\bf 0};t)$ and the renormalized interaction 
$T^{(+)}({\bf 0},{\bf 0},{\bf 0};0)$. To avoid a double 
counting of the effects of the interaction, we should therefore indeed neglect 
the fluctuations when calculating the condensate density. Formally, this is 
equivalent to the requirement that the condensate density fulfills the 
Hugenholtz-Pines theorem \cite{HP}, because Eq.~(\ref{evol}) can be rewritten as 
$\mu(t) = \hbar\Sigma^{(+)}_{11}({\bf 0};t) 
                                        - \hbar\Sigma^{(+)}_{12}({\bf 0};t)$
in terms of the usual normal and anomalous selfenergies in 
$S^{cl}_1[\phi'^*,\phi';n_0,\mu]$. 

Determining the occupation numbers 
$\langle \phi'^*_{\bf k}(t) \phi'_{\bf k}(t) \rangle^{cl}$ requires solving an 
interacting quantum field theory, which cannot be done exactly. An 
approximation is thus called for. Taking only the quadratic terms in 
$S_1^{cl}[\phi'^*,\phi';n_0,\mu]$ into account amounts to the Bogoliubov 
approximation. Indeed, the action for the fluctuations then becomes equal to
\begin{eqnarray}
S_B[\phi'^*,\phi'] = \int dt~ \left\{
   \sum_{{\bf k} \neq {\bf 0}} \phi'^*_{\bf k}(t) 
       \left( i\hbar \frac{\partial}{\partial t} 
                - \epsilon({\bf k}) 
                      - n_0(t) T^{(+)}({\bf 0},{\bf 0},{\bf 0};0) \right) 
                \phi'_{\bf k}(t) \right.        \hspace*{0.8in} \nonumber \\
  \left. - \frac{1}{2} T^{(+)}({\bf 0},{\bf 0},{\bf 0};0) n_0(t)
      \sum_{{\bf k} \neq {\bf 0}} \left( 
         \phi'^*_{\bf k}(t) \phi'^*_{-{\bf k}}(t) +
         \phi'_{-{\bf k}}(t) \phi'_{\bf k}(t) \right)
   \right\}~,
\end{eqnarray}
if we use Eq.~(\ref{e0}) to evaluate the energy difference 
$\epsilon'({\bf k};t) - \epsilon'({\bf 0};t) = \epsilon({\bf k})$ at the long 
wavelengths of interest here. The energies of the Bogoliubov quasiparticles 
thus obey
\begin{equation}
\hbar\omega({\bf k};t) = 
    \sqrt{ \epsilon^2({\bf k})
             + 2n_0(t)T^{(+)}({\bf 0},{\bf 0},{\bf 0};0)\epsilon({\bf k}) }~,
\end{equation}
and are purely real at this level of approximation. However, in the Bogoliubov 
approximation the quasiparticles are noninteracting. This is reasonable 
sufficiently far below the critical temperature when the condensate density is 
large, but not very close to the critical temperature \cite{lee}. In that case 
the interactions between the quasiparticles are very important and cannot be 
neglected. 

\begin{figure}
\psfig{figure=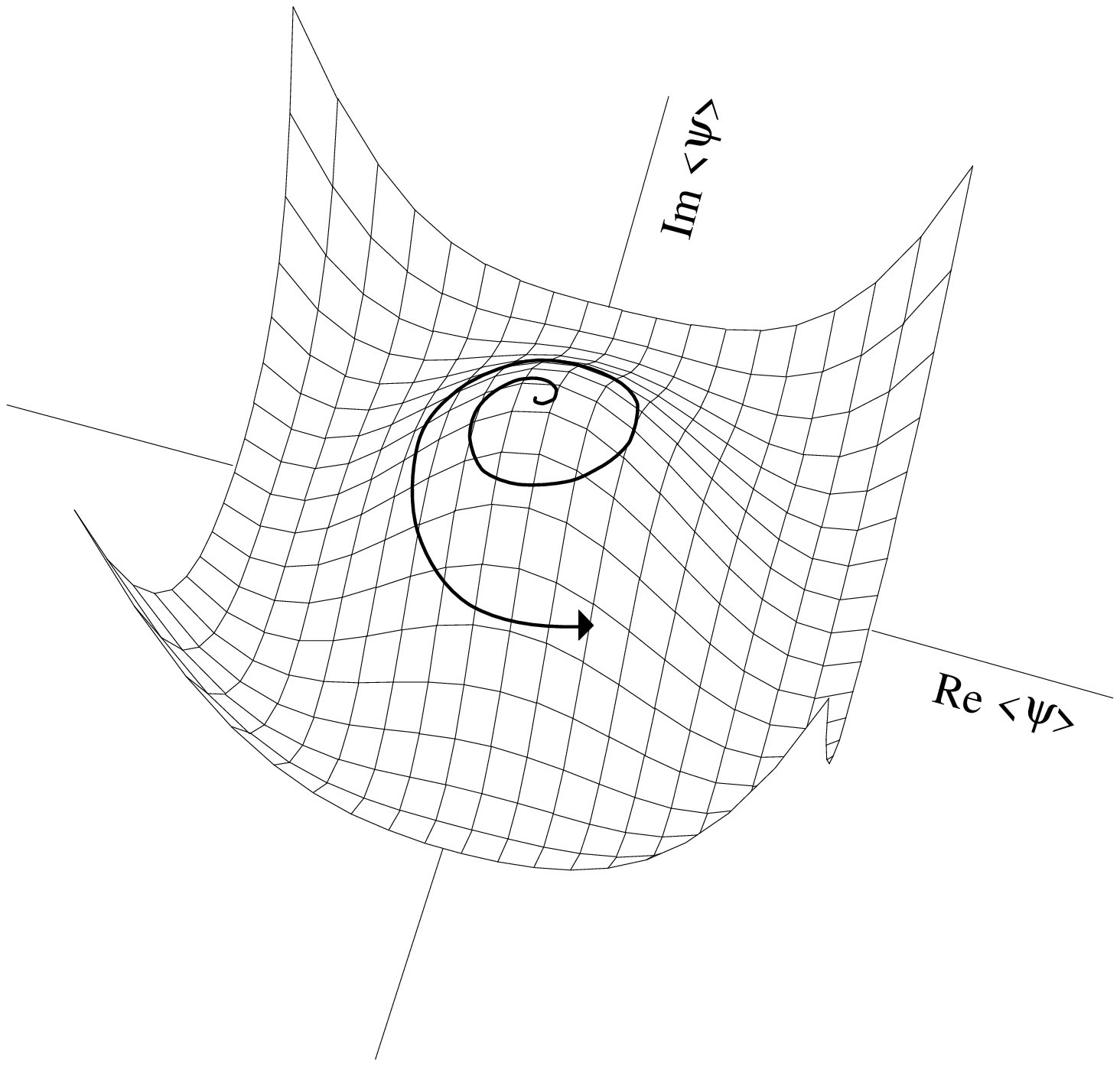}
\vspace*{0.3in}
\caption{Semiclassical picture of the coherent stage of Bose-Einstein   
         condensation 
         \label{semi}}
\end{figure}

A qualitative understanding of the effect of these interactions can be obtained 
by making use of a simple mean-field theory. Since the condensate density is 
relatively small we can approximate the average interaction between the 
quasiparticles by
\begin{eqnarray}
\left\langle \frac{1}{2V} \sum_{{\bf k},{\bf k}',{\bf K} \neq {\bf 0}} 
       T^{(+)}({\bf 0},{\bf 0},{\bf 0};0) 
         \phi'^*_{{\bf K}/2+{\bf k}}(t) \phi'^*_{{\bf K}/2-{\bf k}}(t)
         \phi'_{{\bf K}/2-{\bf k}'}(t) \phi'_{{\bf K}/2+{\bf k}'}(t)
\right\rangle^{cl}                       \hspace*{0.7in} \nonumber \\
\simeq V T^{(+)}({\bf 0},{\bf 0},{\bf 0};0) 
    \left( n^2(t) - n(t)n_0(t) 
           - n_0(t) \frac{1}{V} \sum_{{\bf k} \neq {\bf 0}}
                    \langle \phi'^*_{\bf k}(t) \phi'_{\bf k}(t) \rangle^{cl}
    \right)~.
\end{eqnarray}    
Within the context of a mean-field approximation, we therefore see that the 
interactions effectively renormalize the Bogoliubov theory to
\begin{eqnarray}
S_{RB}[\phi'^*,\phi'] = \int dt~ \left\{
   \sum_{{\bf k} \neq {\bf 0}} \phi'^*_{\bf k}(t) 
       \left( i\hbar \frac{\partial}{\partial t} - \epsilon({\bf k})  \right) 
                \phi'_{\bf k}(t) \right.        \hspace*{2.0in} \nonumber \\
  \left. - \frac{1}{2} T^{(+)}({\bf 0},{\bf 0},{\bf 0};0) n_0(t)
      \sum_{{\bf k} \neq {\bf 0}} \left( 
         \phi'^*_{\bf k}(t) \phi'^*_{-{\bf k}}(t) +
         \phi'_{-{\bf k}}(t) \phi'_{\bf k}(t) \right)
   \right\}
\end{eqnarray}
and lead to quasiparticle energies
\begin{equation}
\hbar\omega({\bf k};t) = 
    \sqrt{ \epsilon^2({\bf k})
           - \left( n_0(t)T^{(+)}({\bf 0},{\bf 0},{\bf 0};0) \right)^2 }
\end{equation}
that are purely imaginary at long wavelengths. The occupation numbers for these 
momentum states thus decay in favor of the condensate density. The typical time 
scale for the decay is of ${\cal O}(\hbar/ n_0T^{(+)}({\bf 0},{\bf 0};0))$ and 
always much slower than $\hbar/ k_BT_c$ because $a/\Lambda_c \ll 1$ for the 
quantum gases of interest. This {\it a posteriori} justifies our neglect of the 
memory effects in the above. In summary, the semiclassical picture of 
Bose-Einstein condensation can then be visualized as in Fig.~\ref{semi}, where 
we have also tried to indicate the dynamics of the phase $\theta_0(t)$ due to 
the Josephson relation. After the condensate density has been formed in this 
fully coherent manner, the distribution function $N({\bf k};t)$ is clearly not 
in equilibrium since all the particles in the low-energy states have been 
transferred to the condensate. In the third and last stage of the condensation 
process this distribution function relaxes to thermal equilibrium due to 
elastic collisions between the Bogoliubov quasiparticles. This can again be 
accurately described by a quantum Boltzmann equation \cite{bob,eckern}. We do 
not discuss this third stage in any detail here (see for example 
Refs.~\cite{ST,eckern}) and just end with a brief overview of the semiclassical 
picture of Bose-Einstein condensation. 

\subsection{Overview} 
In the semiclassical theory the divergence of the time required for 
Bose-Einstein condensation is avoided, because the incoherent dynamics of the 
occupation numbers $N({\bf k};t)$ for the one-particle states is coupled to the 
coherent dynamics of the energies $\epsilon'({\bf k};t)$ of these states, and 
the fluctuations of the condensate are neglected. As a result the process of 
Bose-Einstein condensation now proceeds in three stages. In the first stage the 
gas moves, for example as a response to quickly removing the hottest atoms, 
towards the critical region of the phase transition. This part of the evolution 
can be accurately described by a quantum Boltzmann equation, in which in a 
selfconsistent manner collisions between the atoms lead to changes in the 
occupation numbers $N({\bf k};t)$ and these changes in their turn affect the 
mean-field corrections to the energies $\epsilon'({\bf k};t)$. As long as the 
gas is outside the critical region, however, these mean-field corrections are 
constant and we simply have 
$\epsilon'({\bf k};t) = \epsilon({\bf k}) + 8\pi na \hbar^2/m$. Hence, if 
initially not enough energy is removed, the gas always remains in this regime 
and then typically evolves towards equilibrium on a time scale of 
${\cal O}(\tau_{el})$, which near the critical temperature is of 
${\cal O}( (\Lambda_c/a)^2 \hbar/k_BT_c )$ \cite{jom,wolfgang1,dave,boris,ST}. 

If on the other hand sufficient energy is removed initially, the gas enters 
the critical region and the second stage towards Bose-Einstein condensation 
begins. In this second stage, the gas first develops the required instability 
for the phase transition. Physically, the instability is a result of the fact 
that in the critical region, fluctuations tend to decrease the mean-field 
corrections to the energies of the low-momentum states with $\hbar k < {\cal 
O}(\hbar \sqrt{8\pi na})$  \cite{michel1,michel2}, which makes it 
possible for the zero-momentum state to acquire an energy which is less than 
the instantaneous chemical potential. It is then clearly energetically favorable 
to put a macroscopic number of particles in that state. It is important to note 
that for this instability mechanism, it is crucial that the typical time scale 
on which the mean-field corrections are established is only of 
${\cal O}(\hbar/k_BT_c)$ and therefore much faster than the time scale on which 
elastic collisions can change the occupations numbers. In that sense it can be 
said that Bose-Einstein condensation is `nucleated' on a time scale of ${\cal 
O}(\hbar/k_BT_c)$ \cite{henk2}.

As we have seen above, incoherent collisions alone cannot lead to a macroscopic 
occupation of the zero-momentum state. After the gas has become unstable, the 
actual growth of the condensate density is therefore in the semiclassical theory 
caused by a coherent depletion of the low-lying excited states in favor of the 
one-particle ground state. At the end of the second stage, which also has a 
duration of ${\cal O}( (\Lambda_c/a)^2 \hbar/k_BT_c )$, the gas has thus 
acquired a highly nonequilibrium energy distribution and has to come to 
equilibrium in a third and final stage. This last stage of the condensation 
process is again of a kinetic nature and can be accurately described by a 
quantum Boltzmann equation for the Bogoliubov quasiparticles \cite{bob,eckern}. 
Due to the linear dispersion of these quasiparticles at long wavelengths the 
typical time scale for relaxation towards equilibrium is exceptionally long and 
of ${\cal O}( (\Lambda_c/a)^3 \hbar/k_BT_c )$, as was shown by Eckern in the 
context of spin-polarized atomic hydrogen \cite{eckern}.    
 
For completeness sake, we should mention that Kagan, Svistunov, and Shlyapnikov 
have put forward a different semiclassical picture and assert that during the 
second, coherent stage of the evolution, the gas does not form a condensate 
but, just as in a two dimensional gas, only a quasicondensate \cite{kagan1}. 
Mathematically, this implies that the gas does not acquire a nonzero average of 
the annihilation operator of the zero-momentum state 
$\hat{\psi}_{\bf 0}(t)$ alone, but only of the sum 
$\sum_{k<k_0} \hat{\psi}_{\bf k}(t) e^{i{\bf k} \cdot {\bf x}}$, where 
$\hbar k_0$ is a momentum cut-off of ${\cal O}(\hbar \sqrt{8\pi na})$ 
\cite{popov}. This semiclassical picture is however not confirmed by our 
microscopic calculation, because from the above discussion of the instability 
mechanism it is clear that a possible instability for a quasicondensate is 
always preceeded by an instability for a real condensate since a removal of the 
long-wavelength fluctuations with $k<k_0$ always increases the mean-field 
corrections to the energies $\epsilon'({\bf k};t)$. 

The semiclassical theory neglects in particular the effect of thermal 
fluctuations on the condensate dynamics, which are nevertheless anticipated to 
be very important for conditions near the critical temperature when a large 
fraction of the gas is still uncondensed. For a quantitative comparison with 
experiments that are presently being performed \cite{wolfgang2}, the inclusion 
of these thermal fluctuations is therefore clearly of interest. There are also 
several theoretical reasons why we need to consider the effect of thermal 
fluctuations. 
Although the semiclassical theory of Bose-Einstein condensation gives important 
insight into the dynamics of this phase transition, it has two important 
drawbacks. First, it makes a sharp distinction between the kinetic and coherent 
stages of the evolution, which in principle of course overlap with each other. 
Second, the discussion of the growth of the condensate actually makes use of the 
presence of fluctuations, without taking these explicitly into account. This is 
due to the fact that at the semiclassical level we always have a solution to the 
equations of motion in which the gas remains at all times in the metastable, 
uncondensed phase. By considering the effect of thermal fluctuations, both these 
problems can indeed be resolved and we finally arrive at a single Fokker-Planck 
equation that describes all the three stages of the condensation process 
simultaneously. How this is achieved is reviewed next.

\section{QUANTUM THEORY OF BOSE-EINSTEIN CONDENSATION}
\label{quant}
In the semiclassical theory of the previous section the main emphasis is only 
on the average $\langle \phi_{\bf 0} \rangle(t)$, whereas the aim of the quantum 
theory is to consider also the fluctuations and determine the full probability 
distribution $P[\phi^*,\phi;t]$. Combining the results of Secs.~\ref{CL} and 
\ref{semicl}, we find that this probability distribution can be expressed as a 
functional integral with an effective action that in the many-body T-matrix 
approximation is given by
\begin{eqnarray}
S^{eff}[\psi^*,\psi]
  &=& \int_{{\cal C}^t} dt' \int_{{\cal C}^t} dt''~
       \left\{ \sum_{\bf k_{}} \psi^*_{\bf k}(t') \left[ 
            \left( i\hbar \frac{\partial}{\partial t'} 
                   - \epsilon({\bf k}) + \mu(t') \right) \delta(t',t'')
                   - \hbar \Sigma({\bf k};t',t'') 
               \right] \psi_{\bf k}(t'') \right. \hspace*{-0.1in} \nonumber \\
& & \hspace*{-0.4in} - \left. \frac{1}{2V} \sum_{{\bf k},{\bf k}',{\bf K}} 
     V({\bf k},{\bf k}',{\bf K};t',t'') 
       \psi^*_{{\bf K}/2+{\bf k}}(t') \psi^*_{{\bf K}/2-{\bf k}}(t')
       \psi_{{\bf K}/2-{\bf k}'}(t'') \psi_{{\bf K}/2+{\bf k}'}(t'') \right\}~.
\end{eqnarray} 
We next have to project the field $\psi({\bf x},t)$ onto the real axis by means 
of $\psi({\bf x},t_{\pm}) = \phi({\bf x},t) \pm \xi({\bf x},t)/2$. Up to 
quadratic terms in the fluctuations $\xi({\bf x},t)$ and neglecting 
again the memory effects, which is also known as the Markovian approximation 
\cite{mandel,nico} and is justified here because the memory time is at most of 
${\cal O}(\hbar/k_BT)$, we then easily obtain
\begin{eqnarray}
\label{full}
S^{eff}[\phi^*,\phi;\xi^*,\xi] 
  &=& \sum_{{\bf k},{\bf k'}} \int_{-\infty}^t dt'~
      \phi^*_{\bf k}(t') \left\{
            \left( i\hbar \frac{\partial}{\partial t'} 
                - \epsilon({\bf k}) + \mu(t') 
                - \hbar \Sigma^{(-)}({\bf k}) \right) \delta_{{\bf k},{\bf k'}}
            \right.                                         \nonumber \\
  & & \hspace*{1.3in} \left. 
       + \frac{1}{V} \sum_{{\bf k}''} V^{(-)}({\bf k},{\bf k'},{\bf k}'')
              \phi^*_{{\bf k'}+{\bf k}''-{\bf k}}(t') \phi_{{\bf k}''}(t')
            \right\} \xi_{\bf k'}(t')                       \nonumber \\
  &+& \sum_{{\bf k},{\bf k'}} \int_{-\infty}^t dt'~
      \xi^*_{\bf k}(t') \left\{
            \left( i\hbar \frac{\partial}{\partial t'} 
                - \epsilon({\bf k}) + \mu(t') 
                - \hbar \Sigma^{(+)}({\bf k}) \right) \delta_{{\bf k},{\bf k'}}
            \right.                                          \nonumber \\
  & & \hspace*{1.3in} \left. 
       + \frac{1}{V} \sum_{{\bf k}''} V^{(+)}({\bf k},{\bf k'},{\bf k}'')
              \phi^*_{{\bf k'}+{\bf k}''-{\bf k}}(t') \phi_{{\bf k}''}(t')
            \right\} \phi_{\bf k'}(t')                       \nonumber \\
  &-& \frac{1}{2} \sum_{{\bf k},{\bf k'}} \int_{-\infty}^t dt'~
      \xi^*_{\bf k}(t')
          \left\{ \hbar \Sigma^K({\bf k}) \delta_{{\bf k},{\bf k'}} 
              {}^{ {}^{ {}^{ {}^{ {}^{ {}^{} } } } } }\right. \nonumber \\
  & & \hspace*{1.3in} \left.
         + \frac{2}{V} \sum_{{\bf k}''} V^K({\bf k},{\bf k'},{\bf k}'')
              \phi^*_{{\bf k'}+{\bf k}''-{\bf k}}(t') \phi_{{\bf k}''}(t')
            \right\} \xi_{\bf k'}(t')~, 
\end{eqnarray}
introducing the shorthand notation $\Sigma^{(\pm),K}({\bf k})$ for the 
quantities $\Sigma^{(\pm),K}({\bf k};\epsilon'({\bf k})-\mu)$ that, just like 
the matrix elements $V^{(\pm),K}({\bf k},{\bf k'},{\bf k}'')$, implicitly also 
depend on time due to the time dependence of the occupation numbers. At this 
point it is possibly not immediately clear why we do not need to consider cubic 
and quartic terms in the fluctuations. The reason for that has been explicitly 
mentioned in Ref.~\cite{henk1} and boils down to the general problem of working 
with an effective theory, i.e., we must be careful not to double count the 
effects of the interaction. This almost happened in Sec.~\ref{sdyn} when we 
derived the semiclassical equation of motion for the condensate density and it 
would also happen here if we do not restrict ourselves to the linear and 
quadratic terms in the fluctuations.

As long as all the one-particle states are stable, we can in first instance 
neglect the nonlinear terms in $S^{eff}[\phi^*,\phi;\xi^*,\xi]$ that are 
proportional to $V^{(\pm),K}({\bf k},{\bf k'},{\bf k}'')$. Physically, this 
implies that we can treat the interacting atomic Bose gas as an (almost) 
noninteracting gas of dressed quasiparticles, i.e., as a `Bose liquid'. As a 
result the effective action becomes essentially identical to that of our 
Caldeira-Leggett toy model in Sec.~\ref{BQDR} and we can immediately write down 
the corresponding Fokker-Planck equation. Denoting the imaginary part of 
$\hbar\Sigma^{(\pm)}({\bf k})$ again by $\mp R({\bf k};t)$, we have
\begin{eqnarray}
\label{FPn}
i\hbar \frac{\partial}{\partial t} P[\phi^*,\phi;t]
 &=& - \sum_{\bf k} \frac{\partial}{\partial \phi_{\bf k}} 
       (\epsilon'({\bf k};t) - iR({\bf k};t) - \mu(t))
                                      \phi_{\bf k} P[\phi^*,\phi;t]
                                                        \nonumber \\ 
 &+&  \sum_{\bf k} \frac{\partial}{\partial \phi^*_{\bf k}} 
     (\epsilon'({\bf k};t) + iR({\bf k};t) - \mu(t))
                                    \phi^*_{\bf k} P[\phi^*,\phi;t]
                                                        \nonumber \\
 &-&  \frac{1}{2} \sum_{\bf k} 
    \frac{\partial^2}{\partial \phi^*_{\bf k} \partial \phi_{\bf k}} 
       \hbar \Sigma^K({\bf k}) P[\phi^*,\phi;t]~.                                                                                                       
\end{eqnarray}
However, to extract the physics from this Fokker-Planck equation, we now need 
to determine the quantities $R({\bf k};t)$ and $\hbar \Sigma^K({\bf k})$. Using 
our results from the semiclassical theory, in particular Eqs.~(\ref{splus}), 
(\ref{ot}) and (\ref{tsmall}), it is not difficult to show that the imaginairy 
part of the retarded and advanced selfenergies obeys
\begin{eqnarray}
R({\bf k};t) = 2\pi
     \int \frac{d{\bf k}'}{(2\pi)^3}
     \int \frac{d{\bf k}''}{(2\pi)^3}~
             \delta\left(\epsilon'({\bf k}';t) + \epsilon'({\bf k}'';t) -
                 \epsilon'({\bf k}' + {\bf k}'' - {\bf k};t) -
                 \epsilon'({\bf k};t) \right)   \hspace*{0.3in} \nonumber \\
\times \left| V^{(+)}({\bf k},{\bf k}',{\bf k}'') \right|^2 
   \left[ (1 + N({\bf k}';t) + N({\bf k}'';t)) 
                                        N({\bf k}' + {\bf k}''- {\bf k};t)
              - N({\bf k}';t) N({\bf k}'';t) 
          \right]~,
\end{eqnarray}  
which has recently also been obtained by an equation-of-motion method instead 
of the Schwinger-Keldysh formalism that is employed here \cite{nick}. To obtain 
the Keldysh component of the selfenergy we first have to realize that from the 
general Hartree-Fock-like relation for the selfenergy 
$\hbar\Sigma({\bf k};t,t')$ on the Schwinger-Keldysh contour it follows that 
\begin{eqnarray}
\hbar\Sigma^K({\bf k};\epsilon'({\bf k})-\mu) = 
  2 \int \frac{d{\bf k}'}{(2\pi)^3}~
      V^K({\bf k}-{\bf k}',{\bf k}-{\bf k}',{\bf k}+{\bf k}';
           \epsilon'({\bf k}) + \epsilon'({\bf k}') - 2\mu) N({\bf k}')
                                                                 \nonumber \\
 + 2 \int \frac{d{\bf k}'}{(2\pi)^3}~                                                          
      V^<({\bf k}-{\bf k}',{\bf k}-{\bf k}',{\bf k}+{\bf k}';
           \epsilon'({\bf k}) + \epsilon'({\bf k}') - 2\mu) 
                                                        (1 + 2N({\bf k}'))~,
\end{eqnarray}
after a Fourier transformation of the time difference $t-t'$. Subsituting now  
Eqs.~(\ref{tsmall}) and (\ref{tkel}) we then finally obtain   
\begin{eqnarray}
\hbar\Sigma^K({\bf k}) = - 4\pi i
     \int \frac{d{\bf k}'}{(2\pi)^3}
     \int \frac{d{\bf k}''}{(2\pi)^3}~
               \delta\left(\epsilon'({\bf k}';t) + \epsilon'({\bf k}'';t) -
                 \epsilon'({\bf k}' + {\bf k}''-{\bf k};t) -
                 \epsilon'({\bf k};t) \right)
                                   \nonumber \\
  \times  \left| V^{(+)}({\bf k},{\bf k}',{\bf k}'') \right|^2
      \left[ (1 + N({\bf k}';t) + N({\bf k}'';t))
                                N({\bf k}' + {\bf k}''- {\bf k};t)
                                   \right. \hspace*{1.0in} \nonumber \\
  \hspace*{1.8in} + \left. N({\bf k}';t) N({\bf k}'';t)
                      (1 + 2N({\bf k}' + {\bf k}'' - {\bf k};t))
          \right]~.
\end{eqnarray}

What does the above imply for the dynamics of the gas? Considering first the 
equation of motion of $\langle \phi_{\bf k} \rangle(t)$, the Fokker-Planck 
equation tells us that
\begin{equation}
i\hbar \frac{\partial}{\partial t} \langle \phi_{\bf k} \rangle(t)
  = (\epsilon'({\bf k};t) - iR({\bf k};t) - \mu(t))
                                   \langle \phi_{\bf k} \rangle(t)~.
\end{equation}
Hence the rate of decay of a particle with momentum $\hbar{\bf k}$ due to 
collisions with other particles in the gas equals 
$\Gamma({\bf k};t) = 2R({\bf k};t)/\hbar$. This indeed precisely agrees with a 
Fermi's Golden Rule calculation, because the rate for a particle to scatter out 
of the momentum state $\chi_{\bf k}({\bf x})$ is
\begin{eqnarray}
\label{out}
\Gamma^{out}({\bf k};t) = \frac{4\pi}{\hbar}
     \int \frac{d{\bf k}'}{(2\pi)^3}
     \int \frac{d{\bf k}''}{(2\pi)^3}~
             \delta\left(\epsilon'({\bf k}';t) + \epsilon'({\bf k}'';t) -
                 \epsilon'({\bf k}' + {\bf k}'' - {\bf k};t) -
                 \epsilon'({\bf k};t) \right)   \hspace*{0.1in} \nonumber \\
\times \left| V^{(+)}({\bf k},{\bf k}',{\bf k}'') \right|^2 
          (1 + N({\bf k}';t))(1 + N({\bf k}'';t)) 
                                        N({\bf k}' + {\bf k}''- {\bf k};t)~,
\end{eqnarray}
whereas the rate to scatter into that state is
\begin{eqnarray}
\label{in}
\Gamma^{in}({\bf k};t) = \frac{4\pi}{\hbar}
     \int \frac{d{\bf k}'}{(2\pi)^3}
     \int \frac{d{\bf k}''}{(2\pi)^3}~
             \delta\left(\epsilon'({\bf k}';t) + \epsilon'({\bf k}'';t) -
                 \epsilon'({\bf k}' + {\bf k}'' - {\bf k};t) -
                 \epsilon'({\bf k};t) \right)   \hspace*{0.1in} \nonumber \\
\times \left| V^{(+)}({\bf k},{\bf k}',{\bf k}'') \right|^2 
          N({\bf k}';t)N({\bf k}'';t)(1 + N({\bf k}' + {\bf k}''- {\bf k};t))~.
\end{eqnarray}
Note that both the in and the out rate are a factor of two larger than might 
have been expected in first instance due to the fact that the direct and 
exchange contributions interfere constructively in the collision process. 

Although the value of the decay rate 
$\Gamma({\bf k};t) = \Gamma^{out}({\bf k};t) - \Gamma^{in}({\bf k};t)$ is 
clearly a correct consequence of our Fokker-Planck equation, what is missing at 
this point is the dynamics of the occupation numbers $N({\bf k};t)$ themselves. 
This omission can be repaired by considering also the average 
$\langle |\phi_{\bf k}|^2 \rangle(t) = N({\bf k};t) + 1/2$. For that quantity 
the Fokker-Planck equation gives
\begin{equation}
i\hbar \frac{\partial}{\partial t} \langle |\phi_{\bf k}|^2 \rangle(t)
  = -2iR({\bf k};t) \langle |\phi_{\bf k}|^2 \rangle(t)
    - \frac{1}{2} \hbar \Sigma^K({\bf k})~.
\end{equation} 
As expected, this leads to the appropriate quantum Boltzmann equation 
\begin{eqnarray}
\frac{\partial}{\partial t} N({\bf k};t) &=& 
      - (\Gamma^{out}({\bf k};t) - \Gamma^{in}({\bf k};t)) 
        \left( N({\bf k};t) + \frac{1}{2} \right)
      + \frac{1}{2} i\Sigma^K({\bf k})                       \nonumber \\
  &=& - \Gamma^{out}({\bf k};t) N({\bf k};t) 
      + \Gamma^{in}({\bf k};t) (1 + N({\bf k};t))~,
\end{eqnarray}
if we make use of the observation that 
$i\Sigma^K({\bf k}) = \Gamma^{out}({\bf k};t) + \Gamma^{in}({\bf k};t)$. The 
gas, therefore, relaxes for $t \rightarrow \infty$ to an equilibrium in which 
the occupation numbers $N({\bf k})$ are equal to the Bose distribution 
evaluated at $\epsilon'({\bf k}) - \mu$, where both $\epsilon'({\bf k})$ and 
$\mu$ are determined selfconsistently by the retarded selfenergy in the same way 
as in the semiclassical theory. An identical conclusion is reached by noting 
that in equilibrium the fluctuation-dissipation theorem is satisfied, i.e., 
$\hbar\Sigma^K({\bf k}) = -2i(1 + 2N({\bf k}))R({\bf k})$. Therefore, the 
stationary solution of the Fokker-Planck equation equals
\begin{equation}
P[\phi^*,\phi;\infty] = \prod_{\bf k} \frac{1}{N({\bf k}) + 1/2}
     \exp \left\{ - \frac{1}{N({\bf k}) +1/2} |\phi_{\bf k}|^2
          \right\}~.
\end{equation}
In the degenerate regime and at long wavelengths the probability distribution 
for $|\phi_{\bf k}|$ is thus proportional to the Boltzmann factor 
$e^{-\beta(\epsilon'({\bf k}) - \mu) |\phi_{\bf k}|^2}$. Once again, this shows 
that the Landau free energy for the zero-momentum state 
$F(|\phi_{\bf 0}|) = (\epsilon'({\bf 0}) - \mu) |\phi_{\bf 0}|^2$ is unstable 
when $\epsilon'({\bf 0}) < \mu$. If that happens we are no longer allowed to 
neglect the nonlinear terms in $S^{eff}[\phi^*,\phi;\xi^*,\xi]$, because they 
are essential to ultimately stabilize the large fluctuations in the amplitude of 
$\phi({\bf x})$. 

The linearized version of the full quantum theory given by 
$S^{eff}[\phi^*,\phi;\xi^*,\xi]$ thus accurately describes the first kinetic 
stage of the condensation process in which the gas is quenched into the 
unstable region of the phase diagram. The next question is: How does the quantum 
theory deal with the subsequent coherent stage? To answer that question we can 
neglect all the imaginary parts in the effective action, which are associated 
with the kinetic processes as we have seen, but now retain the nonlinear terms. 
The Fokker-Planck equation for the long-wavelength dynamics of interest then 
reads
\begin{eqnarray}
\label{FPcoh}
i\hbar \frac{\partial}{\partial t} P[\phi^*,\phi;t] = \hspace*{0.2in} && 
                                                      \nonumber \\
 && \hspace*{-1.3in}
   - \sum_{\bf k} \frac{\partial}{\partial \phi_{\bf k}}
     \left( 
       (\epsilon'({\bf k};t) - \mu(t)) \phi_{\bf k} 
       + \frac{1}{V} \sum_{{\bf k}',{\bf k}''} 
              T^{(+)}({\bf 0},{\bf 0},{\bf 0};0)
              \phi^*_{{\bf k'}+{\bf k}''-{\bf k}} 
              \phi_{{\bf k}''}  \phi_{{\bf k}'}                          
      \right) P[\phi^*,\phi;t]                       \nonumber \\ 
 && \hspace*{-1.3in}
   + \sum_{\bf k} \frac{\partial}{\partial \phi^*_{\bf k}} 
     \left(
       (\epsilon'({\bf k};t) - \mu(t)) \phi^*_{\bf k}
       + \frac{1}{V} \sum_{{\bf k}',{\bf k}''} 
              T^{(+)}({\bf 0},{\bf 0},{\bf 0};0)
              \phi^*_{{\bf k}'}  \phi^*_{{\bf k}''}  
              \phi_{{\bf k'}+{\bf k}''-{\bf k}} 
      \right) P[\phi^*,\phi;t]~.                                                                                                       
\end{eqnarray}       
Since it contains only streaming terms the solution is immediately seen to be
\begin{equation}
\label{sol}
P[\phi^*,\phi;t] = 
      \int d[\phi^*_1]d[\phi_1]~ P[|\phi_1|^2;t_1]
        \prod_{\bf k} \delta \left( |\phi_{\bf k} - \phi^{cl}_{\bf k}(t)|^2
                              \right)~,
\end{equation}
where $P[|\phi_1|^2;t_1]$ is the probability distribution after the kinetic 
stage, $\phi^{cl}({\bf x},t)$ is the solution of the nonlinear Schr\"odinger 
equation 
\begin{equation}
i\hbar \frac{\partial \phi^{cl}({\bf x},t)}{\partial t}
  = \left( - \frac{\hbar^2 \nabla^2}{2m} + S({\bf 0};t) - \mu(t)
            + T^{(+)}({\bf 0},{\bf 0},{\bf 0};0) |\phi^{cl}({\bf x},t)|^2
    \right) \phi^{cl}({\bf x},t)
\end{equation}
with the initial condition $\phi^{cl}({\bf x},t_1) = \phi_1({\bf x})$, and 
$S({\bf k};t)$ denotes the real part of the retarded selfenergy that is well 
approximated by the constant $S({\bf 0};t)$ for small momenta 
$\hbar k \ll \hbar /\Lambda$ and given explicitly in Eq.~(\ref{e0}).

The behavior of the solution in Eq.~(\ref{sol}) has recently been investigated 
numerically by Damle, Majumdar, and Sachdev \cite{subir}. In agreement with the 
semiclassical picture, they find that the nonlinear Schr\"odinger equation has 
the property of depleting the states with momenta 
$\hbar k < \hbar \sqrt{8\pi n_0a}$ in favor of the growth of the condensate. In 
contrast with the semiclassical theory, however, the time it actually takes to 
form the condensate in a truly infinite system is infinite also. More 
precisely, the time to form a condensate in a system with linear system size $L$ 
diverges as 
\begin{eqnarray}
\frac{\hbar}{n_0 T^{(+)}({\bf 0},{\bf 0};0)} 
   \left( \frac{L}{\xi} \right)^z~,                            \nonumber
\end{eqnarray}
introducing the correlation length of the condensate $\xi = 1/\sqrt{8\pi n_0a}$ 
and the usual dynamical exponent $z$ \cite{bray} that was shown to be equal to 
$1$ in this case \cite{kagan2}. 
From this outcome of the quantum theory we conclude that the 
fluctuations with wavelengths larger than the correlation length $\xi$ play a 
crucial role. In the semiclassical theory the effect of these fluctuations is 
only accounted for in a mean-field sense, which allows the condensation to take 
place in a finite time. Put differently, not taking fully account of the 
fluctuations on the longest length scales has the same effect as considering a 
finite-size system. For a truly infinite system the thermodynamic limit in fact 
prevents the formation of the condensate even though the gas becomes unstable 
in a finite amount of time.  
    
To understand the last remarks in a somewhat different manner, let us now 
consider the gas in a box with volume $V=L^3$. Because of the energy gap of 
$(2\pi\hbar)^2/2mL^2$ between the one-particle ground state and the first 
excited state, there is always a regime (the details are given below) in which 
it is allowed to single out the zero-momentum state. In that regime the 
Fokker-Planck equation following from the zero-momentum part of 
$S^{eff}[\phi^*,\phi;\xi^*,\xi]$ is
\begin{eqnarray}
\label{FPcond}
i\hbar \frac{\partial}{\partial t} P[\phi^*_{\bf 0},\phi_{\bf 0};t] &=&
   - \frac{\partial}{\partial \phi_{\bf 0}} 
       \left( S({\bf 0};t) - i R({\bf 0};t) - \mu(t) 
         + \frac{1}{V} T^{(+)}({\bf 0},{\bf 0},{\bf 0};0)
              |\phi_{\bf 0}|^2 \right) \phi_{\bf 0} 
                                P[\phi^*_{\bf 0},\phi_{\bf 0};t] \nonumber \\ 
 &+& \frac{\partial}{\partial \phi^*_{\bf 0}} 
      \left( S({\bf 0};t) + iR({\bf 0};t) - \mu(t)
        + \frac{1}{V} T^{(+)}({\bf 0},{\bf 0},{\bf 0};0)
             |\phi_{\bf 0}|^2 \right) \phi^*_{\bf 0} 
                                P[\phi^*_{\bf 0},\phi_{\bf 0};t] \nonumber \\
 &-& \frac{1}{2} 
     \frac{\partial^2}{\partial \phi^*_{\bf 0} \partial \phi_{\bf 0}} 
       \hbar \Sigma^K({\bf 0}) P[\phi^*_{\bf 0},\phi_{\bf 0};t]~.                                                                                                       
\end{eqnarray}    
Anticipating now that the condensate dynamics is much slower than the dynamics 
of the noncondensed part of the gas, we can take for the selfeneries 
$\hbar\Sigma^{(\pm),K}({\bf 0})$ and the chemical potential their equilibrium 
values. There is, however, again a slight complication because the streaming 
terms in the right-hand side of Eq.~(\ref{FPcond}) show that the energy of the 
condensate is 
$\epsilon'({\bf 0}) + T^{(+)}({\bf 0},{\bf 0},{\bf 0};0)|\phi_{\bf 0}|^2/V$. To 
be fully consistent we must therefore evaluate 
$\hbar\Sigma^{(\pm),K}({\bf 0};\epsilon)$ at these energies to obtain the 
correct equilibrium values. Making then also use of the fluctuation-dissipation 
theorem that now becomes 
\begin{equation}
iR({\bf 0}) = - \frac{\beta}{4} \hbar \Sigma^K({\bf 0}) 
     \left( \epsilon'({\bf 0}) - \mu
            + \frac{1}{V} T^{(+)}({\bf 0},{\bf 0},{\bf 0};0) |\phi_{\bf 0}|^2
                                                        \right)
\end{equation}
in the degenerate regime of interest, we find that the Fokker-Planck equation 
for a $U(1)$ invariant probability distribution $P[|\phi_{\bf 0}|;t]$ is
\begin{eqnarray}
\label{laser}
i\hbar \frac{\partial}{\partial t} P[|\phi_{\bf 0}|;t] &=&
 - \frac{\beta}{4} \hbar \Sigma^K({\bf 0}) 
   \frac{\partial}{\partial \phi_{\bf 0}} 
       \left( \epsilon'({\bf 0}) - \mu 
         + \frac{1}{V} T^{(+)}({\bf 0},{\bf 0},{\bf 0};0)
              |\phi_{\bf 0}|^2 \right) \phi_{\bf 0} P[|\phi_{\bf 0}|;t] 
                                                                 \nonumber \\ 
&-& \frac{\beta}{4} \hbar \Sigma^K({\bf 0}) 
    \frac{\partial}{\partial \phi^*_{\bf 0}} 
      \left( \epsilon'({\bf 0}) - \mu
        + \frac{1}{V} T^{(+)}({\bf 0},{\bf 0},{\bf 0};0)
             |\phi_{\bf 0}|^2 \right) \phi^*_{\bf 0} P[|\phi_{\bf 0}|;t]
                                                                 \nonumber \\
&-& \frac{1}{2} \hbar \Sigma^K({\bf 0}) 
     \frac{\partial^2}{\partial \phi^*_{\bf 0} \partial \phi_{\bf 0}} 
       P[|\phi_{\bf 0}|;t]~.                                                                                                       
\end{eqnarray}     
The stationary solution therefore obeys
\begin{eqnarray}
P[|\phi_{\bf 0}|;\infty] \propto \exp \left\{
   - \beta \left( \epsilon'({\bf 0}) - \mu 
         + \frac{T^{(+)}({\bf 0},{\bf 0},{\bf 0};0)}{2V} |\phi_{\bf 0}|^2
               \right) |\phi_{\bf 0}|^2 \right\}~, \nonumber
\end{eqnarray} 
and shows explicitly that the nonlinear terms in the effective action 
$S^{eff}[\phi^*,\phi;\xi^*,\xi]$ have indeed stabilized the Landau free energy 
for the condensate 
$F(|\phi_{\bf 0}|) = (\epsilon'({\bf 0}) - \mu) |\phi_{\bf 0}|^2
                   +T^{(+)}({\bf 0},{\bf 0},{\bf 0};0)|\phi_{\bf 0}|^4/2V$ 
in the case that $\epsilon'({\bf 0}) < \mu$. Note that this result also shows 
why the nonlinearities can be neglected in the stable regime when 
$\epsilon'({\bf 0}) > \mu$, because the most important feature of this 
stationary solution is that if we consider the probability distribution for the 
condensate density 
$n_0 = V|\phi_{\bf 0}|^2$, i.e.,
\begin{eqnarray}
P[n_0;\infty] \propto \exp \left\{
   - \beta V \left( \epsilon'({\bf 0}) - \mu 
         + \frac{T^{(+)}({\bf 0},{\bf 0},{\bf 0};0)}{2} n_0
               \right) n_0 \right\}~, \nonumber
\end{eqnarray} 
it is for a large volume $V$ very sharply peaked around zero if 
$\epsilon'({\bf 0}) > \mu$. On the other hand, if $\epsilon'({\bf 0}) < \mu$ it 
is very sharply peaked around the semiclassical solution 
$n_0 = (\mu - \epsilon'({\bf 0}))/T^{(+)}({\bf 0},{\bf 0},{\bf 0};0)$. This 
represents the phenomenon of spontaneous symmetry breaking in the context of 
Bose-Einstein condensation. 

To extract the time scale for the probability distribution to relax to 
equilibrium, it is convenient to realize that Eq.~(\ref{laser}) is equivalent 
to the well-known Fokker-Planck equation for a single-mode laser \cite{mandel}. 
Indeed, introducing the dimensionless time variable
$\tau = t (i\Sigma^K({\bf 0})/8)
                      (2\beta T^{(+)}({\bf 0},{\bf 0},{\bf 0};0)/V)^{1/2}$ 
and in addition the dimensionless quantities 
$I = |\phi_{\bf 0}|^2 (2\beta T^{(+)}({\bf 0},{\bf 0},{\bf 0};0)/V)^{1/2}$
and $a = 2\beta(\mu - \epsilon'({\bf 0})) 
                      (V/2\beta T^{(+)}({\bf 0},{\bf 0},{\bf 0};0))^{1/2}$, 
that in the laser theory correspond respectively to the `laser intensity' and 
the so-called `pump parameter', it acquires the standard form
\begin{equation}
\label{sml}
\frac{\partial}{\partial \tau} P[I;\tau] =
  \frac{\partial}{\partial I} \left\{
    2(I-a)I + 4I \frac{\partial}{\partial I} \right\} P[I;\tau]
\end{equation}  
that has been studied numerically by Risken \cite{risken}. From the linear part 
of this equation we observe that initially the average `intensity' behaves as
\begin{equation}
\langle I \rangle(\tau) \simeq \frac{2}{a} \left( e^{2a\tau} - 1 \right)~.
\end{equation}
Hence, $\langle I \rangle(\tau)$ is of the order of its equilibrium value, 
i.e., 
of ${\cal O}(a)$, when $\tau$ is of ${\cal O}(\ln(a)/a)$. The time it takes to 
form a condensate is therefore at least of 
${\cal O}(\tau_{el} \ln(L/\Lambda_c))$ and again diverges for 
$L \rightarrow \infty$. At this point it is important to remember that the same 
time scale for the formation of a condensate was obtained in Sec.~\ref{semicl} 
from the quantum Boltzmann equation. This is not surprising because by 
considering only the dynamics of the zero-momentum state, a condensate can only 
be formed by incoherent collisions. Physically this is, however, only a good 
approximation as long as the mean-field interaction $n_0T^{(+)}({\bf 0},{\bf 
0},{\bf 0};0)$ due to the condensate is small compared to the energy splitting 
of the one-particle states in the box or if
\begin{eqnarray}
\mu - \epsilon'({\bf 0}) \ll 
     \left( \frac{2\pi}{L} \right)^2 \frac{\hbar^2}{2m}~. \nonumber
\end{eqnarray}
If this condition is not fulfilled, which is generally the case for a large 
system size, we are no longer allowed to single out the zero-momentum state 
since all the states with momenta $\hbar k < \hbar \sqrt{8\pi n_0a}$ are now 
coupled by the mean-field interaction. As we have seen in Eq.~(\ref{sol}), and 
in an approximate sense also in the semiclassical theory, the condensate can 
then also be formed in a fully coherent manner. In the language of laser theory 
this means that Bose-Einstein condensation is then analogous to a multi-mode 
laser instead of a single-mode laser described by Eq.~(\ref{sml}).

The competition between the coherent and incoherent processes trying to form a 
condensate is of course contained in the Fokker-Planck equation associated with 
the full effective action $S^{eff}[\phi^*,\phi;\xi^*,\xi]$ given in 
Eq.~(\ref{full}). By performing a Hubbard-Stratonovich transformation or by 
integrating over the fluctuations and quantizing the resulting theory, we 
ultimately find that \cite{henk3}
\begin{eqnarray}
\label{FPfull}
i\hbar \frac{\partial}{\partial t} P[\phi^*,\phi;t] = && \nonumber \\
&& \hspace{-1.1in}
 - \sum_{\bf k}
   \frac{\partial}{\partial \phi_{\bf k}}
       \left( (\epsilon({\bf k}) 
            + \hbar\Sigma^{(+)}({\bf k}) - \mu(t)) \phi_{\bf k}
            + \frac{1}{V} \sum_{{\bf k}',{\bf k}''}
               V^{(+)}({\bf k},{\bf k}',{\bf k}'')
                   \phi^*_{{\bf k}'+ {\bf k}''-{\bf k}}
                   \phi_{{\bf k}''} \phi_{{\bf k}'}
       \right) P[\phi^*,\phi;t]                         \nonumber \\
&& \hspace{-1.1in}
+ \sum_{\bf k}
  \frac{\partial}{\partial \phi^*_{\bf k}}
       \left( (\epsilon({\bf k}) 
            + \hbar\Sigma^{(-)}({\bf k}) - \mu(t)) \phi^*_{\bf k}
            + \frac{1}{V} \sum_{{\bf k}',{\bf k}''}
               V^{(-)}({\bf k},{\bf k}',{\bf k}'')
                   \phi^*_{{\bf k}'} \phi^*_{{\bf k}''}
                   \phi_{{\bf k}'+ {\bf k}''-{\bf k}}
       \right) P[\phi^*,\phi;t]                         \nonumber \\
&& \hspace{-1.1in}
- \frac{1}{2} \sum_{{\bf k},{\bf k}'}
  \frac{\partial^2}{\partial \phi_{\bf k}
                    \partial \phi^*_{{\bf k}'}}
       \left( \hbar\Sigma^K({\bf k}) \delta_{{\bf k},{\bf k}'}
            + \frac{2}{V} \sum_{{\bf k}''}
               V^K({\bf k},{\bf k}',{\bf k}'')
                   \phi^*_{{\bf k}'+ {\bf k}''-{\bf k}}
                   \phi_{{\bf k}''}
       \right) P[\phi^*,\phi;t]~.
\end{eqnarray}
We have already seen in detail how this equation describes the first and second 
stages of Bose-Einstein condensation. What remains to be shown is that it in 
fact also describes the third kinetic stage in which the condensed and 
noncondensed parts of the gas come to equilibrium. To do so, we make the ansatz 
that the probability distribution $P[\phi^*,\phi;t]$ separates as 
$P_0[\phi^*_{\bf 0},\phi_{\bf 0};t] P_1[\phi'^*,\phi';t]$. As will be clear 
shortly, this essentially amounts to a Hartree-Fock approximation. Substituting 
this ansatz in Eq.~(\ref{FPfull}) and integrating over the field 
$\phi'({\bf x})$ that is again associated with the noncondensed part of the 
gas, we recover Eq.~(\ref{FPcond}) for the probability distribution of the 
condensate
\begin{eqnarray}
i\hbar \frac{\partial}{\partial t} P_0[\phi^*_{\bf 0},\phi_{\bf 0};t] &=&
   - \frac{\partial}{\partial \phi_{\bf 0}} 
       \left( \hbar\Sigma^{(+)}({\bf 0}) - \mu(t) 
         + \frac{1}{V} T^{(+)}({\bf 0},{\bf 0},{\bf 0};0)
              |\phi_{\bf 0}|^2 \right) \phi_{\bf 0} 
                                P_0[\phi^*_{\bf 0},\phi_{\bf 0};t] \nonumber \\ 
 &+& \frac{\partial}{\partial \phi^*_{\bf 0}} 
      \left( \hbar\Sigma^{(-)}({\bf 0}) - \mu(t)
        + \frac{1}{V} T^{(+)}({\bf 0},{\bf 0},{\bf 0};0)
             |\phi_{\bf 0}|^2 \right) \phi^*_{\bf 0} 
                                P_0[\phi^*_{\bf 0},\phi_{\bf 0};t] \nonumber \\
 &-& \frac{1}{2} 
     \frac{\partial^2}{\partial \phi^*_{\bf 0} \partial \phi_{\bf 0}} 
       \hbar \Sigma^K({\bf 0}) P_0[\phi^*_{\bf 0},\phi_{\bf 0};t]~,                                                                                                       
\end{eqnarray}    
if we are careful not to double count the effects of the noncondensed part of 
the gas that are already included in the selfenergies 
$\hbar\Sigma^{(\pm),K}({\bf 0})$. Integrating over $\phi_{\bf 0}$ instead, we 
also arrive at a Fokker-Planck equation for the probability distribution 
$P_1[\phi'^*,\phi';t]$ that reads
\begin{eqnarray}
\label{FPHF}
i\hbar \frac{\partial}{\partial t} P_1[\phi'^*,\phi';t] &=&  \nonumber \\
&& \hspace*{-0.8in}
- \sum_{{\bf k} \neq {\bf 0}}
   \frac{\partial}{\partial \phi'_{\bf k}}
       \left( (\epsilon({\bf k}) 
           + \hbar\Sigma^{(+)}({\bf k}) 
           + 2 V^{(+)}({\bf k},{\bf k},{\bf 0}) n_0(t) 
           - \mu(t)) \phi'_{\bf k}
       \right) P_1[\phi'^*,\phi';t]                         \nonumber \\
&& \hspace*{-0.8in}
+ \sum_{{\bf k} \neq {\bf 0}}
  \frac{\partial}{\partial \phi'^*_{\bf k}}
       \left( (\epsilon({\bf k}) 
           + \hbar\Sigma^{(-)}({\bf k}) 
           + 2 V^{(-)}({\bf k},{\bf k},{\bf 0}) n_0(t) 
           - \mu(t)) \phi'^*_{\bf k}
       \right) P_1[\phi'^*,\phi';t]                         \nonumber \\
&& \hspace*{-0.8in}
- \frac{1}{2} \sum_{{\bf k} \neq {\bf 0}}
  \frac{\partial^2}{\partial \phi'^*_{\bf k}
                    \partial \phi'_{{\bf k}}}
       \left( \hbar\Sigma^K({\bf k})
            + V^K({\bf k},{\bf k},{\bf 0}) n_0(t) 
       \right) P_1[\phi'^*,\phi';t]~.
\end{eqnarray}
From the real part of the streaming terms we observe that in the presence of a 
condensate the renormalized energies of the states $\chi_{\bf k}({\bf x})$ with 
${\bf k} \neq {\bf 0}$ are well approximated by
\begin{equation}
\epsilon''({\bf k};t) = \epsilon({\bf k}) 
  + S({\bf k};t) + 2n_0(t) T^{(+)}({\bf 0},{\bf 0},{\bf 0};0)~,
\end{equation}
as expected from a Hartree-Fock approximation. Furthermore, the quantum 
Boltzmann equation that follows from the imaginary terms in the right-hand side 
of Eq.~(\ref{FPHF}) is again of the form
\begin{equation}
\frac{\partial}{\partial t} N({\bf k};t) 
    = - \Gamma^{out}({\bf k};t) N({\bf k};t) 
      + \Gamma^{in}({\bf k};t) (1 + N({\bf k};t))~,
\end{equation}
but $\Gamma^{out}$ and $\Gamma^{in}$ are now given by Eqs.~(\ref{out}) and 
(\ref{in}) with $N({\bf k}' + {\bf k}''- {\bf k};t)$ replaced by 
$N({\bf k}' + {\bf k}''- {\bf k};t) 
   + n_0(t) (2\pi)^3 \delta({\bf k}' + {\bf k}''- {\bf k})$ and 
$\epsilon'({\bf k};t)$ by $\epsilon''({\bf k};t)$
to account for the macroscopic occupation of the zero-momentum state. Apart 
from the use of the renormalized one-particle energies in the expressions for 
the rates for particles to scattering out of and into the momentum state 
$\chi_{\bf k}({\bf x})$, this kinetic equation is identical to the one studied 
by Semikoz and Tkachev \cite{ST}.

In equilibrium the occupation numbers are thus equal to a Bose distribution 
evaluated at  
$\epsilon({\bf k}) 
 + S({\bf k}) + 2n_0 T^{(+)}({\bf 0},{\bf 0},{\bf 0};0)  - \mu$.
However, for a large system size we have seen that the chemical potential is 
given by the Hugenholtz-Pines relation
$\mu = S({\bf k}) + n_0 T^{(+)}({\bf 0},{\bf 0},{\bf 0};0)$. 
Substituting this we obtain that the Bose distribution is in effect evaluated 
at
$\hbar\omega({\bf k}) = 
   \epsilon({\bf k}) + T^{(+)}({\bf 0},{\bf 0},{\bf 0};0) n_0$,
which precisely corresponds to the high-momentum (or Hartree-Fock) limit of the 
expected Bogoliubov dispersion
$\hbar\omega({\bf k}) = 
   ( \epsilon^2({\bf k})+ 2n_0 
                 T^{(+)}({\bf 0},{\bf 0},{\bf 0};0) \epsilon({\bf k}) )^{1/2}$.
For the quantum gases of interest the thermal momenta at the critical 
temperature are always much larger than $\hbar \sqrt{8\pi n_0 a}$ and the 
Hartree-Fock approximation presented above is very accurate. It can thus safely 
be used to describe the third stage of the condensation process. For 
applications of Eq.~(\ref{FPfull}) at temperatures very far below the critical 
temperature, it is in principle also of interest to knows how we can recover the 
Bogoliubov approximation from our Fokker-Planck equation \cite{yvan1}. Such 
extreme low temperatures have experimentally, however, only been obtained 
because of the inhomogeneity of the magnetically trapped atomic gases. 
Therefore, we do not discuss the Bogoliubov approximation here and return to 
this point in the next section when we have generalized the quantum theory to 
the inhomogeneous case. For now we only mention that the derivation of the 
Bogoliubov theory is not difficult in principle but unnecessary for a 
homogeneous atomic Bose gas, because our previous ways of looking at the 
Fokker-Planck equation in Eq.~(\ref{FPfull}) have already shown that it can 
accurately describe all the coherent and incoherent processes during 
Bose-Einstein condensation in that case.  
 
\section{BOSE-EINSTEIN CONDENSATION IN AN EXTERNAL POTENTIAL}
\label{inhom}
\begin{figure}
\psfig{figure=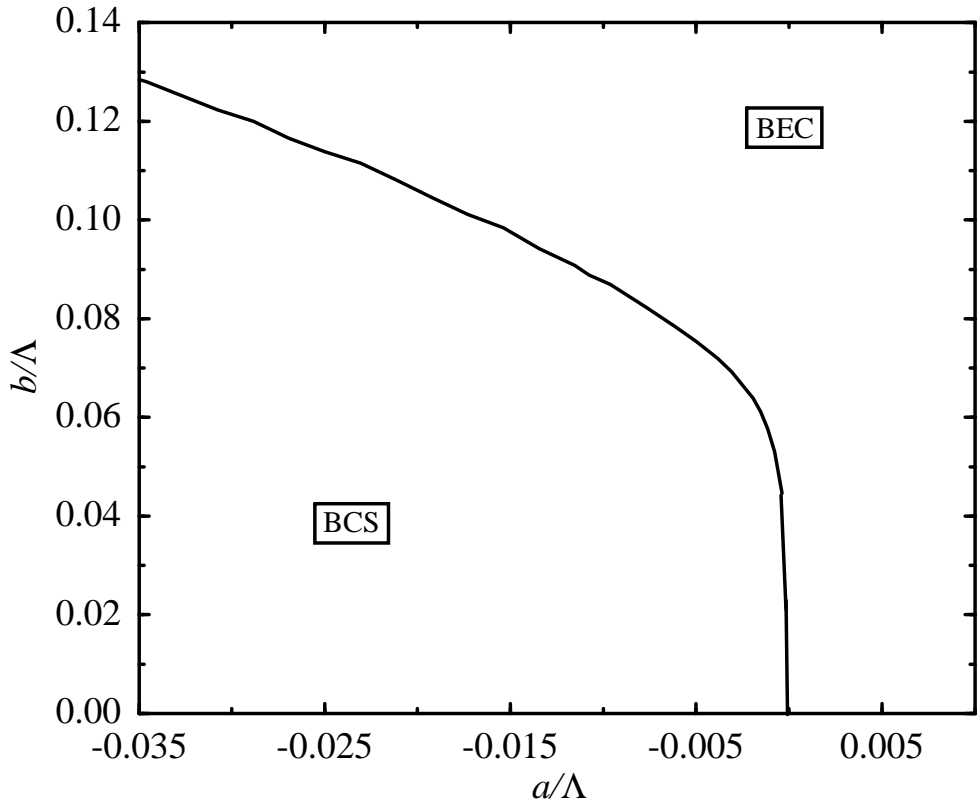}
\caption{Renormalization group calculation for the (meta)stable phases of a
         homogeneous Bose gas with two and three-body interactions. The
         two-body T matrix is equal to $4\pi\hbar^2 a/m$, whereas the
         three-body T matrix equals $4\pi\hbar^2 b^4/m$.
         \label{RG}}
\end{figure}

Up to now we have only considered atomic Bose gases with effectively repulsive 
interactions, i.e., a positive scattering length $a$. The reason for this 
restriction is that a homogeneous gas with effectively attractive interactions 
never develops an instability towards Bose-Einstein condensation as we have 
seen \cite{neg}. Instead it develops an instability towards a BCS-like 
transition that for bosons is known as an Evans-Rashid transition \cite{evans}. 
More precisely, this state of affairs depends on the strength of the three-body 
interaction, which in principle can stabilize a condensate if it is sufficiently 
repulsive. Using the renormalization group approach of Ref.~\cite{michel2}, this 
can easily be studied quantitatively and the result is shown here in 
Fig.~\ref{RG}. It is important to mention that the renormalization group 
calculation predicts the Evans-Rashid transition to take place in the 
mechanically stable region of the phase diagram, in contrast to the mean-field 
treatment \cite{RGref}. Note also that in general the three-body interaction 
must be anomalously large to be able to lead to Bose-Einstein condensation and 
this is therefore not expected to occur in realistic atomic gases with a 
negative scattering length.

Although this essentially rules out the achievement of Bose-Einstein 
condensation in a homogeneous gas with $a < 0$, this is no longer true if the 
gas is confined in an external trapping potential, as was first argued by 
Sackett and Hulet \cite{cass1} and subsequently shown convincingly by 
Ruprecht, Holland, Burnett, and Edwards \cite{keith1}. In fact, the 
quantum theory for Bose-Einstein condensation in a trapped gas with attractive 
interactions is even more simple than for a gas with repulsive interactions, 
because the average mean-field interaction $n|T^{(+)}({\bf 0},{\bf 0};0)|$ can 
never become much larger than the energy splitting of the trap due to the 
intrinsic instability of the gas to collapse to a dense phase. We therefore 
treat this case first and then turn to the discussion of Bose-Einstein 
condensation in a harmonically trapped gas with repulsive interactions, where 
the average mean-field interaction is usually much larger than the energy 
splitting in the trap and we cannot neglect the effect of the interactions on 
the one-particle eigenstates in the external potential. These conditions are in 
particular realized in the recent experiment by H.-J. Miesner {\it et al.} that 
has resulted in the first quantitative data on the dynamics of condensate 
formation \cite{wolfgang2}.

\subsection{Weak-Coupling Limit}
\label{wc}
For a trapped atomic gas the appropriate one-particle states are no longer the 
momentum states $\chi_{\bf k}({\bf x})$ but instead the eigenstates 
$\chi_{\alpha}({\bf x})$ of the external potential $V^{ex}({\bf x})$. As a 
result the Fokker-Planck equation for the interacting gas will now quite 
generally be given by
\begin{eqnarray}
\label{FPinhom}
i\hbar \frac{\partial}{\partial t} P[\phi^*,\phi;t] = && \nonumber \\
&& \hspace{-0.7in}
 - \sum_{\alpha,\alpha'}
   \frac{\partial}{\partial \phi_{\alpha}}
       \left\{ \left( (\epsilon_{\alpha} - \mu(t)) \delta_{\alpha,\alpha'} 
            + \hbar\Sigma^{(+)}_{\alpha,\alpha'} \right) \phi_{\alpha'}
            + \sum_{\beta,\beta'}
               V^{(+)}_{\alpha,\beta;\alpha',\beta'}
                   \phi^*_{\beta}
                   \phi_{\beta'} \phi_{\alpha'}
       \right\} P[\phi^*,\phi;t]                         \nonumber \\
&& \hspace{-0.7in}
+ \sum_{\alpha,\alpha'}
  \frac{\partial}{\partial \phi^*_{\alpha}}
       \left\{ \left( (\epsilon_{\alpha} - \mu(t)) \delta_{\alpha,\alpha'}
            + \hbar\Sigma^{(-)}_{\alpha',\alpha} \right) \phi^*_{\alpha'}
            + \sum_{\beta,\beta'}
               V^{(-)}_{\alpha',\beta';\alpha,\beta}
                   \phi^*_{\alpha'} \phi^*_{\beta'}
                   \phi_{\beta}
       \right\} P[\phi^*,\phi;t]                         \nonumber \\
&& \hspace{-0.7in}
- \frac{1}{2} \sum_{\alpha,\alpha'}
  \frac{\partial^2}{\partial \phi_{\alpha}
                    \partial \phi^*_{\alpha'}}
       \left\{ \hbar\Sigma^K_{\alpha,\alpha'}
            + \sum_{\beta,\beta'}
               V^K_{\alpha,\beta;\alpha',\beta'}
                   \phi^*_{\beta}\phi_{\beta'}
       \right\} P[\phi^*,\phi;t]~.
\end{eqnarray}
For this equation to be useful, however, we now need to determine the various 
selfenergies and interactions. To do so, let us for simplicity first consider 
the normal state of the gas which is usually in the weak-coupling limit 
$|\hbar\Sigma^{(+)}_{\alpha,\alpha'}| 
                                  \ll |\epsilon_{\alpha} - \epsilon_{\alpha'}|$ 
for the experiments with atomic alkali gases that are of interest 
here. In this regime we can neglect the nondiagonal elements of the 
selfenergies and also the nonlinear terms in the right-hand side of 
Eq.~(\ref{FPinhom}). The Fokker-Planck equation then reduces to
\begin{eqnarray}
i\hbar \frac{\partial}{\partial t} P[\phi^*,\phi;t] = && \nonumber \\
&& \hspace{-1.0in}
- \sum_{\alpha}
  \frac{\partial}{\partial \phi_{\alpha}}
       (\epsilon_{\alpha} + \hbar\Sigma^{(+)}_{\alpha} - \mu(t)) \phi_{\alpha}
             P[\phi^*,\phi;t] 
+ \sum_{\alpha}
  \frac{\partial}{\partial \phi^*_{\alpha}}
       (\epsilon_{\alpha} + \hbar\Sigma^{(-)}_{\alpha} - \mu(t)) 
\phi^*_{\alpha}
            P[\phi^*,\phi;t]                         \nonumber \\
&& \hspace{-1.0in}
- \frac{1}{2} \sum_{\alpha}
  \frac{\partial^2}{\partial \phi^*_{\alpha}
                    \partial \phi_{\alpha}}
        \hbar\Sigma^K_{\alpha} P[\phi^*,\phi;t]~,
\end{eqnarray} 
with $\hbar\Sigma^{(\pm),K}_{\alpha} 
           = \hbar\Sigma^{(\pm),K}_{\alpha,\alpha}(\epsilon'_{\alpha} - \mu)$
and $\epsilon'_{\alpha} = \epsilon_{\alpha} + S_{\alpha}$. This equation looks 
identical to the one we derived for our Caldeira-Leggett toy model but 
describes in fact quite different physics because now the selfenergies are not 
due to an interaction with a reservoir but due to the interactions between the 
atoms themselves. This is, however, completely hidden in the selfenergies as we 
show now explicitly. Performing the same manipulations as in the homogeneous 
case, we find in the many-body T-matrix approximation first of all that
\begin{eqnarray}
\hbar\Sigma^{(+)}_{\alpha} = \left( \hbar\Sigma^{(-)}_{\alpha} \right)^* =
  2 \sum_{\beta} V^{(+)}_{\alpha,\beta;\alpha,\beta}
                         (\epsilon'_{\alpha} + \epsilon'_{\beta} - 2\mu)
                                    N_{\beta}    \hspace*{1.5in} \nonumber \\
  + 2i \sum_{\beta} \int \frac{d\epsilon}{2\pi\hbar}~
       V^<_{\alpha,\beta;\alpha,\beta}(\epsilon) 
 \frac{\hbar}{\epsilon^- - (\epsilon'_{\alpha} + \epsilon'_{\beta} - 2\mu)}
\end{eqnarray}  
and similarly that
\begin{equation}
\hbar\Sigma^K_{\alpha} = 
  2 \sum_{\beta} V^K_{\alpha,\beta;\alpha,\beta}
                         (\epsilon'_{\alpha} + \epsilon'_{\beta} - 2\mu)
                                    N_{\beta} 
  + 2 \sum_{\beta} V^<_{\alpha,\beta;\alpha,\beta}
                           (\epsilon'_{\alpha} + \epsilon'_{\beta} - 2\mu) 
        (1 + 2N_{\beta})~.
\end{equation}

Our next task is therefore the evaluation of the various many-body T-matrix 
elements 
$V^{(+),<,K}_{\alpha,\beta;\alpha',\beta'}(\epsilon) = 
    (T^{(+),<,K}_{\alpha,\beta;\alpha',\beta'}(\epsilon) +    
                       T^{(+),<,K}_{\beta,\alpha;\alpha',\beta'}(\epsilon))/2$
that occur in these expressions, which is easily carried out with our 
experience of the homogeneous gas and by noting that the Lippmann-Schwinger 
equation in Eq.~(\ref{LS}) is in fact an operator equation that only needs to 
be expressed in a new basis. In terms of the appropriate matrix elements of the 
interatomic interaction, i.e.,
\begin{equation}
V_{\alpha,\beta;\alpha',\beta'} = \int d{\bf x} \int d{\bf x}'~
   \chi^*_{\alpha}({\bf x}) \chi^*_{\beta}({\bf x}') V({\bf x}-{\bf x}')
        \chi_{\alpha'}({\bf x}) \chi_{\beta'}({\bf x}')~,
\end{equation}
the retarded and advanced components of the many-body T matrix thus obey the 
Bethe-Salpeter equation \cite{nick}
\begin{equation}
T^{(\pm)}_{\alpha,\beta;\alpha',\beta'}(\epsilon)
  = V_{\alpha,\beta;\alpha',\beta'} 
      + \sum_{\alpha'',\beta''} V_{\alpha,\beta;\alpha'',\beta''}
      \frac{1 + N_{\alpha''} + N_{\beta''} }
        {\epsilon^{\pm} - (\epsilon'_{\alpha''} + \epsilon'_{\beta''} - 2\mu)}
    T^{(\pm)}_{\alpha'',\beta'';\alpha',\beta'}(\epsilon) 
\end{equation}
and the optical theorem
\begin{eqnarray}
T^{(+)}_{\alpha,\beta;\alpha',\beta'}(\epsilon)
  - T^{(-)}_{\alpha,\beta;\alpha',\beta'}(\epsilon) =
                                                  \hspace*{3.7in} \nonumber \\
- 2\pi i \sum_{\alpha'',\beta''} 
     \delta(\epsilon - (\epsilon'_{\alpha''} + \epsilon'_{\beta''} - 2\mu))
        T^{(+)}_{\alpha,\beta;\alpha'',\beta''}(\epsilon)
            (1 + N_{\alpha''} + N_{\beta''}) 
                   T^{(-)}_{\alpha'',\beta'';\alpha',\beta'}(\epsilon)~.
\end{eqnarray}
Finally, we also need
\begin{equation}
T^<_{\alpha,\beta;\alpha',\beta'}(\epsilon) =
- 2\pi i \sum_{\alpha'',\beta''} 
    \delta(\epsilon - (\epsilon'_{\alpha''} + \epsilon'_{\beta''} - 2\mu))
       T^{(+)}_{\alpha,\beta;\alpha'',\beta''}(\epsilon)
          N_{\alpha''} N_{\beta''} 
               T^{(-)}_{\alpha'',\beta'';\alpha',\beta'}(\epsilon)
\end{equation}
and the Keldysh component
\begin{eqnarray}
T^K_{\alpha,\beta;\alpha',\beta'}(\epsilon) =  
- 2\pi i \sum_{\alpha'',\beta''} 
  \delta(\epsilon - (\epsilon'_{\alpha''} + \epsilon'_{\beta''} - 2\mu))
                                               \hspace*{2.5in} \nonumber \\
\times T^{(+)}_{\alpha,\beta;\alpha'',\beta''}(\epsilon)
    (1 + N_{\alpha''} + N_{\beta''} + 2 N_{\alpha''} N_{\beta''}) 
                   T^{(-)}_{\alpha'',\beta'';\alpha',\beta'}(\epsilon)~.
\end{eqnarray}
Notice that the appearance of a $\delta$ function in the last three equations 
in practice never leads to any problems, because in the experiments with 
trapped atomic gases the number of particles is always so large that the 
critical temperature of the gas is much larger than the energy splitting of the 
trap and the discrete sum over states is well approximated by an integral over 
a continuous spectrum.

Making use of the above relations we again find that for the states with 
thermal energies the renormalized energy is given by the pseudopotential result
\begin{equation}
\epsilon'_{\alpha}(t) = \epsilon_{\alpha} 
   + \frac{8\pi a \hbar^2}{m} 
        \int d{\bf x}~
             \chi^*_{\alpha}({\bf x}) n({\bf x},t) \chi_{\alpha}({\bf x})~,
\end{equation}
with $n({\bf x},t)$ the density profile of the gas cloud. For the states near 
the one-particle ground state there is, however, a correction given by
\begin{eqnarray}
-2 \sum_{\beta} \sum_{\alpha'',\beta''} 
    |V^{(+)}_{\alpha,\beta;\alpha'',\beta''}(t)|^2
          N_{\alpha''}(t) N_{\beta''}(t)
           \frac{ {\cal P} } 
                { \epsilon'_{\alpha}(t) + \epsilon'_{\beta}(t) 
                 - \epsilon'_{\alpha''}(t) - \epsilon'_{\beta''}(t) } \nonumber
\end{eqnarray}
and 
$V^{(+)}_{\alpha,\beta;\alpha'',\beta''}(t) = 
    V^{(+)}_{\alpha,\beta;\alpha'',\beta''}
                      (\epsilon'_{\alpha''} + \epsilon'_{\beta''} - 2\mu(t))$. 
In the weak-coupling limit, the renormalization of the one-particle energies 
plays by definition not such an important role and we are much more interested 
in the imaginary part of the retarded and advanced selfenergies. They are given 
by
\begin{eqnarray}
R_{\alpha}(t) = 2\pi \sum_{\beta} \sum_{\alpha'',\beta''}
  \delta(\epsilon'_{\alpha''}(t) + \epsilon'_{\beta''}(t) 
         - \epsilon'_{\alpha}(t) - \epsilon'_{\beta}(t)) 
                                               \hspace*{2.0in} \nonumber \\
  \times |V^{(+)}_{\alpha,\beta;\alpha'',\beta''}(t)|^2
            [(1 + N_{\alpha''}(t) + N_{\beta''}(t)) N_{\beta}(t) 
                                   - N_{\alpha''}(t) N_{\beta''}(t)]~.  
\end{eqnarray}
Furthermore, the Keldysh component of the selfenergies equals
\begin{eqnarray}
\hbar\Sigma^K_{\alpha} = - 4\pi \sum_{\beta} \sum_{\alpha'',\beta''}
  \delta(\epsilon'_{\alpha''}(t) + \epsilon'_{\beta''}(t) 
         - \epsilon'_{\alpha}(t) - \epsilon'_{\beta}(t)) 
                                               \hspace*{2.0in} \nonumber \\
  \times |V^{(+)}_{\alpha,\beta;\alpha'',\beta''}(t)|^2
            [(1 + N_{\alpha''}(t) + N_{\beta''}(t)) N_{\beta}(t) 
                + N_{\alpha''}(t) N_{\beta''}(t)(1 + 2N_{\beta}(t))]~.  
\end{eqnarray}
As a result the quantum Boltzmann equation for the gas becomes
\begin{equation}
\label{BEinhom}
\frac{\partial}{\partial t} N_{\alpha}(t) =
  - \Gamma^{out}_{\alpha}(t) N_{\alpha}(t) 
     + \Gamma^{in}_{\alpha}(t) (1 + N_{\alpha}(t))~,
\end{equation}
with the rate to scatter out of the state $\chi_{\alpha}({\bf x})$ given by 
\begin{eqnarray}
\Gamma^{out}_{\alpha}(t) =  \frac{4\pi}{\hbar} 
 \sum_{\beta} \sum_{\alpha'',\beta''}
  \delta(\epsilon'_{\alpha''}(t) + \epsilon'_{\beta''}(t) 
         - \epsilon'_{\alpha}(t) - \epsilon'_{\beta}(t)) 
                                               \hspace*{2.0in} \nonumber \\
  \times |V^{(+)}_{\alpha,\beta;\alpha'',\beta''}(t)|^2
            (1 + N_{\alpha''}(t))(1 + N_{\beta''}(t)) N_{\beta}(t)  
\end{eqnarray}
and the rate to scatter into this state by
\begin{eqnarray}
\Gamma^{in}_{\alpha}(t) =  \frac{4\pi}{\hbar} 
 \sum_{\beta} \sum_{\alpha'',\beta''}
  \delta(\epsilon'_{\alpha''}(t) + \epsilon'_{\beta''}(t) 
         - \epsilon'_{\alpha}(t) - \epsilon'_{\beta}(t)) 
                                               \hspace*{2.0in} \nonumber \\
  \times |V^{(+)}_{\alpha,\beta;\alpha'',\beta''}(t)|^2
            N_{\alpha''}(t) N_{\beta''}(t) (1 + N_{\beta}(t))~,  
\end{eqnarray}
as expected from Fermi's Golden Rule.

The quantum Boltzmann equation found above, in combination with the expressions 
for the shift in the energy levels, fully describes the dynamics of the gas in 
the normal state. However, close to the critical temperature when the 
occupation numbers of the one-particle groundstate $\chi_g({\bf x})$ start to 
become large, we are no longer allowed to neglect the nonlinear terms in the 
Fokker-Planck equation in Eq.~(\ref{FPinhom}). Applying again the Hartree-Fock 
approximation, which is essentially exact in the weak-coupling limit as we have 
seen in the homogeneous case, and substituting the ansatz 
$P[\phi^*,\phi;t] = P_0[\phi^*_g,\phi_g;t] P_1[\phi'^*,\phi';t]$ into 
Eq.~(\ref{FPinhom}), we obtain for the probability distribution of the 
condensate
\begin{eqnarray}
\label{FPCinhom}
i\hbar \frac{\partial}{\partial t} P_0[\phi^*_g,\phi_g;t] &=&
   - \frac{\partial}{\partial \phi_g} 
       \left( \epsilon_g + \hbar\Sigma^{(+)}_g - \mu(t) 
         + T^{(+)}_{g,g;g,g}(t)
              |\phi_g|^2 \right) \phi_g 
                                P_0[\phi^*_g,\phi_g;t] \nonumber \\ 
 &+& \frac{\partial}{\partial \phi^*_g} 
      \left( \epsilon_g + \hbar\Sigma^{(-)}_g - \mu(t) 
        + T^{(+)}_{g,g;g,g}(t)
             |\phi_g|^2 \right) \phi^*_g 
                                P_0[\phi^*_g,\phi_g;t] \nonumber \\
 &-& \frac{1}{2} 
     \frac{\partial^2}{\partial \phi^*_g \partial \phi_g} 
       \hbar \Sigma^K_g P_0[\phi^*_g,\phi_g;t]~,                                                                                                       
\end{eqnarray}     
where it should be remembered that for consistency reasons the selfenergies 
need to be evaluated at the energies 
$\epsilon'_g(t) + T^{(+)}_{g,g;g,g}(t) |\phi_g|^2$. In equilibrium the 
fluctuation-dissipation theorem then reads
\begin{equation}
iR_{\alpha} = - \frac{\beta}{4} \hbar \Sigma^K_g 
   \left( \epsilon'_g - \mu + T^{(+)}_{g,g;g,g} |\phi_g|^2 \right)~,
\end{equation}
which guarantees that the probability distribution for the condensate relaxes 
to the correct equilibrium distribution
\begin{eqnarray}
P_0[\phi^*_g,\phi_g;\infty] \propto \exp \left\{ -\beta
   \left( \epsilon'_g - \mu + \frac{T^{(+)}_{g,g;g,g}}{2} |\phi_g|^2 \right)
                            |\phi_g|^2 \right\}~.          \nonumber
\end{eqnarray}
The Fokker-Planck equation, and hence the Boltzmann equation, for the 
noncondensed part of the gas is the same as in the normal state because for a 
discrete system the contributions from the condenstate to the selfenergies is 
already included in our expressions for $\hbar \Sigma^{(\pm),K}_{\alpha}$ and 
needs not be separated out explicitly. This implies of course, that if we 
evaluate the rates $\Gamma^{out}_{\alpha}(t)$ and $\Gamma^{in}_{\alpha}(t)$ by 
replacing the sum over states by an integral over a continuum, we first need to 
separate the contribution from the condensate to obtain the correct result. 

In summary we have thus found that in the weak-coupling limit the dynamics of 
Bose-Einstein condensation in an external trapping potential is described by 
the coupled equations in Eq.~(\ref{BEinhom}) and Eq.~(\ref{FPCinhom}), i.e., by 
a Boltzmann equation for the noncondensed part of the gas and a Fokker-Planck 
equation for the condensate, respectively. If we apply these equations to a 
Bose gas with a positive scattering length, we essentially arrive at the theory 
recently put forward by Gardiner and coworkers \cite{peter1}. 
Indeed, neglecting the energy 
shifts and assuming that the noncondensed cloud is in equilibrium with a 
chemical potential $\mu > \epsilon_g$, our Fokker-Planck equation for a $U(1)$ 
invariant probability distribution of the condensate becomes
\begin{eqnarray}
i\hbar \frac{\partial}{\partial t} P_0[|\phi_g|;t] =
   - \frac{\beta}{4} \hbar \Sigma^K_g \frac{\partial}{\partial \phi_g} 
       \left( \epsilon_g - \mu 
         + T^{(+)}_{g,g;g,g}
              |\phi_g|^2 \right) \phi_g P_0[|\phi_g|;t]
                                               \hspace*{1.3in} \nonumber \\ 
 - \frac{\beta}{4} \hbar \Sigma^K_g \frac{\partial}{\partial \phi^*_g} 
      \left( \epsilon_g - \mu 
        + T^{(+)}_{g,g;g,g}
             |\phi_g|^2 \right) \phi^*_g 
                                P_0[|\phi_g|;t] 
 - \frac{1}{2} \hbar \Sigma^K_g
     \frac{\partial^2}{\partial \phi^*_g \partial \phi_g} 
        P_0[|\phi_g|;t]~.                                                                                                       
\end{eqnarray}     
It is again equivalent to the Fokker-Planck equation for a single-mode laser 
and thus incorporates in a slightly different language exactly the same physics 
as the theory of Gardiner {\it et al.} A comparison with the experiments of 
H.-J. Miesner {\it et al.} shows that the theory of Gardiner and coworkers is 
unfortunately not fully satisfactory \cite{wolfgang2}. The reason for this is 
presumably that these experiments are really in the strong-coupling limit, 
where the above theory does not apply. As mentioned previously, the theory is 
however applicable to gases with a negative scattering length. Since we know 
from our experience with homogeneous gases that the chemical potential will 
never be larger than $\epsilon'_g$ in this case, we clearly see from the 
equilibrium probability distribution $P_0[\phi^*_g,\phi_g;\infty]$ that a 
condensate in a gas with effectively attractive interactions is only metastable 
and will ultimately collapse to a dense state after it has overcome an energy 
barrier, either by thermal fluctuations or by quantum mechanical tunneling. In 
the next section we study the dynamics of this collapse by an extension of the 
method that has been devised in Ref.~\cite{henk5} and that is more accurate 
than the weak-coupling theory discussed sofar. A brief review of this approach 
also turns out to be a useful intermediate step towards the discussion of the 
strong-coupling limit presented in Sec.~\ref{sc}.
  
\subsection{Negative Scattering Length}
\label{neg}
For definiteness we now consider an external potential in the form of an 
isotropic harmonic oscillator. Thus $V^{ex}({\bf x}) = m\omega^2 {\bf x}^2/2$ 
and the weak-coupling limit then requires that 
$8\pi\hbar^2|a|n({\bf 0},t)/m \ll \hbar\omega$. During the growth of the 
condensate this condition is essentially always fulfilled but during the 
collapse, that occurs after the number of condensate particles has become too 
large, the density in the center of the trap quickly increases and we leave 
the weak-coupling regime. To study also the dynamics of the collapse requires 
therefore a more accurate theory. However, an important simplification occurs 
because the typical time scale for the collapse turns out to be of 
${\cal O}(1/\omega)$ which is much shorter than the time between collisions 
$\tau_{el}$. Indeed, their ratio is of order
\begin{eqnarray}
\frac{n({\bf 0},t)T^{(+)}({\bf 0},{\bf 0};0)}{\hbar \omega} 
      \left( \frac{a}{\Lambda} \right) \ll 1                     \nonumber
\end{eqnarray} 
and the dynamics of the condensate is thus collisionless. From our general 
Fokker-Planck equation in Eq.~(\ref{FPinhom}) we then see that the condensate 
wavefunction obeys
\begin{eqnarray}
i\hbar \frac{\partial}{\partial t} \langle \phi_{\alpha} \rangle(t) =
                                          \hspace*{4.9in} \nonumber \\
   \sum_{\alpha'} \left\{ 
       \left( (\epsilon_{\alpha} - \mu) \delta_{\alpha,\alpha'} 
            + {\rm Re}[\hbar\Sigma^{(+)}_{\alpha,\alpha'}] \right) 
                        \langle \phi_{\alpha'} \rangle(t)
            + \sum_{\beta,\beta'}
               V^{(+)}_{\alpha,\beta;\alpha',\beta'}
                   \langle \phi^*_{\beta} \rangle(t)
                   \langle \phi_{\beta'} \rangle(t) 
                   \langle \phi_{\alpha'} \rangle(t) \right\}~. 
\end{eqnarray}
In coordinate space this corresponds, in a reasonable first approximation, to 
the expected nonlinear Schr\"odinger equation at nonzero temperatures 
\begin{eqnarray}
i\hbar \frac{\partial}{\partial t} \langle \phi({\bf x}) \rangle(t) =
                                          \hspace*{4.5in} \nonumber \\
   \left\{ - \frac{\hbar^2 \nabla^2}{2m} + V^{ex}({\bf x}) - \mu 
            + \frac{8\pi a\hbar^2}{m} n'({\bf x},t)  
            + \frac{4\pi a\hbar^2}{m}
                   |\langle \phi({\bf x}) \rangle(t)|^2 \right\}
                         \langle \phi({\bf x}) \rangle(t)~, 
\end{eqnarray}
that was first studied by Goldman, Silvera, and Leggett \cite{tony2} and  
Huse and Siggia \cite{HS} for a gas with positive scattering length, because 
these authors were at that time interested in spin-polarized atomic hydrogen. 
It was recently applied to atomic $^7$Li by Houbiers and Stoof \cite{marianne1} 
and Bergeman \cite{tom}. Notice also that we have evaluated the sum over states 
by an integral over a continuum to obtain the noncondensate density 
$n'({\bf x},t)$ in the right-hand side of this equation.

Neglecting the variation of the noncondensate density on the size of the 
condensate \cite{marianne1}, we conclude that the dynamics of the collapse is, 
apart from an unimportant shift in the chemical potential, determined by the 
Gross-Pitaevskii equation \cite{GP}
\begin{equation}
\label{GPeq}
i\hbar \frac{\partial}{\partial t} \langle \phi({\bf x}) \rangle(t) =
   \left\{ - \frac{\hbar^2 \nabla^2}{2m} + V^{ex}({\bf x}) - \mu   
            + \frac{4\pi a\hbar^2}{m}
                   |\langle \phi({\bf x}) \rangle(t)|^2 \right\}
                         \langle \phi({\bf x}) \rangle(t)~.
\end{equation}         
More precisely, this determines only the semiclassical dynamics. If we also 
want to study the quantum fluctuations, which is necessary if we are also 
interested in how the condensate tunnels through the macroscopic energy 
barrier, it is most convenient to calculate the grand canonical partition 
function of the condensate \cite{tunnel}. Quantizing the Gross-Pitaevskii 
equation we obtain for this partition function the functional integral
\begin{equation}
Z_{gr}(\mu) = \int d[\psi^*]d[\psi]~ 
                    \exp \left\{ - \frac{1}{\hbar} S_E[\psi^*,\psi] \right\}~,
\end{equation}
over the complex field $\psi({\bf x},\tau)$ and with the Euclidian action
\begin{eqnarray}
S_E[\psi^*,\psi] =                              \hspace*{4.9in} \nonumber \\
   \int_0^{\hbar\beta} d\tau \int d{\bf x}~ 
     \psi^*({\bf x},\tau) \left( \hbar \frac{\partial}{\partial \tau}
            + \frac{\hbar^2 \nabla^2}{2m} + V^{ex}({\bf x}) - \mu   
            + \frac{2\pi a\hbar^2}{m}
                   |\psi({\bf x},\tau)|^2 \right) \psi({\bf x},\tau)~.
\end{eqnarray}
As always for Bose systems, the integration is only over fields that are 
periodic on the imaginary time axis.

Although it has recently been shown by Freire and Arovas that the tunneling 
process can also be studied in terms of the complex field $\psi({\bf x},\tau)$ 
\cite{arovas}, we believe that it leads to somewhat more physical insight if we 
use instead the fields $\rho({\bf x},\tau)$ and $\theta({\bf x},\tau)$ that 
correspond to the density and phase fluctuations of the condensate, 
respectively. They are introduced by performing the canonical variable 
transformation \cite{popov}
\begin{eqnarray}
\psi({\bf x},\tau) = \sqrt{\rho({\bf x},\tau)} e^{i \theta({\bf x},\tau)}
                                                             \nonumber 
\end{eqnarray}
in the functional integral for the partition function. As a result we find
\begin{equation}
\label{gr}
Z_{gr}(\mu) = \int d[\rho]d[\theta]~ 
              \exp \left\{ - \frac{1}{\hbar} S_E[\rho,\theta;\mu] \right\}~,
\end{equation} 
with
\begin{eqnarray}
S_E[\rho,\theta;\mu] = 
   \int_0^{\hbar\beta} d\tau \int d{\bf x}~ 
   \left( i\hbar \rho({\bf x},\tau) 
                 \frac{\partial \theta({\bf x},\tau)}{\partial \tau}
            + \frac{\hbar^2 \rho({\bf x},\tau)}{2m} 
                        (\nabla \theta({\bf x},\tau))^2 \right.
                                        \hspace*{1.1in} \nonumber \\
     \left. + \frac{\hbar^2}{8m\rho({\bf x},\tau)} 
                        (\nabla \rho({\bf x},\tau))^2
            + V^{ex}({\bf x})\rho({\bf x},\tau) - \mu\rho({\bf x},\tau)  
            + \frac{2\pi a\hbar^2}{m} \rho^2({\bf x},\tau) \right)~.
\end{eqnarray}
Next, we notice that this action is only quadratic in the phase fluctuations. 
The field $\theta({\bf x},\tau)$ can therefore be integrated over exactly, 
because it only involves the evaluation of a gaussian integral. 

Compared to ordinary gaussian integrals there is, however, one slight 
complication which is associated with the fact that $\theta({\bf x},\tau)$ are 
phase variables. This implies that the periodicity of the original field 
$\psi({\bf x},\tau)$ only constraints the phase field $\theta({\bf x},\tau)$ to 
be periodic up to a multiple of $2\pi$. To evaluate the grand canonical 
partition function in Eq.~(\ref{gr}) we must therefore first integrate over all 
fields $\theta({\bf x},\tau)$ that obey the boundary condition 
$\theta({\bf x},\hbar\beta) = \theta({\bf x},0) + 2\pi j$ and subsequently sum 
over all possible integers $j$. Because these different boundary conditions 
only affect the zero-momentum part of $\theta({\bf x},\tau)$ we first have to 
evaluate the sum
\begin{eqnarray}
\sum_{j} \int^{\theta_0(\hbar\beta)=\theta_0(0)+2\pi j}   
 d[\theta_0]~ 
   \exp \left\{ -i \int_0^{\hbar\beta} d\tau~ N_0(\tau)
                        \frac{\partial \theta_0(\tau)}{\partial \tau}
        \right\}~,                                             \nonumber
\end{eqnarray}
with $N_0(\tau) = \int d{\bf x}~ \rho({\bf x},\tau)$ the number of condensate 
particles. After performing a partial integration on the integral in the 
exponent, we can carry out the path integration over $\theta_0(\tau)$ to obtain
\begin{eqnarray}
\sum_{j} e^{2\pi i N_0 j} \delta \left[ 
               \frac{\partial N_0(\tau)}{\partial \tau} \right]~. 
                                         \nonumber
\end{eqnarray}
As expected, the integration over the global phase of the condensate leads to 
the constraint of a constant number of condensate particles, i.e., 
$N_0(\tau)=N_0$. Moreover, we have $\sum_{j} e^{2\pi i N_0 j} = \sum_{j} 
\delta(N_0 - j)$, which restricts the number of condensate particles to an 
integer. Putting all these results together, we see that the integration over 
the zero-momentum part of $\rho({\bf x},\tau)$ is only a sum over the number of 
condensate particles and we have that
\begin{equation}
Z_{gr}(\mu) = \sum_{N_0} e^{\beta\mu N_0} Z_{N_0}~.
\end{equation}
Here we introduced the canonical partition function of the condensate, which is 
apparently equal to the functional integral  
\begin{equation}
Z_{N_0} = \int d[\rho]d[\theta]~ 
                 \exp \left\{ - \frac{1}{\hbar} S_E[\rho,\theta;0] \right\}
\end{equation} 
over all the nonzero momentum components of the density and phase fields.

The integration over the nonzero momentum components of the phase field 
$\theta({\bf x},\tau)$ is easily performed, because it now involves an ordinary 
gaussian integral. Introducing the Green's function for the phase fluctuations 
$G({\bf x},{\bf x}';\rho)$ by
\begin{equation}
\label{geq}
\frac{\hbar}{m} \left( (\nabla \rho) \cdot \nabla
      + \rho \nabla^2 \right) G({\bf x},{\bf x}';\rho) =
                                 \delta({\bf x}-{\bf x}')~,
\end{equation}
we immediately obtain the desired effective action for the density field
\begin{eqnarray}
S_E[\rho] = \int_0^{\hbar\beta} d\tau \int d{\bf x} \int d{\bf x}'~
    \left(-\frac{\hbar}{2} \frac{\partial \rho({\bf x},\tau)}
                                 {\partial \tau}
             G({\bf x},{\bf x}';\rho)
                            \frac{\partial \rho({\bf x}',\tau)}
                                 {\partial \tau}
    \right)                       \hspace*{1.4in}  \nonumber \\
        + \int_0^{\hbar\beta} d\tau \int d{\bf x}~
             \left(
               \frac{\hbar^2}{8m\rho({\bf x},\tau)} 
                                 (\nabla \rho({\bf x},\tau))^2
               + V^{ex}({\bf x}) \rho({\bf x},\tau)
               + \frac{2\pi a\hbar^2}{m} \rho^2({\bf x},\tau)
             \right)~.
\end{eqnarray}
Being an action for the density fluctuations of the condensate, 
$S_E[\rho]$ also describes all the collisionless modes of the condensate. This 
is important for our purposes, because the mode which becomes unstable first, 
determines precisely how the condensate collapses. Moreover, it determines the 
probability with which the collapse is going to take place, both for quantum and 
thermal fluctuations, since the energy barrier is smallest in that direction of 
the configuration space. It should be noted that as long as we can neglect the 
interaction between the condensate and the thermal cloud, the action $S_E[\rho]$ 
describes also the collective modes of a gas with positive scattering length. 
For various other theoretical approaches that have been applied under these 
conditions see, for example, 
Refs.~\cite{sandro,singh,keith2,yvan2,peter3,keith3,juha,li,stig}. The actual 
measurements of the collective mode frequencies have been performed by 
Jin {\it et al.} \cite{coll1} and Mewes {\it et al.} \cite{coll2} and are at 
sufficiently low temperatures indeed in good agreement with the theoretical 
predictions \cite{eugene1,keith4}. We expect the same to be true for a gas with 
effectively attractive interactions and, therefore, the action $S_E[\rho]$ to be 
a good starting point for the following discussion. 

To obtain the collisionless modes explicitly we consider first the case of an 
ideal Bose gas by putting $a=0$. For the ideal Bose gas we expect the gaussian 
profile 
\begin{equation}
\label{gprof}
\rho({\bf x};q(\tau)) = N_0 \left( \frac{1}{\pi q^2(\tau)} \right)^{3/2}
                       \exp \left( - \frac{{\bf x}^2}{q^2(\tau)} \right)
\end{equation}
to describe an exact mode of the condensate. The reason is that in the 
noninteracting case we can make a density fluctuation by taking one particle 
from the condensate and putting that in one of the excited states of the 
external potential. The corresponding density fluctuation obeys
\begin{eqnarray}
\delta\rho({\bf x},t) \propto e^{-i(\epsilon_{\alpha}-\epsilon_g)t/\hbar}
    \chi^*_{\alpha}({\bf x}) \chi_g({\bf x})~.                      \nonumber
\end{eqnarray} 
For our isotropic harmonic oscillator $\alpha$ combines the two angular 
momentum quantum numbers $\ell$ and $m$ with the quantum number $n$ that counts 
the number of nodes in the radial wavefunction $\chi_{n\ell}(x)$. The density 
fluctuation then becomes
\begin{eqnarray}
\delta\rho({\bf x},t) \propto e^{-i(2n + \ell)\omega t}
    \chi_{n\ell}(x) Y_{\ell m}^*({\bf \hat{x}})
                          \frac{e^{-x^2/2l^2}}{(\pi l^2)^{3/4}}~,    \nonumber
\end{eqnarray} 
with $\epsilon_{n\ell m} - \epsilon_{000} = (2n + \ell) \hbar \omega$ the 
excitation energy and $l=(\hbar/m\omega)^{1/2}$ the size of the condensate 
wavefunction. Comparing this now with the expansion of the gaussian profile in 
Eq.~(\ref{gprof}) around the groundstate density profile, which is obtained by 
substituting $q(\tau)=l + \delta q(\tau)$, we find that 
\begin{equation}
\delta\rho({\bf x},\tau) = 
    - \sqrt{6} N_0 \frac{\delta q(\tau)}{l}
        \chi_{10}(x) Y_{00}^*({\bf \hat{x}})
                             \frac{e^{-x^2/2l^2}}{(\pi l^2)^{3/4}}
\end{equation}
has precisely the same form as a density fluctuation in which one particle is 
taken from the condensate and put into the harmonic oscillator state with 
quantum numbers $(n\ell m) = (100)$. The frequency of the so-called `breathing' 
mode described by the gaussian density profile must therefore be equal to 
$2\omega$.

To proof that this is indeed correct, we need to evaluate the effective action 
$S_E[\rho]$, and hence the Green's function $G({\bf x},{\bf x}';\rho)$, for a 
gaussian density profile. Substituting such a profile in Eq.~(\ref{geq})  
immediately leads to 
$G({\bf x},{\bf x}';\rho) = G({\bf x},{\bf x}';q)/\rho({\bf x}';q)$, with
\begin{equation}
\frac{\hbar}{m} \left( - \frac{2}{q^2} {\bf x} \cdot \nabla
      + \nabla^2 \right) G({\bf x},{\bf x}';q) =
                                 \delta({\bf x}-{\bf x}')~.
\end{equation} 
The latter equation can be solved if we can solve the eigenvalue problem
\begin{equation}
\left( \nabla^2 - \frac{2x}{q^2} \frac{\partial}{\partial x} \right) 
    \xi({\bf x}) = \lambda \xi({\bf x})~.
\end{equation}
This turns out to be an easy task, because substituting 
\begin{equation}
\xi_{n\ell m}({\bf x}) 
    = \xi_{n\ell}(x) \frac{e^{x^2/2q^2}}{x} Y_{\ell m}({\bf \hat{x}})
\end{equation}
gives essentially the radial Schr\"odinger equation for an isotropic harmonic 
oscillator with frequency $\omega_q = \hbar/mq^2$, i.e.,
\begin{equation}
- \frac{2m}{\hbar^2} 
  \left( - \frac{\hbar^2}{2m} \frac{\partial^2}{\partial x^2} 
         + \frac{1}{2} m\omega_q^2 x^2 + \frac{\hbar^2 \ell(\ell+1)}{2mx^2}
         - \frac{3}{2} \hbar\omega_q \right) \xi_{n\ell}(x) 
                                   = \lambda_{n\ell} \xi_{n\ell}(x)~.
\end{equation}
The desired eigenfunctions are therefore 
$\xi_{n\ell m}({\bf x};q) = \varphi_{n\ell m}({\bf x}) e^{x^2/2q^2}$, with 
$\varphi_{n\ell m}({\bf x})$ the usual (properly normalized) harmonic 
oscillator states with the energies $(2n + \ell + 3/2)\hbar\omega_q$, and the 
corresponding eigenvalues are $\lambda_{n\ell}(q) = - 2 (2n + \ell)/q^2$. 
Introducing finally the `dual' eigenfunctions 
$\bar{\xi}_{n\ell m}({\bf x};q)
                          \equiv \varphi^*_{n\ell m}({\bf x}) e^{-x^2/2q^2}$,
the Green's function $G({\bf x},{\bf x}';q)$ is given by
\begin{equation}
\label{gq}
G({\bf x},{\bf x}';q) 
  = {\sum_{n\ell m}}' \xi_{n\ell m}({\bf x};q) 
       \frac{m}{\hbar \lambda_{n\ell}(q)} \bar{\xi}_{n\ell m}({\bf x}';q)~.
\end{equation}
Note that prime on the summation sign indicates that the sum is over all 
quantum numbers except $(n\ell m) = (000)$. The latter is excluded because the 
associated eigenfunction $\xi_{000}({\bf x};q)$ is just a constant and thus 
does not contribute to $G({\bf x},{\bf x}';\rho)$, which is defined as the 
Green's function for all phase fluctuations with nonvanishing momenta.

Putting all these results together, we see that the dynamics of the collective 
variable $q(\tau)$ is determined by the action
\begin{eqnarray}
S_E[q] = \int_0^{\hbar\beta} d\tau~
   \left\{ \frac{3mN_0}{4} \left( \frac{dq}{d\tau} \right)^2 
          + N_0 \left( \frac{3\hbar^2}{4mq^2} + \frac{3}{4}m\omega^2q^2
            \right) \right\}                   \hspace*{1.2in} \nonumber \\
   \equiv \int_0^{\hbar\beta} d\tau~
   \left\{ \frac{1}{2} m^* \left( \frac{dq}{d\tau} \right)^2 + V(q) \right\}~,
\end{eqnarray}
that is equivalent to the action of a particle with effective mass 
$m^* = 3mN_0/2$ in a potential 
$V(q) = N_0(3\hbar^2/4mq^2 + 3m\omega^2q^2/4)$. As expected from our previous 
remarks, this potential has a minimum for $q=l$ and can be expanded near its 
minimum as
\begin{equation}
V(q) \simeq \frac{3}{2}N_0\hbar\omega 
               + \frac{1}{2} m^* (2\omega)^2 (\delta q)^2~.
\end{equation}
It thus comfirms that the gaussian profile describes a breathing mode with 
frequency $2\omega$ around an equilibrium density profile that is given by 
$\rho({\bf x};l) = N_0 |\chi_{000}({\bf x})|^2$. 

Up to now we have only considered the breathing mode of the condensate. 
However, the explicit expression for the Green's function 
$G({\bf x},{\bf x}';q)$ in Eq.~(\ref{gq}), together with our argument for the 
form of the density fluctuations in the various modes, presents us with a clue 
on how to obtain all the collisionless modes. It suggests that we should expand 
the density profile of the condensate as
\begin{equation}
\label{exp}
\rho({\bf x},\tau) = \rho({\bf x};q(\tau)) 
  + {\sum_{n\ell m}}'' C_{n\ell m}(\tau) 
      \frac{\bar{\xi}_{n\ell m}({\bf x};q(\tau))}{(\pi q^2(\tau))^{3/4}}~,
\end{equation}
where the sum is now over all quantum numbers except $(n\ell m) = (000)$ and 
$(n\ell m) = (100)$. Substituting this in the effective action $S_E[\rho]$ and 
expanding up to second order in the complex coefficients $C_{n\ell m}$, we find 
after somewhat tedious but straightforward algebra \cite{richard} that
\begin{equation}
\label{collac}
S_E[\rho] \simeq S_E[q] 
   + {\sum_{n\ell m}}'' \int_0^{\hbar\beta} d\tau~
        \frac{1}{4N_0} 
          \left( \frac{m q^2}{2n + \ell} 
                              \left| \frac{d C_{n\ell m}}{d\tau} \right|^2
                 + \frac{\hbar^2}{m q^2} (2n + \ell) |C_{n\ell m}|^2
          \right)~.
\end{equation}
Neglecting then the interaction between the breathing mode and all the other 
modes of the condensate, we conclude that the coefficients $C_{n\ell m}$ are 
harmonic oscillators with the frequencies 
\begin{equation}
\omega_{n\ell m} = (2n + \ell) \frac{\hbar}{ml^2} = (2n + \ell) \omega~.
\end{equation}
Hence, we have indeed succeeded in correctly describing also all the other 
collective modes of the condensate. At this point, we might wonder why we have 
chosen in Eq.~(\ref{exp}) to still treat the breathing mode with the collective 
variable $q(\tau)$ and not with an additional coefficient $C_{100}(\tau)$ and 
the expansion
\begin{equation}
\label{expa}
\rho({\bf x},\tau) = \rho({\bf x};l) 
  + {\sum_{n\ell m}}' C_{n\ell m}(\tau) 
       \frac{\bar{\xi}_{n\ell m}({\bf x};l)}{(\pi l^2)^{3/4}}.
\end{equation}
Doing that is clearly also a valid procedure and results at the quadratic level 
in
\begin{equation}
S_E[\rho] \simeq \hbar\beta \left( \frac{3}{2} N_0 \hbar\omega \right) 
  + {\sum_{n\ell m}}' \int_0^{\hbar\beta} d\tau~
        \frac{1}{4N_0} 
          \left( \frac{m l^2}{2n + \ell} 
                              \left| \frac{d C_{n\ell m}}{d\tau} \right|^2
                 + \frac{\hbar^2}{m l^2} (2n + \ell) |C_{n\ell m}|^2
          \right)~.
\end{equation}
The reason for using the former, more general, choice is related to the fact 
that when we include the attractive interaction, the breathing mode turns out to 
be the unstable mode of the condensate. It therefore plays a special role in the 
dynamics of the collapse, as we will see now.

Considering first again only gaussian density profiles, the action $S_E[q]$ is 
again that of a particle with effective mass $m^* = 3mN_0/2$ but now in the 
potential \cite{BP}
\begin{equation}
V(q) =  N_0 \left( \frac{3\hbar^2}{4mq^2} + \frac{3}{4}m\omega^2q^2
              - \frac{N_0}{\sqrt{2\pi}} \frac{\hbar^2 |a|}{m q^3} \right)~.
\end{equation} 
The physically most important feature of this potential is that it is unbounded 
from below, since $V(q) \rightarrow -\infty$ if $q \downarrow 0$. Hence, the 
condensate always has the tendency to collapse to the high-density state 
$\lim_{q \downarrow 0} \rho({\bf x};q) = N_0 \delta({\bf x})$. However, if the 
number of condensate particles is sufficiently small, or more precisely if 
\cite{fetter2}
\begin{equation}
N_0 < \frac{2\sqrt{2\pi}}{5^{5/4}} \frac{l}{|a|} \simeq 0.68 \frac{l}{|a|}~,
\end{equation}
the condensate has to overcome a macroscopic energy barrier before it can 
collapse. Under these conditions the condensate is therefore really metastable 
and can in principle be observed experimentally. The most important question in 
this respect is of course: How metastable is the condensate? Within the 
gaussian approximation this question is easily answered, because then the 
dynamics of the condensate is equivalent to the dynamics of a particle in an 
unstable potential, as we have just seen. We therefore only need to evaluate 
the WKB-expression for the tunnneling rate \cite{shuryak} and compare this to 
the rate of decay due to thermal fluctuations by calculating also the height of 
the energy barrier. The outcome of this comparison for the conditions of the 
experiment with atomic $^7$Li is presented in Ref.~\cite{cass2} and shows that 
for the relatively high temperatures $T \gg \hbar\omega/k_B$ that have been 
obtained thusfar \cite{Rice} the decay by means of thermal fluctuations over 
the energy barrier is the dominant decay mechanism of the condensate.

\begin{figure}
\vspace*{4.5in}
\psfig{figure=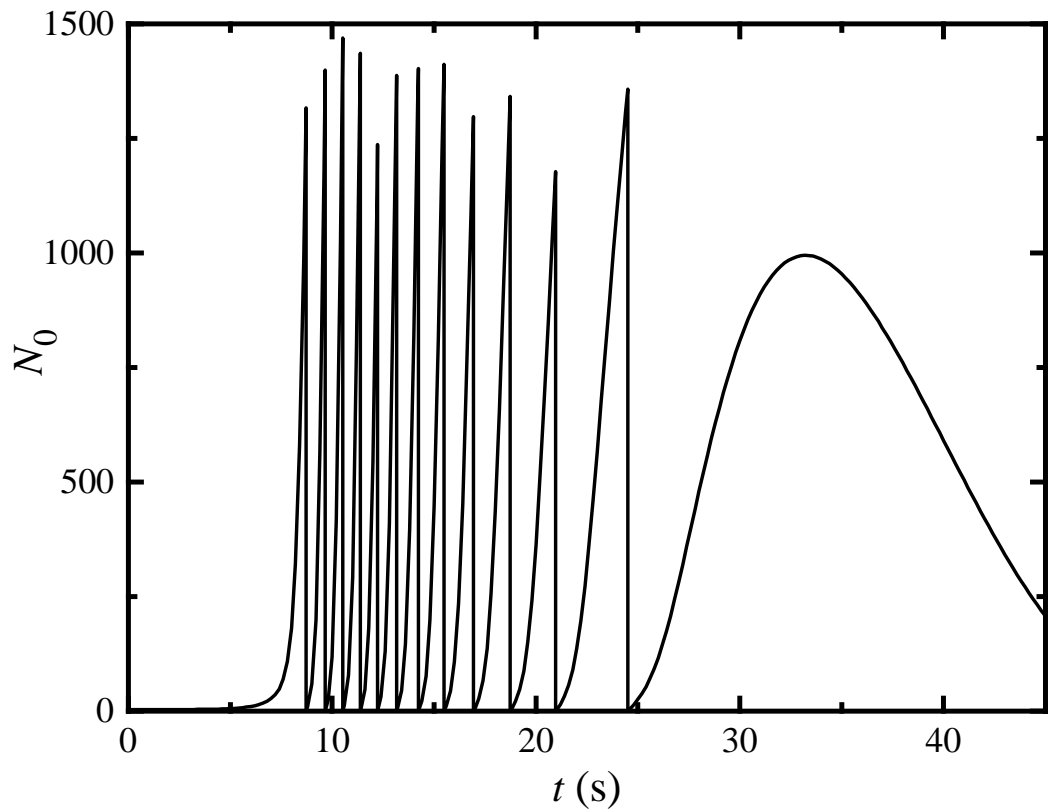}
\caption{Typical evolution of condensate number $N_0$ in response to
         evaporative cooling. The long-time decay of the condensate is due
         to two and three-body inelastic collisions.
         \label{N0}}
\end{figure}

More important for our purposes, however, is that sufficiently close to the 
maximum number of condensate particles $N_m$ the collective decay of the 
condensate discussed above is always much more probable than the decay due to 
two and three-body collisions that lead to a spin-flip or the formation of 
$^7$Li molecules, respectively. As a result the collapse of the condensate 
should be observable within the finite lifetime of the gas. In fact, on the 
basis of this separation of time scales we expect the condensate to go through 
a number of growth and collapse cycles \cite{cass2,cass3}. Theoretically this 
physical picture arises as follows. Starting from a gas with a number of atoms 
$N \gg N_m$, the initial growth of the condensate as a response to evaporative 
cooling is described by the Boltzmann equation in Eq.~(\ref{BEinhom}) and the 
Fokker-Planck equation for the condensate in Eq.~(\ref{FPCinhom}). However, if 
the number of condensate atoms starts to come close to $N_m$, the condensate 
fluctuates over the energy barrier and collapses in a very short time of 
${\cal O}(1/\omega)$ \cite{lev}. During the collapse the condensate density 
increases rapidly and two and three-body inelastic processes quickly remove 
almost all the atoms from the condensate. After this has occurred the 
condensate grows again from the noncondensed part of the gas and a new growth 
and collapse cycle begins. It is only after many of these cycles that enough 
atoms are removed for the gas to relax to an equilibrium with a number of 
condensate particles that is less than $N_m$. This is shows quantitatively in 
Fig.~\ref{N0} for the experimental conditions of interest.

A final issue which needs to be addressed at this point is the actual dynamics 
of the collapse and in particular how we must include the effect of the 
inelastic decay processes on this dynamics. From our Caldeira-Leggett toy model 
we know that the inclusion of inelastic processes in first instance means that 
we must replace the collisionless Gross-Pitaevskii equation for the condensate 
by a dissipative nonlinear Schr\"odinger equation with noise. Considering for 
simplicity only two-body relaxation since three-body recombination processes 
can be treated analogously, we now have to deal with the stochastic equation
\begin{equation}
\left\{ i\hbar \frac{\partial}{\partial t} 
    + \frac{\hbar^2 \nabla^2}{2m} - V^{ex}({\bf x}) + \mu   
            -\left( \frac{4\pi a\hbar^2}{m} - i\hbar G_0 \right)
                   |\phi({\bf x},t)|^2 \right\} \phi({\bf x},t)
    = \eta({\bf x},t)~,
\end{equation}  
where the strength of the noise $\eta({\bf x},t)$ is such that 
\begin{equation}
\frac{d N_0(t)}{dt} \equiv
  \frac{d}{dt} \int d{\bf x}~ n_0({\bf x},t) =
  \frac{d}{dt} \int d{\bf x}~ \langle |\phi({\bf x})|^2 \rangle(t) 
                       = -2G_0 \int d{\bf x}~ n_0^2({\bf x},t)~. 
\end{equation}
The latter equation shows that the condensate density $n_0({\bf x},t)$ obeys 
the expected rate equation
\begin{equation}
\frac{\partial}{\partial t} n_0({\bf x},t) =
              -2G_0 n_0^2({\bf x},t)
\end{equation}
irrespective of the fact that for atomic $^7$Li we have that 
$G_0 \ll 4\pi |a|\hbar/m$ and hence that the equation for 
$\langle \phi({\bf x}) \rangle(t)$ is in a good approximation still equal to 
the Gross-Pitaevskii equation in Eq.~(\ref{GPeq}). On the basis of these 
arguments, which are in fact quite similar to the ones used for the 
strong-coupling limit of our Caldeira-Leggett toy model, we then conclude that 
the dynamics of the collapse can still be studied by the effective action 
$S_E[\rho]$. We only need to add the constraint that
\begin{equation}
\frac{d N_0(t)}{dt} = -2G_0 \int d{\bf x}~ \rho^2({\bf x},t)~. 
\end{equation}  
In the gaussian approximation this simply becomes
\begin{equation}
\frac{d N_0(t)}{dt} = -2G_0 \frac{1}{(2\pi q^2(t))^{3/2}} N_0^2(t)~, 
\end{equation}  
which needs to be solved in combination with the dynamics of the condensate 
`width' $q(t)$ obeying Newton's law 
\begin{equation}
m^* \frac{d^2 q(t)}{dt^2} = - \frac{dV}{dq}(q(t))~.
\end{equation}
A numerical solution of these two coupled equations for $N_0(t)$ and $q(t)$ 
shows that, in contrast to the case when there is no decay \cite{lev}, the 
collapse is not complete and is finally arrested when the number of condensate 
particles becomes too small \cite{cass2}. In the gaussian approximation we find 
that the final number of condensate atoms is of ${\cal O}(1)$. Preliminary 
experimental results indicate that this number is presumably too small and 
about 1\% of the condensate atoms remain after a single collapse \cite{randy}.  
This difference between theory and experiment is presumably due to the fact 
that the gaussian approximation to the density profile of the condensate 
overestimates the densities in the final stage of the collapse when the decay 
of the condensate is most severe \cite{kagan3}.

Fortunately, our discussion of the ideal Bose gas shows how one can 
systematically improve on the gaussian approximation by including also the 
effect of the other collective variables $C_{n\ell m}(t)$ in the density 
expansion in Eq.~(\ref{exp}). Of course, if one is only interested in the 
dynamics of the collapse it is also possible to (numerically) solve the 
Gross-Pitaevskii equation together with 
\begin{equation}
\frac{dN_0(t)}{dt} = -2G_0 \int d{\bf x}~ 
                          |\langle \phi({\bf x}) \rangle(t)|^4~, 
\end{equation} 
However, to gain more physical insight into the solution of the 
Gross-Pitaevskii equation it is helpful to apply the collective coordinate 
method mentioned above. This is in particularly true because for a gas with 
effectively attractive interactions the breathing mode is the only unstable 
mode of the condensate \cite{singh,keith3}. Therefore, it appears to be an 
excellent approximation to the effective action $S_E[\rho]$ to expand up to 
quadratic terms in the coefficients $C_{n\ell m}(t)$, but to take all orders of 
$q(t)$ into account. Clearly such an approximation is not possible if we 
describe the breathing mode with an additional coefficient $C_{100}(t)$ as in 
Eq.~(\ref{expa}). 

To complete our discusion of the collisionless dynamics of a condensate with 
negative scattering length, we thus need to add to the noninteracting action in 
Eq.~(\ref{collac}) the various contributions from the interaction term
\begin{eqnarray}
\frac{2\pi a\hbar^2}{m} \int_0^{\hbar\beta} d\tau \int d{\bf x}~
                                             \rho^2({\bf x},\tau)~ \nonumber
\end{eqnarray}
in the action $S_E[\rho]$. In lowest order in the coefficients 
$C_{n\ell m}(\tau)$ 
this only adds the mean-field or Hartree term 
$- N_0^2 \hbar^2|a|/\sqrt{2\pi}mq^3$ to $V(q)$ as we have already discussed in 
detail above. The contributions linear in $C_{n\ell m}(\tau)$ are
\begin{eqnarray}
\frac{4\pi a\hbar^2}{m} {\sum_n}'' \int_0^{\hbar\beta} d\tau~
   C_{n00} \int d{\bf x}~ \rho({\bf x};q) 
              \frac{\bar{\xi}_{n00}({\bf x};q)}{(\pi q^2)^{3/4}} = 
                                             \hspace*{2.0in} \nonumber \\
  N_0 \frac{4\pi a\hbar^2}{m}{\sum_n}'' \int_0^{\hbar\beta} d\tau~
      C_{n00} \frac{1}{(\pi q^2)^{3/2}}
            \frac{(-1)^n\sqrt{(2n+1)!}}{2^{2n+3/2} n!}~.       \nonumber
\end{eqnarray}  
Note that we have only linear terms for the coefficients $C_{n00}(\tau)$. This 
is as expected because in the metastable state the average density profile of 
the condensate is given by
\begin{equation}
\langle \rho({\bf x}) \rangle = \rho({\bf x};\langle q \rangle) 
  + {\sum_{n\ell m}}'' \langle C_{n\ell m} \rangle
      \frac{\bar{\xi}_{n\ell m}({\bf x};\langle q \rangle)}
           {(\pi \langle q \rangle^2)^{3/4}}~,
\end{equation}
where due to the rotational symmetry of the isotropic harmonic oscillator 
potential the average value of the coefficients $C_{n\ell m}(\tau)$ must be zero 
if the angular momentum variables $(\ell m) \neq (00)$. The average value of 
$C_{n00}(\tau)$ can be calculated if we know the quadratic terms in 
$C_{n00}(\tau)$, which become 
\begin{eqnarray}
\frac{2\pi a\hbar^2}{m} {\sum_{n,n'}}'' \int_0^{\hbar\beta} d\tau~
   C_{n00} C_{n'00}
   \frac{1}{(\pi q^2)^{3/2}}
        \frac{\Gamma(n+n'+3/2)}{\sqrt{2\pi(2n+1)!(2n'+1)!}}~. \nonumber
\end{eqnarray} 
For the calculation of all the static properties of the condensate and in 
particular a more accurate value of the maximum number of atoms $N_m$, the 
latter two contributions are all that we need. However, for a study of its 
dynamics we require the full quadratic expression
\begin{eqnarray}
\frac{2\pi a\hbar^2}{m} {\sum_{n,n'}}'' \sum_{\ell m}
  \int_0^{\hbar\beta} d\tau~
   C^*_{n\ell m} C_{n'\ell m}
    \frac{1}{(\pi q^2)^{3/2}} \int d{\bf x}~
      \bar{\xi}_{n\ell m}^*({\bf x};q) \bar{\xi}_{n'\ell m}({\bf x};q) =
                                                    \hspace{1.55in} \nonumber \\
\frac{2\pi a\hbar^2}{m} {\sum_{n,n'}}'' \sum_{\ell m}
  \int_0^{\hbar\beta} d\tau~
      C^*_{n\ell m} C_{n'\ell m} \frac{1}{(\pi q^2)^{3/2}}
        \frac{\Gamma(n+n'+\ell+3/2)}
             {2^{n+n'+\ell+3/2}
               \sqrt{n! \Gamma(n+\ell+3/2) n'! \Gamma(n'+\ell+3/2)}}~. \nonumber
\end{eqnarray} 

\begin{figure}
\psfig{figure=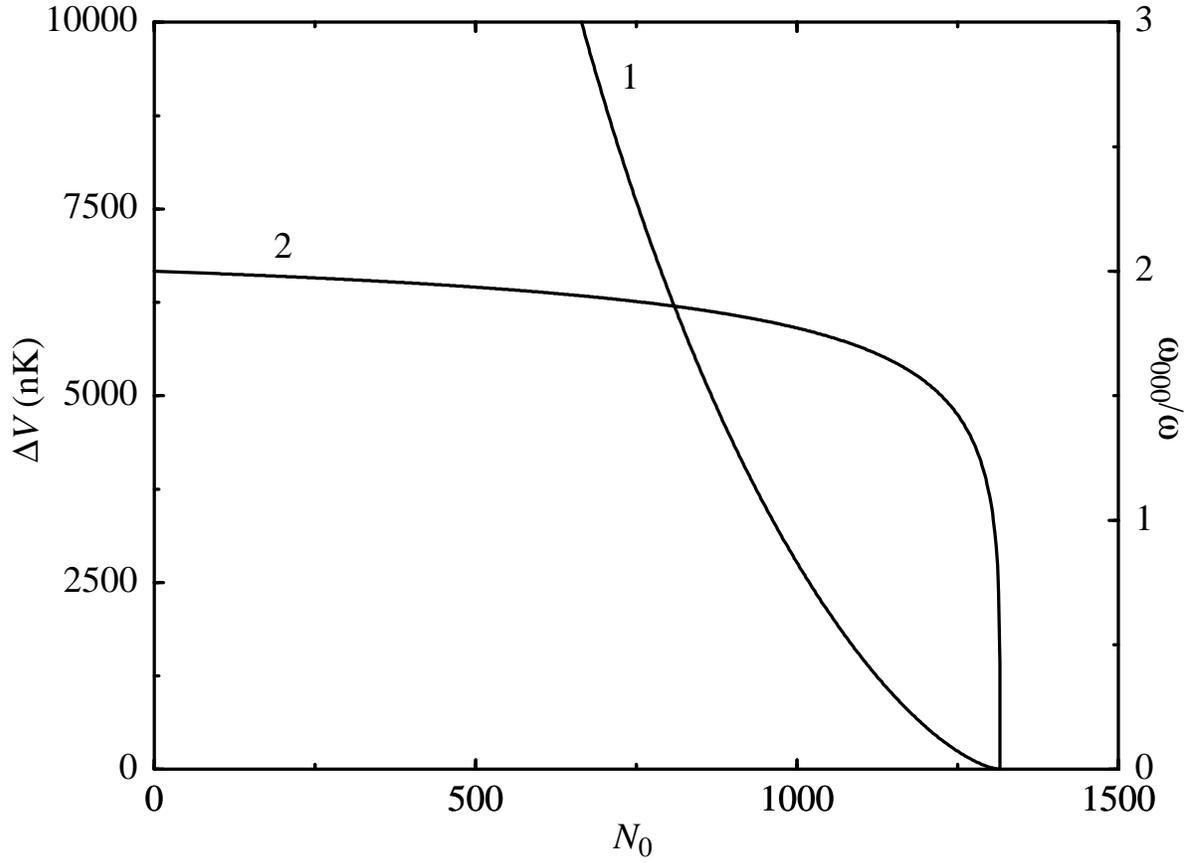}
\caption{Dynamical properties of a $^7$Li condensate. Curve 1 gives the 
         energy barrier $\Delta V$ for thermal fluctuations and curve 2 the
         frequency for the unstable breathing mode. Together these quantities
         determine the decay rate of the metastable condensate at the 
         relatively high temperatures achieved in current experiments.
         \label{Li7}}
\end{figure}

As an application of the above approach, we consider the decay rate of the 
condensate due to thermal fluctuations. This rate equals \cite{henk5}
\begin{equation}
\Gamma(N_0,T) = \frac{\omega_{000}}{2\pi} 
                       \exp \left\{ - \frac{\Delta V}{k_BT} \right\}~,
\end{equation}
where $\Delta V$ represents the macroscopic energy barrier for the condensate. 
Calculating the average $\langle C_{n\ell m} \rangle$ as a function of $q$ and 
substituting the result back into the action $S_E[\rho]$, we obtain a more 
accurate potential $V(q)$ that now incorporates the fact that the average 
density profile of the condensate is not gaussian near the point of instability. 
With this more accurate potential we can again calculate the frequency of the 
`breathing' mode $\omega_{000}$ and the macroscopic energy barrier $\Delta V$ as 
a function of the number of condensate particles. For the experimental 
conditions of interest the results are given in Fiq.~\ref{Li7}. We see that the 
maximum number of condensate particles is $N_m \simeq 1316$ in this case, which 
corresponds to $N_m \simeq 0.60 l/|a|$ and should be compared to the exact 
result $N_m \simeq 0.57 l/|a|$ obtained by Ruprecht {\it et al.} \cite{keith1}. 
The difference is due to the fact that we have included in the action only terms 
up to quadratic order in $C_{n\ell m}(\tau)$. By including higher order terms we 
should in principle be able to reproduce the exact value of $N_m$, although this 
will be rather tedious in practice. The advantage of the present approach is 
that we can systematically incorporate the effect of fluctuations of the 
condensate, which cannot be achieved by a solution of the Gross-Pitaevskii 
equation. Hereby, we conclude our treatment of the dynamics of Bose-Einstein 
condensation in an interacting Bose gas with negative scattering length. From 
now on we will again focus on a gas with effectively repulsive interactions, in 
which case the experiments are in general in the more complicated 
strong-coupling limit.

\subsection{Phase `Diffusion'}
\label{diff}
As we have just seen explicitly, a particularly interesting 
consequence of the finite size of the gas is that quantum fluctuations play a 
much more important role. Although this is especially true for the case of 
attractive interactions, it is also true for a Bose gas with repulsive 
interactions. A striking example in this respect is the phenomenon of phase 
`diffusion', which was recently discussed by Lewenstein and You \cite{maciek}. 
We rederive their results for a trapped Bose gas in a moment, but first 
consider also the same phenomenon for a neutral and homogeneous superconductor. 
Not only is it possible in this manner to bring out the physics involved more 
clearly, but we are then also able to point out an important distinction  
between a fermionic and a bosonic superfluid that is also crucial for a good 
understanding of the concept of a quasicondensate in weakly-interacting Bose 
gases \cite{popov}.

At zero temperature the dynamics of the superconducting order parameter, i.e., 
the BCS gap parameter $\Delta({\bf x},t)$ that is proportional to the 
wavefunction of the condensate of Cooper pairs, is in a good approximation 
determined by a time-dependent Ginzburg-Landau theory 
\cite{anderson,abrahams,hagen1} with the action 
\begin{equation}
S[\Delta^*,\Delta] =
 \frac{N(0)}{4} \int dt \int d{\bf x}~ 
   \left\{ \frac{\hbar^2}{|\Delta_0|^2}
      \left| \frac{\partial \Delta}{\partial t} \right|^2
      - \frac{\hbar^2 v_F^2}{3|\Delta_0|^2} 
      \left| \nabla \Delta \right|^2
      + 2 |\Delta|^2 \left( 1 - \frac{|\Delta|^2}{2|\Delta_0|^2} \right)
   \right\}~, 
\end{equation}
where $N(0)$ is the density of states for one spin projection at the Fermi 
energy $\epsilon_F = mv_F^2/2$ and $\Delta_0$ is the equilibrium value of the 
order parameter \cite{micheal}. Writing the complex order parameter in terms of 
an amplitude and a phase, we immediately observe that the amplitude 
fluctuations are gapped \cite{gap} and can, therefore, be safely neglected at 
large length scales. The long-wavelength dynamics of the superconductor is thus 
dominated by the phase fluctuations, according to the action
\begin{equation}
S[\theta] =
 \frac{N(0)\hbar^2}{4} \int dt \int d{\bf x}~ 
   \left\{ \left( \frac{\partial \theta}{\partial t} \right)^2
      - \frac{v_F^2}{3} 
      \left( \nabla \theta \right)^2
   \right\}~. 
\end{equation}
This also implies that the global phase 
$\theta_0(t) = \int d{\bf x}~ \theta({\bf x},t)/V$ of the superconductor has a 
dynamics that is governed by 
\begin{equation}
\label{phase}
S[\theta_0] =
  \frac{N(0)N\hbar^2}{4n} \int dt~ 
             \left( \frac{d \theta_0}{dt} \right)^2~,
\end{equation}
using the fact that the total volume $V$ of the system is given by $N/n$. 

Up to now our discussion has again been semiclassical. To consider also the 
quantum fluctuations, we have to quantize this theory by applying the usual 
rules of quantum mechanics. Doing so, we find that the wavefunction of the 
overall phase obeys a Schr\"odinger equation
\begin{equation}
i\hbar \frac{\partial}{\partial t} \Psi(\theta_0;t) =
   - \frac{n}{N(0)N}~ \frac{\partial^2}{\partial \theta_0^2}
                                                   \Psi(\theta_0;t)~,
\end{equation}
with a `diffusion' constant that can easily be shown to be equal to 
$(2/\hbar) \partial \epsilon_F/\partial N$ \cite{dan} and is, most importantly 
for our purposes, proportional to $1/N$. In the thermodynamic limit $N 
\rightarrow \infty$ a state with a well defined stationary phase is clearly a 
solution and we are then dealing with a system having a spontaneously broken 
$U(1)$ symmetry. However, for a finite (and fixed) number of particles the 
global phase cannot be well defined at all times and always has to `diffuse' in 
accordance with the above Schr\"odinger equation. Note also that in the 
groundstate the phase is fully undetermined and 
$|\Psi(\theta_0;t)|^2 = 1/2\pi$. Maybe surprisingly, the same 
calculation is somewhat more complicated for a Bose gas because the amplitude 
fluctuations of the order parameter cannot be neglected even at the largest 
length scales. However, taking these amplitude fluctuations into account 
properly, we nevertheless arrive at an action that is equivalent to 
Eq.~(\ref{phase}) and hence again leads to the phenomenon of phase `diffusion'. 

We start again from the action $S_E[\rho,\theta;\mu]$ for the condensate. The 
difference with the previous subsection is, however, that now we are not so much 
interested in the dynamics of the density but in the phase dynamics instead. 
Therefore we now want to integrate over the density field $\rho({\bf x},\tau)$. 
This cannot be done exactly and we therefore here consider only the 
strong-coupling limit, which was also treated by Lewenstein and You 
\cite{maciek}. In that limit we are allowed to neglect the gradient of the 
average density profile \cite{tony2} and the action $S_E[\rho,\theta;\mu]$ is 
for the longest wavelengths well approximated by
\begin{eqnarray}
S_E[\rho,\theta;\mu] = 
   \int_0^{\hbar\beta} d\tau \int d{\bf x}~ 
   \left( i\hbar \rho({\bf x},\tau) 
                 \frac{\partial \theta({\bf x},\tau)}{\partial \tau} \right.
                                                \hspace*{2.5in} \nonumber \\
     \left. + V^{ex}({\bf x})\rho({\bf x},\tau) - \mu\rho({\bf x},\tau)  
            + \frac{2\pi a\hbar^2}{m} \rho^2({\bf x},\tau) \right)~.
\end{eqnarray}
In equilibrium the average density profile of the condensate thus obeys
\begin{equation}
\langle \rho({\bf x}) \rangle = \frac{m}{4\pi a \hbar^2}
   \left( \mu - V^{ex}({\bf x}) \right) \Theta( \mu - V^{ex}({\bf x}) )~.
\end{equation}
Performing now the shift 
$\rho({\bf x},\tau) 
           = \langle \rho({\bf x}) \rangle + \delta\rho({\bf x},\tau)$, we find 
for the zero-momentum part of the action \cite{top}
\begin{equation}
S_E[\delta N_0,\theta_0;\mu] =  \hbar\beta E_0(\mu)  
 + \int_0^{\hbar\beta} d\tau  
   \left( i\hbar \delta N_0 \frac{d \theta_0}{d\tau}
          + \frac{2\pi a\hbar^2}{mV_0(\mu)} (\delta N_0)^2 \right)~,
\end{equation}
where $E_0(\mu)$ and $V_0(\mu)$ correspond, respectively, to the energy and the 
volume of the condensate in the so-called Thomas-Fermi approximation \cite{TF}. 
Moreover, $\delta N_0(\tau) = \int d{\bf x}~ \delta\rho({\bf x},\tau)$ 
represents the fluctuations in the total number of condensate particles in that 
same approximation, implying that the density fluctuations 
$\delta\rho({\bf x},\tau)$ are only nonzero in that region of space where the 
condensate density does not vanish.

Performing now the integration over the number fluctuations $\delta N_0(\tau)$ 
and the usual Wick rotation to real times, we immediately see that the 
effective action for the global phase of the condensate has precisely the same 
form as in Eq.~(\ref{phase}), i.e.,
\begin{equation}
S[\theta_0;\mu] = \frac{mV_0(\mu)}{8\pi a} \int dt~
                                 \left( \frac{d\theta_0}{dt} \right)^2~.
\end{equation}
The appropriate `diffusion' constant is therefore equal to 
$2\pi a\hbar/mV_0(\mu)$, which can easily be shown to be equal to 
$(1/2\hbar) \partial \mu/\partial N_0$ if we make use of the fact that in the 
Thomas-Fermi approximation the chemical potential obeys
$\mu = m\omega^2R_{TF}^2/2$ and the radius of the condensate is given by
$R_{TF} = (15a\hbar^2 N_0/m^2\omega^2)^{1/5}$ \cite{BP}. Hence, the `diffusion' 
constant is proportional to $1/N_0^{3/5}$. Note that if the condensate where 
contained in a box the `diffusion' constant would be proportional to $1/N_0$ 
instead. It is important to note also that, in contrast to the case of a 
fermionic superfluid, we have to integrate over the amplitude fluctuations of 
the order parameter to arrive at a quadratic action for the phase fluctuations. 
This leads to the important conclusion that for a bosonic superfluid it is 
impossible to be in a state with only phase fluctuations and no density 
fluctuations, even at the largest length scales. A more implicit way to arrive 
at the same conclusion was in fact already presented in Sec.~\ref{quant}, since 
it is precisely the coupling between the amplitude and phase fluctuations that 
is required to understand that the dynamical exponent $z$ is equal to $1$.

\subsection{Strong-Coupling Limit}
\label{sc}
The difficulty with the strong-coupling limit is that the appropriate 
one-particle states of the gas are no longer the states $\chi_{\alpha}({\bf x})$ 
of the trapping potential because these states are now strongly coupled by the 
mean-field interaction. For our Caldeira-Leggett model this problem could be 
easily resolved by introducing the states $\chi'_{\alpha}({\bf x})$ that 
incorporate the average interaction with the reservoir. However, for a trapped 
Bose gas the same procedure would lead to an expansion into states 
$\chi'_{\alpha}({\bf x};t)$ that parametrically depend on time since the 
mean-field interaction depends on time also. In principle, we thus have to deal 
with the situation that not only the occupation numbers vary with time but also 
the states to which these occupation numbers belong. This results in highly 
nontrivial dynamics since it is {\it a priori} clear that an adiabatic 
approximation, in which we neglect the nondiagonal matrix elements 
$\int d{\bf x}~\chi'^*_{\alpha}({\bf x};t) (\partial/\partial t) 
                                                 \chi'_{\alpha'}({\bf x};t)$,
is not valid, because of the lack of a separation of time scales. Notice that we 
were able to circumvent the above problems for a Bose gas with effectively 
attractive interactions due to the fact that on the one hand the growth of the 
condensate essentially takes place in the weak-coupling regime and that on the 
other hand the collapse, which certainly occurs in the strong-coupling regime, 
requires in first instance only knowledge of the collisionless dynamics of the 
condensate.                                                 

Notwithstanding the above remarks, we can make progress in the strong-coupling 
limit if we realize that in this limit the typical length scale on which the 
(noncondensate) density of the gas varies is always much larger than the trap 
size. We are therefore justified in performing a gradient expansion. To do so in 
a systematic fashion, it is convenient to express the various selfenergies and 
interactions in the coordinate representation, i.e., by introducing
\begin{equation}
\hbar\Sigma^{(\pm),K}({\bf x},{\bf x}';\epsilon) =
 \sum_{\alpha,\alpha'} \chi_{\alpha}({\bf x})
   \hbar\Sigma^{(\pm),K}_{\alpha,\alpha'}(\epsilon) \chi^*_{\alpha'}({\bf x}')
\end{equation}
and similarly
\begin{equation}
V^{(\pm),K}({\bf x},{\bf x}';{\bf y},{\bf y}';\epsilon) =
 \sum_{\alpha,\alpha'} \sum_{\beta,\beta'}
   \chi_{\alpha}({\bf x}) \chi_{\beta}({\bf y})
         V^{(\pm),K}_{\alpha,\beta;\alpha',\beta'}(\epsilon)
   \chi^*_{\alpha'}({\bf x}') \chi^*_{\beta'}({\bf y}')~.
\end{equation}
In this coordinate representation, our Fokker-Planck equation acquires in all 
generality the form
\begin{eqnarray}
\label{FPgen}
i\hbar \frac{\partial}{\partial t} P[\phi^*,\phi;t] = && \nonumber \\
&& \hspace{-1.0in}
 - \int d{\bf x} d{\bf x}'~
   \frac{\delta}{\delta \phi({\bf x})}
       \left\{ \left( \left(
                        - \frac{\hbar^2\nabla^2}{2m} + V^{ex}({\bf x}) - \mu(t)
                      \right) \delta({\bf x}-{\bf x}')  
          + \hbar\Sigma^{(+)}({\bf x},{\bf x}') \right) \phi({\bf x}')
            \right. \nonumber \\
&& \hspace{1.1in}   \left. + \int d{\bf y} d{\bf y}'~
               V^{(+)}({\bf x},{\bf x}';{\bf y},{\bf y}') 
                   \phi^*({\bf y})
                   \phi({\bf y}') \phi({\bf x}')
     {}^{{}^{{}^{{}^{{}^{{}^{}}}}}}  \right\} P[\phi^*,\phi;t] \nonumber \\
&& \hspace{-1.0in}
+ \int d{\bf x} d{\bf x}'~
  \frac{\delta}{\delta \phi^*({\bf x})}
       \left\{ \left( \left(
                        - \frac{\hbar^2\nabla^2}{2m} + V^{ex}({\bf x}) - \mu(t)
                      \right) \delta({\bf x}-{\bf x}')  
          + \hbar\Sigma^{(-)}({\bf x}',{\bf x}) \right) \phi^*({\bf x}')
            \right. \nonumber \\
&& \hspace{1.1in}   \left. + \int d{\bf y} d{\bf y}'~
               V^{(-)}({\bf x}',{\bf x};{\bf y}',{\bf y})
                   \phi^*({\bf x}') \phi^*({\bf y}')
                   \phi({\bf y})
     {}^{{}^{{}^{{}^{{}^{{}^{}}}}}} \right\} P[\phi^*,\phi;t] \nonumber \\
&& \hspace{-1.0in}
- \frac{1}{2} \int d{\bf x} d{\bf x}'~
  \frac{\delta^2}{\delta \phi({\bf x})
                    \delta \phi^*({\bf x}')}
       \left\{ \hbar\Sigma^K({\bf x},{\bf x}')
                   {}^{{}^{{}^{{}^{{}^{}}}}}  \right. \nonumber \\
&& \hspace{1.1in}   \left. + \int d{\bf y} d{\bf y}'~
               V^K({\bf x},{\bf x}';{\bf y},{\bf y}') 
                   \phi^*({\bf y}) \phi({\bf y}') 
       \right\} P[\phi^*,\phi;t]~.
\end{eqnarray}
To study its physical content and to familiarize ourselves with how to perform 
the desired gradient expansion, we first analyze again the normal state of the 
gas in which we can neglect the nonlinear terms representing the effect of the 
condensate.

In the normal state the condensate is absent and we are only interested in the 
behavior of the quantity $\langle \phi({\bf x}) \phi^*({\bf x}') \rangle(t)$. 
From the above Fokker-Planck equation we see that it obeys the equation of 
motion
\begin{eqnarray}
i\hbar \frac{\partial}{\partial t}
           \langle \phi({\bf x}) \phi^*({\bf x}') \rangle(t) = && \nonumber \\
&& \hspace{-1.3in}
  \int d{\bf x}''~
     \left( \left( - \frac{\hbar^2}{2m} \frac{\partial^2}{\partial {\bf x}^2}
                   + V^{ex}({\bf x}) - \mu(t)
            \right) \delta({\bf x}-{\bf x}'')  
            + \hbar\Sigma^{(+)}({\bf x},{\bf x}'') \right)
                       \langle \phi({\bf x}'') \phi^*({\bf x}') \rangle(t)
                                                                  \nonumber \\
&& \hspace{-1.3in}
 - \int d{\bf x}''~
     \left( \left( - \frac{\hbar^2}{2m} \frac{\partial^2}{\partial {\bf x}'^2}
                   + V^{ex}({\bf x}') - \mu(t)
            \right) \delta({\bf x}''-{\bf x}')  
            + \hbar\Sigma^{(-)}({\bf x}'',{\bf x}') \right)
                       \langle \phi({\bf x}) \phi^*({\bf x}'') \rangle(t)                                                                  
                                                                  \nonumber \\
&& \hspace{-1.3in}
 - \frac{1}{2} \hbar\Sigma^K({\bf x},{\bf x}')~.                                                                  
\end{eqnarray}
Introducing now the Wigner distribution \cite{wigner} by means of
\begin{equation}
N({\bf x},{\bf k};t) + \frac{1}{2} =
  \int d{\bf x}'~ e^{-i {\bf k} \cdot {\bf x}' }
                 \langle \phi({\bf x} + {\bf x}'/2) 
                         \phi^*({\bf x} - {\bf x}'/2) \rangle(t)
\end{equation}
and similarly the selfenergies $\hbar\Sigma^{(\pm),K}({\bf x},{\bf k})$ by
\begin{equation}
\hbar\Sigma^{(\pm),K}({\bf x},{\bf k}) =
  \int d{\bf x}'~ e^{-i {\bf k} \cdot {\bf x}' }
     \hbar\Sigma^{(\pm),K}({\bf x} + {\bf x}'/2,{\bf x} - {\bf x}'/2)~,
\end{equation}
we find after substituting these relations in lowest nonvanishing order in the 
gradient $\partial/\partial{\bf x}$ the expected quantum Boltzmann equation 
\cite{D,kadanoff}
\begin{eqnarray}
\label{qBe}
\frac{\partial}{\partial t} N({\bf x},{\bf k};t) 
  + \frac{1}{\hbar} 
    \frac{\partial \epsilon'({\bf x},{\bf k};t)}{\partial {\bf k}} \cdot 
      \frac{\partial}{\partial {\bf x}} N({\bf x},{\bf k};t)
  - \frac{1}{\hbar}
    \frac{\partial \epsilon'({\bf x},{\bf k};t)}{\partial {\bf x}} \cdot
      \frac{\partial}{\partial {\bf k}} N({\bf x},{\bf k};t)
                                             = \hspace*{0.6in} \nonumber \\
      - \Gamma^{out}({\bf x},{\bf k};t) N({\bf x},{\bf k};t) 
      + \Gamma^{in}({\bf x},{\bf k};t) (1 + N({\bf x},{\bf k};t))~,
\end{eqnarray}
where the energies of the atoms have to be determined selfconsistently from
\begin{equation}
\epsilon'({\bf x},{\bf k};t) = \epsilon({\bf k}) + V^{ex}({\bf x}) 
     + {\rm Re}[\hbar\Sigma^{(+)}({\bf x},{\bf k};
                                      \epsilon'({\bf x},{\bf k};t)-\mu(t))]~.    
\end{equation}
Moreover, the expressions for the retarded energy 
$\hbar\Sigma^{(+)}({\bf x},{\bf k};\epsilon'({\bf x},{\bf k};t)-\mu(t))$,
and the collision rates $\Gamma^{out}({\bf x},{\bf k};t)$ and 
$\Gamma^{in}({\bf x},{\bf k};t)$ turn out to be identical to the ones derived 
for a homogeneous gas in Secs.~\ref{semicl} and \ref{quant}. 
We only need to replace $\epsilon'({\bf k};t)$ by $\epsilon'({\bf x},{\bf k};t)$ 
and $N({\bf k};t)$ by $N({\bf x},{\bf k};t)$. Notice also that the same 
substitution has to be made in the Bethe-Salpeter equation for the many-body T 
matrix, which implies that the effective interaction 
$V^{(+)}({\bf k},{\bf k}',{\bf k}'')$ now implicitly depends both on the 
position ${\bf x}$ and the time $t$. 

As an example, let us consider the gas not too close to the critical 
temperature. Then the effective interaction is well approximated by the two-body 
T matrix and we simply have 
\begin{equation}
\epsilon'({\bf x},{\bf k};t) = \epsilon({\bf k}) + V^{ex}({\bf x}) 
        + \frac{8\pi n({\bf x},t) a \hbar^2}{m}~.
\end{equation}
As a result the equilibrium solution $N({\bf x},{\bf k};\infty)$ of the quantum 
Boltzmann equation equals a Bose distribution evaluated at 
$\epsilon({\bf k}) + V^{ex}({\bf x}) + 8\pi n({\bf x}) a \hbar^2/m - \mu$. 
The equilibrium density profile 
$n({\bf x}) = \int d{\bf k}~N({\bf x},{\bf k};\infty)/(2\pi)^3$ thus corresponds 
to a local-density approximation but is indeed very accurate for the atomic 
gases of interest. Apart from describing the relaxation of the gas towards 
equilibrium, the quantum Boltzmann equation can also be used to study the 
collective modes of the trapped gas above the critical temperature. Using the 
same methods that have recently been applied to inhomogeneous Bose gases by 
Zaremba, Griffin, and Nikuni \cite{eugene2}, we can for example derive the 
hydrodynamic equations of motion the gas. Moreover, the collisionless modes are 
obtained from
\begin{equation}
\label{clqBe}
\frac{\partial}{\partial t} N({\bf x},{\bf k};t) 
  + \frac{\hbar {\bf k}}{m} \cdot
      \frac{\partial}{\partial {\bf x}} N({\bf x},{\bf k};t)
  - \frac{1}{\hbar}
    \frac{\partial}{\partial {\bf x}} 
    \left( V^{ex}({\bf x}) + \frac{8\pi n({\bf x},t) a \hbar^2}{m} \right) \cdot
      \frac{\partial}{\partial {\bf k}} N({\bf x},{\bf k};t) = 0~.
\end{equation}                    
Solving this equation is equivalent to the random-phase approximation 
\cite{kadanoff}. More important is that the above collisionless Boltzmann 
equation can be shown to satisfy the Kohn theorem \cite{kohn} by substituting 
the shifted equilibrium profile 
$N({\bf x}-{\bf x}_{cm}(t),{\bf k}-(m/\hbar)d{\bf x}_{cm}(t)/dt;\infty)$. For an 
harmonic external potential the center-of-mass of the gas cloud then precisely 
follows Newton's law 
$d^2 {\bf x}_{cm}(t)/dt^2 = - \nabla V^{ex}({\bf x}_{cm}(t))$, which shows that 
the gas has three collisionless modes with frequencies that are equal to the 
three frequencies of the external trapping potential.

The full quantum Boltzmann equation in Eq.~(\ref{qBe}) also describes the 
relaxation of the gas towards equilibrium. It can thus be used to study forced  
evaporative cooling by adding an appropriate loss term to account for the 
escaping high-energetic atoms. In agreement with our previous picture we then 
expect that, if the evaporative cooling is performed slowly, the distribution 
function $N({\bf x},{\bf k};t)$ can be well approximated by a truncated 
equilibrium distribution with a time-dependent temperature and chemical 
potential. As a result, we can use the results from Sec.~\ref{semicl} to show 
that in the critical regime the gas again becomes unstable towards Bose-Einstein 
condensation. This is the case when $\epsilon'({\bf x},{\bf 0};t) < \mu(t)$ for 
positions near the center of the trap where the density of the gas is highest. 
If that happens we are no longer allowed to neglect the nonlinear terms in the 
Fokker-Planck equation and some corrections are neccesary. Applying again the 
Hartee-Fock approximation, which is correct sufficiently close to the critical 
temperature where $k_BT \gg n_0({\bf x},t) T^{(+)}({\bf 0},{\bf 0};0)$, we find 
first of all that the quantum Boltzmann equation for the noncondensed part of 
the gas has the same form as in Eq.~(\ref{qBe}). We only have to add the 
mean-field interaction $2 n_0({\bf x},t) T^{(+)}({\bf 0},{\bf 0},{\bf 0};0)$ due 
to the condensate to the renormalized energy $\epsilon'({\bf x},{\bf k};t)$ and 
in the same way as in the homogeneous case also account for the presence of the 
condensate in the scattering rates $\Gamma^{out}({\bf x},{\bf k};t)$ and 
$\Gamma^{in}({\bf x},{\bf k};t)$. 

In addition, the resulting quantum Boltzmann equation for the noncondensed atoms 
is coupled to a Fokker-Planck equation for the condensate. After a gradient 
expansion the latter equals
\begin{eqnarray}
\label{FPsc}
i\hbar \frac{\partial}{\partial t} P_0[\phi_0^*,\phi_0;t] = && \nonumber \\
&& \hspace{-1.0in}
 - \int d{\bf x}~
   \frac{\delta}{\delta \phi_0({\bf x})}
        \left( - \frac{\hbar^2\nabla^2}{2m} + V^{ex}({\bf x}) - \mu(t)
               + \hbar\Sigma^{(+)}({\bf x},{\bf 0})
               \right. \nonumber \\
&& \hspace{1.7in}   \left. {}^{{}^{{}^{{}^{{}^{{}^{}}}}}} 
               + T^{(+)}({\bf 0},{\bf 0},{\bf 0};0) |\phi_0({\bf x})|^2
        \right) \phi_0({\bf x}) P_0[\phi_0^*,\phi_0;t]           \nonumber \\
&& \hspace{-1.0in}
+ \int d{\bf x}~
  \frac{\delta}{\delta \phi_0^*({\bf x})}
        \left( - \frac{\hbar^2\nabla^2}{2m} + V^{ex}({\bf x}) - \mu(t)
               + \hbar\Sigma^{(-)}({\bf x},{\bf 0})
               \right. \nonumber \\
&& \hspace{1.7in}   \left. {}^{{}^{{}^{{}^{{}^{{}^{}}}}}} 
               + T^{(+)}({\bf 0},{\bf 0},{\bf 0};0) |\phi_0({\bf x})|^2
        \right) \phi_0^*({\bf x}) P_0[\phi_0^*,\phi_0;t]           \nonumber \\
&& \hspace{-1.0in}
- \frac{1}{2} \int d{\bf x}~
  \frac{\delta^2}{\delta \phi_0({\bf x}) \delta \phi_0^*({\bf x})}
     \hbar\Sigma^K({\bf x},{\bf 0}) P_0[\phi_0^*,\phi_0;t]~,
\end{eqnarray}
where $\hbar\Sigma^{(\pm),K}({\bf x},{\bf 0})$ denotes  
$\hbar\Sigma^{(\pm),K}({\bf x},{\bf k}={\bf 0})$. In more detail we have for the 
real part of the retarded and advanced selfenergies that 
$S({\bf x},{\bf 0}) = {\rm Re}[\hbar\Sigma^{(\pm)}({\bf x},{\bf 0};
                                 \epsilon''({\bf x},{\bf 0};t) - \mu(t))]$. 
Moreover, using our picture of the evaporative cooling of the gas, the imaginary 
parts follow most conveniently from the fluctuation-dissipation theorem
\begin{equation}
iR({\bf x},{\bf 0}) = - \frac{\beta(t)}{4} \hbar\Sigma^K({\bf x},{\bf 0})
   \left( - \frac{\hbar^2\nabla^2}{2m} + \epsilon'({\bf x},{\bf 0};t) - \mu(t)
          + T^{(+)}({\bf 0},{\bf 0},{\bf 0};0) |\phi_0({\bf x})|^2
        \right) 
\end{equation} 
where $\hbar\Sigma^K({\bf x},{\bf 0}) 
               = \hbar\Sigma^{(\pm),K}({\bf x},{\bf 0};
                                       \epsilon''({\bf x},{\bf 0};t) - \mu(t))$ 
can in a good approximation be considered as independent of $|\phi_0({\bf x})|$. 
Near equilibrium the Fokker-Planck equation thus simplifies to
\begin{eqnarray}
i\hbar \frac{\partial}{\partial t} P_0[|\phi_0|;t] = && \nonumber \\
&& \hspace{-1.0in}
 - \frac{\beta}{4} \int d{\bf x}~ \hbar\Sigma^K({\bf x},{\bf 0})
   \frac{\delta}{\delta \phi_0({\bf x})}
        \left( - \frac{\hbar^2\nabla^2}{2m} 
               + \epsilon'({\bf x},{\bf 0}) - \mu
        \right. \nonumber \\
&& \hspace{1.8in}   \left. {}^{{}^{{}^{{}^{{}^{{}^{}}}}}} 
               + T^{(+)}({\bf 0},{\bf 0},{\bf 0};0) |\phi_0({\bf x})|^2
        \right) \phi_0({\bf x}) P_0[|\phi_0|;t]           \nonumber \\
&& \hspace{-1.0in}
- \frac{\beta}{4} \int d{\bf x}~ \hbar\Sigma^K({\bf x},{\bf 0})
  \frac{\delta}{\delta \phi_0^*({\bf x})}
        \left( - \frac{\hbar^2\nabla^2}{2m} 
               + \epsilon'({\bf x},{\bf 0}) - \mu
        \right. \nonumber \\
&& \hspace{1.8in}   \left. {}^{{}^{{}^{{}^{{}^{{}^{}}}}}} 
               + T^{(+)}({\bf 0},{\bf 0},{\bf 0};0) |\phi_0({\bf x})|^2
        \right) \phi_0^*({\bf x}) P_0[|\phi_0|;t]           \nonumber \\
&& \hspace{-1.0in}
- \frac{1}{2} \int d{\bf x}~ \hbar\Sigma^K({\bf x},{\bf 0}) 
  \frac{\delta^2}{\delta \phi_0({\bf x}) \delta \phi_0^*({\bf x})}
     P_0[|\phi_0|;t]
\end{eqnarray}
and the probability distribution of the condensate relaxes to the expected 
stationary state
\begin{eqnarray}
P_0[|\phi_0|;\infty] \propto                    \hspace*{5.3in} \nonumber \\
\exp \left\{ - \beta \int d{\bf x}~ \phi_0^*({\bf x})
  \left( - \frac{\hbar^2\nabla^2}{2m} 
         + \epsilon'({\bf x},{\bf 0}) - \mu
         + \frac{T^{(+)}({\bf 0},{\bf 0},{\bf 0};0)}{2} |\phi_0({\bf x})|^2
  \right) \phi_0({\bf x}) \right\}~. \nonumber
\end{eqnarray}
From this result we conclude that Eqs.~(\ref{qBe}) and (\ref{FPsc}) give an 
accurate discription of the dynamics of Bose-Einstein condensation in the 
strong-coupling regime. Unfortunately, the solution of these coupled equations 
is not possible analytically and we therefore have to refer to a future 
publication for a numerical study and a comparison of this theory with 
experiment. We here only remark that the semiclassical approximation in this 
case amounts to the Thomas-Fermi result for the condensate density 
\begin{equation}
n_0({\bf x};t) = \frac{1}{T^{(+)}({\bf 0},{\bf 0},{\bf 0};0)}
   \left( \mu(t) - \epsilon'({\bf x},{\bf 0};t) \right) 
       \Theta( \mu(t) - \epsilon'({\bf x},{\bf 0};t) )~.
\end{equation}
Such an approximation implicitly assumes, however, that the dominant mechanism 
for the formation of the condensate is by means of a coherent evolution and not 
by incoherent collisions. This might be expected not to be very accurate on the 
basis of the homogeneous result that the second mechanism only diverges 
logarithmically with the system size, whereas the first diverges linearly. 
Nevertheless it seems that only a full solution of Eqs.~(\ref{qBe}) and 
(\ref{FPsc}) can settle this important issue.

\subsection{Collective Modes}
\label{cm}
Although we have up to now primarily been focussed on the dynamics of 
Bose-Einstein condensation itself, it is interesting that we have in fact also 
arrived at an accurate discription of the collective modes of a condensed gas at 
nonzero temperatures. Since these collective modes are receiving much attention 
at present, we now briefly want to discuss how such a description emerges from 
our Fokker-Planck equation \cite{michiel1}. Since most experiments at present 
are performed in the collisionless limit, we consider only this limit here. It 
is in principle possible, however, to derive the hydrodynamic equations of 
motions as well \cite{eugene2}. Let us for simplicitly first assume that the 
temperature of the gas is sufficiently far below the critical temperature that 
the many-body T matrix is well approximated by the two-body T matrix, but not so 
low that the Hartree-Fock approximation is no longer valid. In this regime we 
have that $\hbar\Sigma^{(\pm)}({\bf x},{\bf k}) = 8\pi n'({\bf x},t) a\hbar^2/m$ 
and the dynamics of the noncondensed part of the gas is again determined by the 
collisionless quantum Boltzmann equation given in Eq.~(\ref{clqBe}), where the 
total density now consists of two contributions
\begin{equation}
n({\bf x},t) =  n_0({\bf x},t) + n'({\bf x},t) \equiv
         |\langle \phi({\bf x}) \rangle(t)|^2
                     + \int \frac{d{\bf k}}{(2\pi)^3}~N({\bf x},{\bf k};t)~.
\end{equation}
The dynamics of the condensate follows again from the nonlinear Schr\"odinger 
equation
\begin{equation}
\label{nlse}
i\hbar \frac{\partial}{\partial t} \langle \phi({\bf x}) \rangle(t) =                                       
   \left\{ - \frac{\hbar^2 \nabla^2}{2m} + V^{ex}({\bf x}) - \mu 
           + \frac{4\pi a\hbar^2}{m} (2n'({\bf x},t) + n_0({\bf x},t)) \right\}
                         \langle \phi({\bf x}) \rangle(t)~, 
\end{equation}
which can be rewritten as the continuity equation
\begin{equation}
\label{nlse1}
\frac{\partial}{\partial t} n_0({\bf x},t) 
                  = - \nabla \cdot ( {\bf v}_s({\bf x},t) n_0({\bf x},t) )
\end{equation}
together with the Josephson relation for the superfluid velocity 
\begin{equation}
\label{nlse2}
\frac{\partial}{\partial t} {\bf v}_s({\bf x},t) = 
   - \nabla \left( V^{ex}({\bf x}) - \mu 
            + \frac{4\pi a\hbar^2}{m} (2n'({\bf x},t) + n_0({\bf x},t)) 
            + \frac{1}{2} m {\bf v}_s^2({\bf x},t) \right)~.
\end{equation}
Here, we used that $\langle \phi({\bf x}) \rangle(t) 
                           = \sqrt{n_0({\bf x},t)} e^{i\theta({\bf x},t)}$,
${\bf v}_s({\bf x},t) = \hbar \nabla \theta({\bf x},t)/m$, and have applied the 
Thomas-Fermi approximation, which is appropriate in the strong-coupling limit. 
It will presumably not come as a surprise that these coupled equations for 
$N({\bf x},{\bf k};t)$, $n_0({\bf x},t)$, and ${\bf v}_s({\bf x},t)$ can be 
shown to fulfill the Kohn theorem exactly. This is an important constraint on 
any theory for the collective modes of a trapped gas, as stressed recently by 
Zaremba, Griffin, and Nikuni \cite{eugene2}.

As we have mentioned previously, in trapped atomic gases it is possible to cool 
the gas even to such low temperatures that
$k_BT \ll n_0({\bf x};t) T^{(+)}({\bf 0},{\bf 0};0)$. In this regime it is no 
longer allowed to use the Hartree-Fock approximation and we must use the 
Bogoliubov approximation instead. This is achieved as follows. In the 
Hartree-Fock approximation we have taken for the probability distribution 
$P[\phi^*,\phi;t]$ the separable form 
$P_0[\phi_0^*,\phi_0;t] P_1[\phi'^*,\phi';t]$. For such an ansatz the $U(1)$ 
invariance of the theory requires in fact that $P[\phi^*,\phi;t]$ is invariant 
under independent phase transformations on $\phi_0$ and $\phi'$. Clearly this 
ansatz is therefore too symmetric in general, since it is only required that 
$P[\phi^*,\phi;t]$ is invariant under simultaneous phase transformations on 
$\phi_0$ and $\phi'$. For such more general solutions the probability 
distributions for the condensed and noncondensed parts of the gas are then
given by 
$P_0[\phi_0^*,\phi_0;t] \equiv \int d[\phi'^*]d[\phi']~ P[\phi^*,\phi;t]$ and 
$P_1[\phi'^*,\phi';t] \equiv \int d[\phi_0^*]d[\phi_0]~ P[\phi^*,\phi;t]$,
respectively. The Fokker-Planck equations for these probability distributions 
can thus be found from integrating the general Fokker-Planck equation in 
Eq.~(\ref{FPgen}) first over $\phi'({\bf x})$ and subsequently over 
$\phi_0({\bf x})$. Taking care that we do not double count effects of the 
interaction that are already accounted for in the various selfenergies, we find 
after a gradient expansion that the Fokker-Planck equation for the condensate is 
identical to Eq.~(\ref{FPsc}). The collisionless dynamics of the condensate is 
therefore again determined by the nonlinear Schr\"odinger equation in 
Eq.~(\ref{nlse}) or in the Thomas-Fermi approximation by Eqs.~(\ref{nlse1}) and 
(\ref{nlse2}). The Hartree-Fock and Bogoliubov approximations are clearly 
identical in this respect. This is, however, not true if we consider the 
dynamics of the noncondensed part of the gas.

Integrating over $\phi_0({\bf x})$ we now obtain in the collisionless limit that
\begin{eqnarray}
i\hbar \frac{\partial}{\partial t} P_1[\phi'^*,\phi';t] = && \nonumber \\
&& \hspace{-1.0in}
 - \int d{\bf x}~
   \frac{\delta}{\delta \phi'({\bf x})} \left\{
        \left( - \frac{\hbar^2\nabla^2}{2m} + V^{ex}({\bf x}) - \mu
               + \frac{8\pi a\hbar^2}{m} n({\bf x},t) \right) \phi'({\bf x})
               \right. \nonumber \\
&& \hspace{1.3in}   \left. 
               + \frac{4\pi a\hbar^2}{m} n_0({\bf x},t) e^{2i\theta({\bf x},t)}
        \phi'^*({\bf x}) \right\} P_1[\phi'^*,\phi';t]           \nonumber \\
&& \hspace{-1.0in}
+ \int d{\bf x}~
  \frac{\delta}{\delta \phi'^*({\bf x})} \left\{
        \left( - \frac{\hbar^2\nabla^2}{2m} + V^{ex}({\bf x}) - \mu
               + \frac{8\pi a\hbar^2}{m} n({\bf x},t) \right) \phi'^*({\bf x})
               \right. \nonumber \\
&& \hspace{1.3in}   \left. 
               + \frac{4\pi a\hbar^2}{m} n_0({\bf x},t) e^{-2i\theta({\bf x},t)}
        \phi'({\bf x}) \right\} P_1[\phi'^*,\phi';t]~.        
\end{eqnarray}
This implies that, in contrast to the Hartree-Fock treatment, the equation of 
motion for $\langle \phi'({\bf x}) \phi'^*({\bf x}') \rangle(t)$ is coupled to 
the equation of motion of the so-called anomalous average
$\langle \phi'^*({\bf x}) \phi'^*({\bf x}') \rangle(t)$. In the framework of a 
qradient expansion, these equations can however be decoupled by performing a 
local Bogoliubov transformation \cite{bog}. In this manner we can then show that 
the Wigner distribution for the Bogoliubov quasiparticles obeys the expected 
collisionless Boltzmann equation  
\begin{equation}
\frac{\partial}{\partial t} N({\bf x},{\bf k};t) 
  + \frac{\partial \omega({\bf x},{\bf k};t)}{\partial {\bf k}} \cdot
      \frac{\partial}{\partial {\bf x}} N({\bf x},{\bf k};t)
  - \frac{\partial \omega({\bf x},{\bf k};t)}{\partial {\bf x}} \cdot
      \frac{\partial}{\partial {\bf k}} N({\bf x},{\bf k};t) = 0~,
\end{equation}     
with the dispersion given by
\begin{equation}
\hbar\omega({\bf x},{\bf k};t) =
  \sqrt{\epsilon^2({\bf k}) 
            + \frac{8\pi a\hbar^2}{m} n_0({\bf x},t) \epsilon({\bf k})}
  + \hbar {\bf k} \cdot {\bf v}_s~.
\end{equation}
Note that this shows that the distribution function $N({\bf x},{\bf k};t)$ gives 
the number of Bogoliubov quasiparticles with momentum $\hbar {\bf k}$ at 
position ${\bf x}$ in the frame in which the superfluid velocity vanishes 
\cite{baym,bob}. Note also that the above derivation of the Bogoliubov theory is 
particle-number conserving since the probability distribution $P[\phi^*,\phi;t]$ 
is explicitly $U(1)$ invariant. It is therefore equivalent to the approach 
recently put forward by Gardiner \cite{crispin}, and Castin and Dum 
\cite{yvan3}. More important for our purposes is that the coupled equations for 
$N({\bf x},{\bf k};t)$, $n_0({\bf x},t)$, and ${\bf v}_s({\bf x},t)$ can again 
be shown to fulfill the Kohn theorem exactly. We therefore believe that they 
give an accurate description of the collisionless modes of a trapped Bose 
condensed gas, that may be able to explain the existing discrepancy between 
theory and experiment. 

As a first step towards this goal we here consider the time-dependent 
Hartree-Fock theory for the collective modes of the Bose condensed gas, in which 
the collisionless Boltzmann equation in Eq.~(\ref{clqBe}) is coupled to the 
nonlinear Schr\"odinger equation in Eq.~(\ref{nlse}). It turns out that a 
numerical determination of the eigenmodes of the collisionless Boltzmann 
equation is rather difficult. Therefore, we have recently proposed a variational 
approach to solve these coupled equations \cite{michiel2}. Denoting the 
anisotropic harmonic trapping potential used in the experiments 
\cite{coll1,coll2} by $V^{ex}({\bf x}) = \sum_j m\omega_j^2 x_j^2/2$, we first 
of all take for the condensate wavefunction the gaussian {\it ansatz}
\begin{equation}
\langle \phi({\bf x}) \rangle(t) = \sqrt{N_0} 
  \prod_j \left( \frac{1}{\pi q_j^2(t)} \right)^{1/4} 
    \exp \left\{ - \frac{x_j^2}{2q_j^2(t)} 
                      \left( 1 - i \frac{m q_j(t)}{\hbar}
                                       \frac{d q_j(t)}{dt} \right)
         \right\}~,
\end{equation}  
with the functions $q_j(t)$ describing the `breathing' of the condensate in 
three independent directions. Clearly, this is the anisotropic generalization of 
the gaussian {\it ansatz} that we used in Sec.~\ref{neg} to study the stability 
of a condensate with negative scattering length. Although it is not immediately 
obvious that such an {\it ansatz} is also appropriate in the strong-coupling 
regime of interest here, it has been shown by Perez-Garcia {\it et al.} 
\cite{peter3} that at zero temperature the assumption of a gaussian density 
profile nevertheless leads to rather accurate results for the frequencies of the 
collective modes. We therefore anticipate that at nonzero temperatures the same 
will be the case.

Next we also need an {\it ansatz} for the distribution function 
$N({\bf x},{\bf k};t)$ of the thermal cloud. In Ref.~\cite{michiel2} we 
considered the gaussian
\begin{eqnarray}
N({\bf x},{\bf k};t) = N' \prod_j 
    \left( \frac{\hbar\omega_j}{k_BT \langle \alpha_j \rangle^2} \right)
                                                 \hspace*{3.5in} \nonumber \\
 \times \prod_j \exp \left\{ - \frac{1}{k_BT \langle \alpha_j \rangle^2}
               \left[ \frac{\hbar^2 \alpha_j^2(t)}{2m}
                        \left( k_j - \frac{m x_j}{\hbar\alpha_j(t)}
                                     \frac{d \alpha_j(t)}{dt} \right)^2
                      + \frac{m\omega_j^2}{2} \frac{x_j^2}{\alpha_j^2(t)}
               \right]
             \right\}~,
\end{eqnarray}
where $\alpha_j(t)$ are again three scaling parameters, $N'=N-N_0$ is the number 
of noncondensed particles, and $\langle \alpha_j \rangle$ denotes the 
equilibrium value of $\alpha_j(t)$ that in general is greater than $1$ due to 
the repulsive interactions between the atoms. Here, we however also consider the 
more accurate Bose distribution
\begin{eqnarray}
N({\bf x},{\bf k};t) = N' \frac{1}{\zeta(3)} \prod_j 
    \left( \frac{\hbar\omega_j}{k_BT \langle \alpha_j \rangle^2} \right)
                                                 \hspace*{3.4in} \nonumber \\
 \times \left(
        \exp \left\{ \sum_j \frac{1}{k_BT \langle \alpha_j \rangle^2}
               \left[ \frac{\hbar^2 \alpha_j^2(t)}{2m}
                        \left( k_j - \frac{m x_j}{\hbar\alpha_j(t)}
                                     \frac{d \alpha_j(t)}{dt} \right)^2
                      + \frac{m\omega_j^2}{2} \frac{x_j^2}{\alpha_j^2(t)}
               \right]
             \right\} - 1 \right)^{-1}~,
\end{eqnarray}
with $\zeta(3) \simeq 1.202$. In both cases the functional form of the 
distribution function is physically motivated by the following reasons. First, 
by integrating over the momentum $\hbar{\bf k}$ we see that the time-dependent 
density profile of the noncondensed cloud $n'({\bf x},t)$ is a time-independent 
function of $x_j/\alpha_j(t)$, as desired. Second, for such a density profile we 
must have that the local velocity $\hbar\langle k_j \rangle({\bf x},t)/m$ is 
given by $(x_j/\alpha_j(t)) d\alpha_j(t)/dt$. Third, we want the equilibrium 
distribution to be only anisotropic in coordinate space, and not in momentum 
space. This explains the appearance of the factor $1/\langle \alpha_j \rangle^2$ 
in the exponent.

\begin{figure}
\psfig{figure=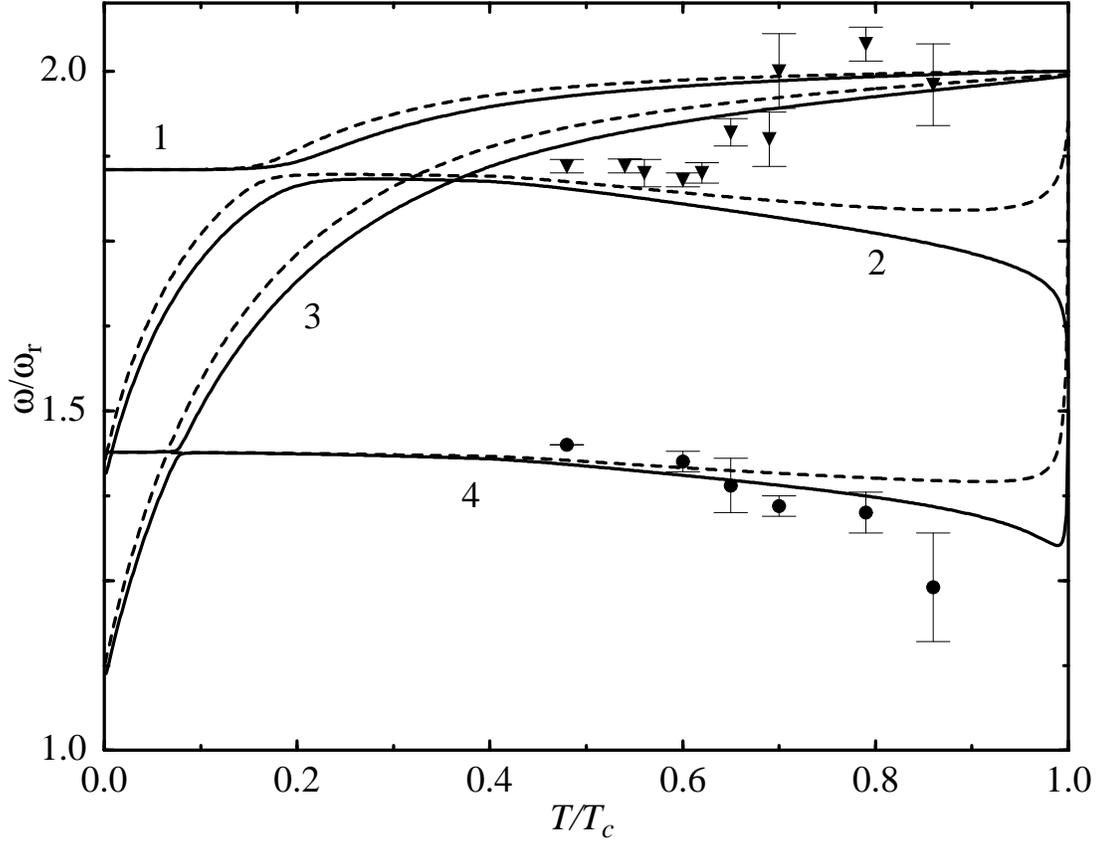}
\caption{Collsionless modes in a Bose condensed $^87$Rb gas. The dashed lines
         give the results with a gaussian {\it ansatz} for the distribution 
         function of the thermal cloud and the full lines for a Bose 
         distribution {\it ansatz}. Curves 1 and 2 correspond to the $m=0$ in 
         and out-of-phase modes, respectively. Similarly, curves 3 and 4 give 
         the frequencies of the $m=2$ in and out-of-phase modes. The 
         experimental data is also shown. Triangles are for a $m=0$ mode and 
         circles for a $m=2$ mode. 
         \label{Rb87}}
\end{figure}

Using the ideal gas result for the number of condensate particles and 
substituting the above {\it ansatz} for the condensate wavefunction and the 
Wigner distribution function into the nonlinear Schr\"odinger and collisionless 
Boltzmann equations, it is straightforward but somewhat tedious to derive the 
six coupled equations of motion for the `breathing' parameters $q_j(t)$ and 
$\alpha_j(t)$. Linearizing these equations around the equilibrium solutions 
$\langle q_j \rangle$ and $\langle \alpha_j \rangle$, respectively, we can then 
also obtain the desired eigenmodes and eigenfrequencies. The results of this 
calculation for the trap parameters of Jin {\it et al.} \cite{coll1} are shown 
in Fig.~\ref{Rb87} together with the experimental data. Note that the trap has a 
disk geometry with $\omega_x = \omega_y \equiv \omega_r$ and 
$\omega_z = \sqrt{8} \omega_r$. Due to the axial symmetry, the modes of the gas 
can be labeled by the usual quantum number $m$ that describes the angular 
dependence around the $z$-axis. On the whole there appears to be a quite 
reasonable agreement between this simple theory and experiment. The only 
exceptions are the two data point almost halfway in between the in and 
out-of-phase $m=0$ modes. A possible explanation for this discrepancy is that 
experimentally both modes are excited simultaneously \cite{eric}. To make sure 
of this, however, we need to determine the oscillator strength of the various 
modes. Work in this direction is in progress and will be presented in a future 
publication.

\section{DISCUSSION AND CONCLUSIONS}
\label{conc}
The purpose of this paper has been twofold. First, we have derived a single 
Fokker-Planck equation for the nonequilibrium dynamics of a weakly interacting 
Bose gas. This Fokker-Planck equation is capable of describing simultaneously 
the coherent and incoherent processes taking place in the gas and can be used 
for homogeneous as well as inhomogeneous situations. It can be derived in a very 
natural way in the Schwinger-Keldysh formalism, to which we have therefore also 
tried to give an introduction. The Fokker-Planck equation is here primarily used 
to study the dynamics of Bose-Einstein condensation in homogeneous and 
inhomogeneous atomic Bose gases. In particular, by treating both the 
weak-coupling and strong-coupling limits of the theory, we have been able to 
consider not only trapped atomic Bose gases with effectively attractive but also 
with effectively repulsive interactions, which leads to a better understanding 
of the fundamental difference between the physics in these two cases. Although 
we have only applied the Schwinger-Keldysh formalism to atomic Bose gases here, 
it can also be used to study nonequilibrium problems in trapped Fermi gases. An 
example is the study of the timescale for the formation of Cooper pairs in 
atomic $^6$Li \cite{marianne2}, which due to the anomalously large and negative 
value of the interatomic scattering length for this species, is of considerable 
interest at present. 

Second, and maybe most importantly, with our Fokker-Planck equation we have been 
able to make some theoretical progress on three topics in the field of Bose 
condensed atomic gases where there exists at present a discrepancy between 
theory and experiment. We refer here to the experiments on the formation of the 
condensate, on the spectrum of the collisionless modes and on the collapse of a 
condensate with negative scattering length. The main difficulty with the first 
topic is that the experiments are performed in the strong-coupling limit. In the 
language of quantum optics this implies that we are dealing with a multi-mode 
situation. Although this complicates the theory considerably, it also makes the 
physics involved much richer because coherent interaction effects now compete 
with the incoherent collisions. In this regime we may therefore hope to study in 
some detail the formation of topological defects in the wake of a phase 
transition, which turns out to be an important problem in cosmology as well 
\cite{james}. Moreover, recent experiments have shown that by a sinusoidal 
modulation of the trapping potential it is possible to adiabatically cycle 
through the phase transition \cite{wolfgang3}. A theoretical description of this 
experiment is easily obtained within our strong-coupling formalism. We only have 
to replace $V^{ex}({\bf x})$ by $V^{ex}({\bf x},t)$. 

The second discrepancy between theory and experiment is associated 
with the eigenfrequencies of the collisionless modes of a Bose condensed gas at 
nonzero temperatures, when also a considerable density of noncondensed atoms is 
present in the trap. The most accurate theory that has recently been applied to 
this problem, is an approximate many-body T-matrix calculation in which the 
noncondensate part of the gas is assumed to have no dynamics \cite{keith5}. Due 
to this last feature the modes of the gas are essentially the modes of a 
condensate in the effective potential 
$V^{ex}({\bf x}) + 2n'({\bf x}) T^{(+)}({\bf 0},{\bf 0},{\bf 0};0)$, which near 
the center of the trap has effectively smaller harmonic oscillator frequencies 
than the external trapping potential. Therefore, the Kohn theorem is not 
satisfied. As we have shown, this problem can be overcome if the dynamics of the 
noncondensed part of the gas is determined by an appropriate collisionless 
Boltzmann equation. Furthermore, by including the collision terms we can also 
find the damping of the long-wavelength modes of interest in a similar way as 
Kavoulakis, Pethick, and Smith have done for the normal state of the gas 
\cite{chris}.

The last theoretical challenge that we have touched upon is an accurate 
determination of the number of atoms that remain after a single collapse of a 
condensate with negative scattering length. We have argued that this in 
principle requires the solution of the Gross-Pitaevskii equation in combination 
with a rate equation for the number of condensate particles. In addition, we 
have indicated how we can find an approximate description of the dynamics of the 
collapse that systematically improves upon our gaussian results. As in the two 
previous problems, work is in progress to numerically investigate if the theory 
presented in this paper is indeed sufficiently accurate to resolve the 
discrepancies with experiment. Although we cannot be sure about the outcome of 
these studies at present, we believe that the many-body T-matrix approximation 
used throughout this paper is an excellent starting point in all three cases, 
and that our Fokker-Planck approach to the dynamics of weakly interacting Bose 
gases presents a convenient framework in which to derive in a straightforward 
way also the necessary corrections if they turn out not to be negligible.     

\section*{ACKNOWLEDGMENTS}
I am very grateful to Keith Burnett, Yvan Castin, Eric Cornell, Crispin 
Gardiner, Steve Girvin, Allan Griffin, Kerson Huang, Randy Hulet, Wolfgang 
Ketterle, Werner Krauth, Tony Leggett, Bill Phillips, Andrei Ruckenstein, Subir 
Sachdev, Cass Sackett, Doug Scalapino, Eite Tiesinga, Bernard de Wit, and Peter 
Zoller for illuminating comments and discussions on various topics reviewed in 
this paper.

\begin{references}
\bibitem{half} This follows from the requirement that we want to be able to
               Fourier transform the Heaviside function.
\bibitem{remark} In agreement with our previous remarks, the notation
                 $\int_{t_0}$ implies that the time integration
                 can be either from $t_0$ to $\infty$ or from $t_0$ to $t$.
\bibitem{gap} The magnitude of the frequency gap is actually not accurately
              described by this time-dependent Ginzburg-Landau theory, because 
              the action $S[\Delta^*,\Delta]$ is implicitly only valid for
              long wavelengths $\hbar v_F k \ll |\Delta_0|$ and low 
              frequencies $\hbar \omega \ll |\Delta_0|$.
\bibitem{top} For simplicity we here again neglect a topological term in the
              action, which essentially plays no role in the following. This is
              also in agreement with our discussion of the superconductor, 
              where exactly the same topological term was not included in the
              effective action for the global phase \cite{micheal}.
\end{references}
\end{document}